\documentclass[fleqn,usenatbib,useAMS]{mnras}

\usepackage{newtxtext,newtxmath}

\usepackage[T1]{fontenc}
\usepackage{ae,aecompl}

\usepackage{graphicx}	
\usepackage{amsmath}	

\usepackage{xcolor}

\newcommand{\hi}{H{\sc i}}

\title[SEDIGISM Cloud Catalogue]{The SEDIGISM survey: Molecular clouds in the inner Galaxy }

\author[Duarte-Cabral et al.]{A. Duarte-Cabral$^{1}$\thanks{E-mail: adc@astro.cf.ac.uk}, 
D. Colombo$^{2}$, J.\,S.\,Urquhart$^{3}$, A. Ginsburg$^{4}$, D. Russeil$^{5}$, 
\newauthor
F.\,Schuller$^{2}$, L.\,D.\,Anderson$^{6}$, P.\,J.\,Barnes$^{7,8}$, M.\,T.\,Beltr\'an$^{9}$, H.\,Beuther$^{10}$,
\newauthor
 S.\,Bontemps$^{11}$, L.\,Bronfman$^{12}$, T.\,Csengeri$^{11}$, C.\,L.\,Dobbs$^{13}$, D.\,Eden$^{14}$, 
\newauthor 
A.\,Giannetti$^{2}$, J.\,Kauffmann$^{15}$, M.\,Mattern$^{2}$, S.-N.\,X.\,Medina$^{2}$, K.\,M.\,Menten$^{2}$, \newauthor
M.-Y.\,Lee$^{2,16}$,
A.\,R.\,Pettitt$^{17}$, M.\,Riener$^{10}$, A.\,J.\,Rigby$^{1}$, A.\,Traficante$^{18}$, V.\,S.\,Veena$^{19}$, 
\newauthor
M.\,Wienen$^{2}$, F.\,Wyrowski$^{2}$,  C.\,Agurto$^{20}$, F.\,Azagra$^{20}$, R.\,Cesaroni$^{9}$, R.\,Finger$^{12}$, 
\newauthor
E.\,Gonzalez$^{20}$, T.\,Henning$^{10}$, A.\,K.\,Hernandez$^{21}$, J.\,Kainulainen$^{2,22}$, S.\,Leurini$^{2,23}$, 
 \newauthor
 S.\,Lopez$^{4}$, F.\,Mac-Auliffe$^{20}$, P.\,Mazumdar$^{2}$, S.\,Molinari$^{18}$, F.\,Motte$^{24}$, E.\,Muller$^{25}$,  
 \newauthor 
 Q.\,Nguyen-Luong$^{15}$, R.\,Parra$^{20}$, J.-P.\,Perez-Beaupuits$^{20}$, F.\,M.\,Montenegro-Montes$^{20}$,  
 \newauthor
 T.\,J.\,T.\,Moore$^{14}$, S.\,E.\,Ragan$^{1}$, A.\,S\'anchez-Monge$^{19}$, A.\,Sanna$^{2}$, P.\,Schilke$^{19}$,  \newauthor 
 E.\,Schisano$^{18}$, N.\,Schneider$^{19}$, S.\,Suri$^{19}$, L.\,Testi$^{19}$, K.\,Torstensson$^{20}$, P.\,Venegas$^{20}$, 
 \newauthor 
 K.\,Wang$^{26}$, and A.\,Zavagno$^{5}$ \\
\\
Affiliations can be found after the references.}

\date{Accepted 2020 May 22. Received 2020 May 22; in original form 2019 October 10.}

\pubyear{2020}

\begin{document}
\label{firstpage}
\pagerange{\pageref{firstpage}--\pageref{lastpage}}
\maketitle

\begin{abstract}

We use the $^{13}$CO\,(2-1) emission from the SEDIGISM (Structure, Excitation, and Dynamics of the Inner Galactic InterStellar Medium) high-resolution spectral-line survey of the inner Galaxy, to extract the molecular cloud population with a large dynamic range in spatial scales, using the Spectral Clustering for Interstellar Molecular Emission Segmentation ({\sc{scimes}}) algorithm. This work compiles a cloud catalogue with a total of 10663 molecular clouds, 10300 of which we were able to assign distances and compute physical properties. We study some of the global properties of clouds using a science sample, consisting of {6664} well resolved sources and for which the distance estimates are reliable. In particular, we compare the scaling relations retrieved from SEDIGISM to those of other surveys, and we explore the properties of clouds with and without high-mass star formation. Our results suggest that there is no single global property of a cloud that determines its ability to form massive stars, although we find combined trends of increasing mass, size, surface density and velocity dispersion for the sub-sample of clouds with ongoing high-mass star formation. We then isolate the most extreme clouds in the SEDIGISM sample (i.e. clouds in the tails of the distributions) to look at their overall Galactic distribution, in search for hints of environmental effects. We find that, for most properties, the Galactic distribution of the most extreme clouds is only marginally different to that of the global cloud population.  The Galactic distribution of the largest clouds, the turbulent clouds and the high-mass star-forming clouds are those that deviate most significantly from the global cloud population. We also find that the least dynamically active clouds (with low velocity dispersion or low virial parameter) are situated further afield, mostly in the least populated areas. However, we suspect that part of these trends may be affected by some observational biases (such as completeness and survey limitations), and thus require further follow up work in order to be confirmed.

\end{abstract}




\section{Introduction}
\label{sec:intro}

The evolution of the gas that makes up the interstellar medium (ISM), and the ultimate means by which that gas gives way to star formation, involve the tight interplay of a wealth of physical processes. Our understanding of those processes has relied upon the statistical characterisation of the molecular gas that is taking part in the star formation process. In particular, the star formation field has relied on a discretisation of the molecular component of the ISM into molecular clouds, across the Galactic disc, either as observed in 2D with dust continuum emission (e.g. the ATLASGAL survey, \citealt{schuller2009}; the Hi-GAL survey, \citealt{Molinari2010}; or the Bolocam Galactic Plane Survey, \citealt{rosolowsky2010}, \citealt{Ginsburg2013}), or with the 3D view of the Galactic plane from spectral-line observations, most commonly using the second-most abundant molecular species in the ISM, the CO molecule (and its isotopologues). Large survey observations of the Galactic plane in CO emission have allowed for a number of statistical studies of molecular clouds across the Galaxy \citep[e.g.][]{Scoville1975,Larson1981,Solomon1987,Heyer2009,Roman-Duval2010,Rice2016,miville-deschenes2017}, and have provided a large-scale view of the distribution of gas in the Milky Way, crucial for our understanding of its spiral structure \citep[e.g.][]{dame2001,vallee2014c,Pettitt2014,Pettitt2015}. 

These Galactic plane surveys, alongside some resolved studies of molecular clouds in nearby spiral galaxies, have also suggested a number of scaling relations \citep[namely between the sizes of clouds, their line-widths, and their mass surface densities, e.g.][]{Larson1981,Solomon1987,Heyer2009,Sun2018}, as well as some differences in the mass spectra of clouds towards different environments \citep[e.g.][]{Colombo2014,Rice2016,miville-deschenes2017}. All of these findings have implications in our interpretation of the global properties of molecular clouds, and how they might evolve. Most of these surveys, however, were finding and describing molecular clouds that had typical sizes close to their resolution element - which can bias the interpretation of the results - and given their lower resolution they could also potentially suffer from severe blending of the emission along the same line of sight, especially with our edge-on perspective of the Milky Way \citep[e.g.][]{dc2015,dc2016}.

With the advent of new high-resolution and large-scale spectroscopic surveys of the Galactic plane (such as the Structure, Excitation, and Dynamics of the Inner Galactic InterStellar Medium survey - SEDIGISM, \citealt{Schuller2017}; the CO High Resolution Survey - COHRS, \citealt{Dempsey2013}; the $^{13}$CO/C$^{18}$O (J=3-2) Heterodyne Inner Milky Way Plane Survey - CHIMPS, \citealt{rigby2016}; the Three-mm Ultimate Mopra Milky Way Survey - ThrUMMS, \citealt{barnes2015}; or the Galactic Census of High and Medium-mass Protostars - CHaMP \citealt{Barnes2011}), not only are these shortcomings now greatly minimised, but we can start to explore the details of the sub-structure within molecular clouds where star formation is actively taking place, and the clouds' link to the large-scale Galactic environment. This opens a new and exciting era in the study of star formation in a Galactic context. Given that molecular clouds are highly hierarchical systems, it is essential to be able to define molecular clouds with a large dynamic range in spatial scales \citep[e.g. as in][]{Colombo2019}, and this is at the heart of this present work. In this paper, we explore the global properties of molecular clouds from the high-resolution $^{13}$CO (2-1) emission from the SEDIGISM survey, covering the inner Galactic plane \citep[from {+300}\degr\, $\leq$ $\ell$ $\leq$ +18\degr,][]{Schuller2017}, which is described in Sect.\,\ref{sec:data}. Section\,\ref{sec:cloud_extraction} contains the details of the method used for the extraction of molecular clouds from this dataset, along with a description of all the derived properties and data-products released with the molecular cloud catalogue. 
In Sect.\,\ref{sec:distances} we describe the methods used to determine the distances and distinguish between derived near/far kinematic distances to all the clouds in the catalogue, essential to derive the physical properties.  In Sect.\,\ref{sec:global_properties} we explore the distributions of the global properties of the SEDIGISM clouds, and also compare these with other samples in the literature. In Sect.\,\ref{sec:extreme}, we explore possible indications of environmental dependency of cloud properties, by isolating the most extreme clouds (i.e. clouds in the tails of the distributions), and comparing their Galactic distribution with that of the entire cloud population. Finally, our findings are summarised in Sect.\,\ref{sec:summary}.


\section{Data}
\label{sec:data}

In this paper, we use data from the SEDIGISM survey conducted with the Atacama Pathfinder Experiment 12\,m submillimetre telescope \citep[APEX,][]{Guesten2006}. In particular, we use the $^{13}$CO (2-1) to extract and characterise the molecular clouds towards the inner Galaxy. The complete details on the observations, data reduction and data-quality checks can be found in the survey overview papers \citep[]{Schuller2017,Schuller2019}. 

In summary, the SEDIGISM survey observed a total of 84\,deg$^2$, covering from $-$60\degr\, $\leq$ $\ell$ $\leq$ +18\degr, and $\vert b \vert$ $\leq$ 0.5\degr, plus a few extensions in $b$ towards some regions, as well as an additional field towards the W43 region (+29\degr\, $\leq$ $\ell$ $\leq$ +31\degr). The $^{13}$CO\,(2-1) data that we use here is the DR1 dataset \citep[fully described in][]{Schuller2019}, which has a typical $1\sigma$ sensitivity of 0.8--1.0\,K (in $T_{\rm{mb}}$) per 0.25\,km\,s$^{-1}$ channel, and a FWHM beam size, $\theta_{\rm MB}$, of 28\arcsec.

\begin{figure*}
\centering
\includegraphics[width=\textwidth]{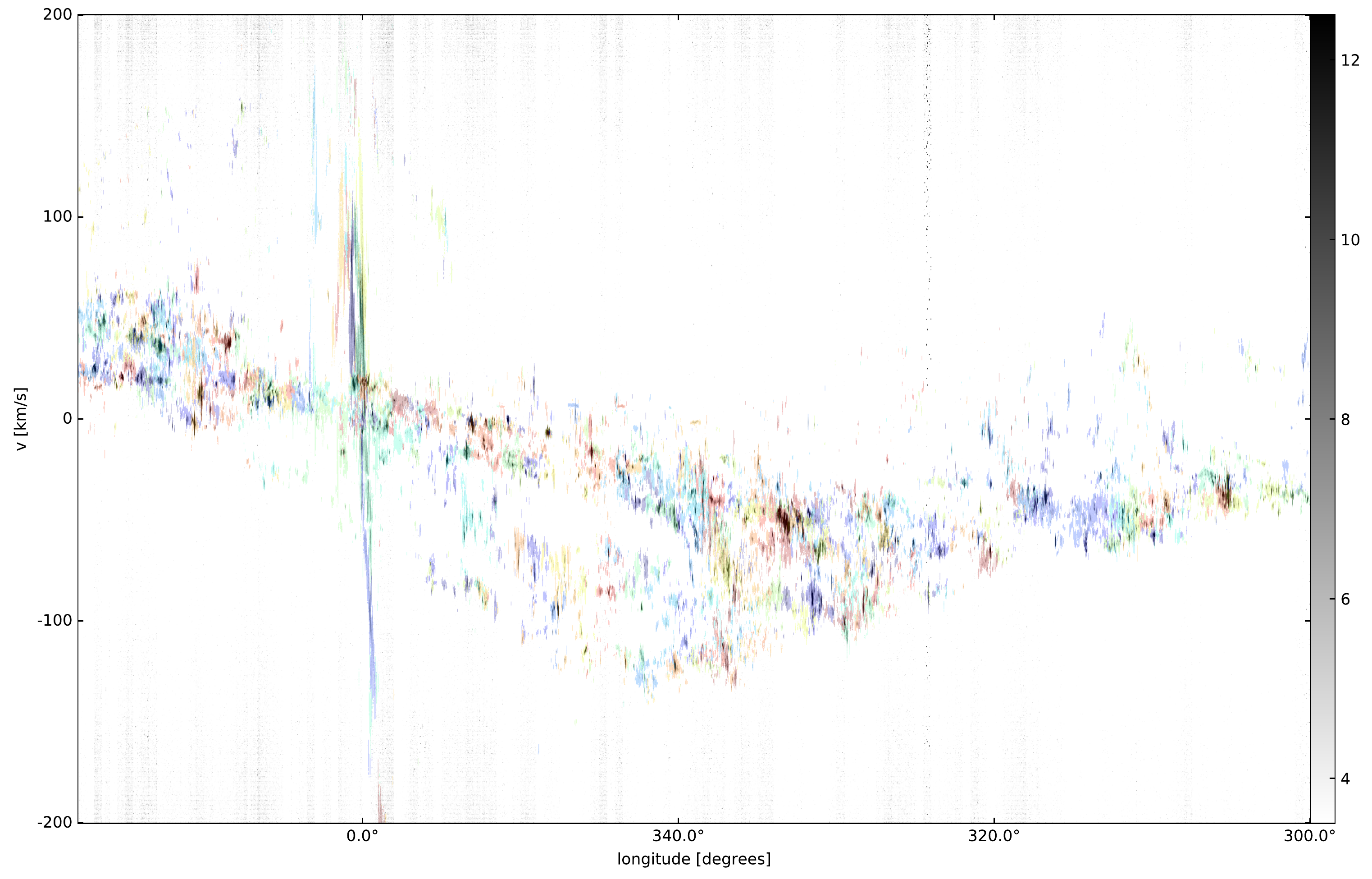}
\vspace{-0.5cm}
\caption{Longitude-velocity ($\ell v$) map of the $^{13}$CO peak intensity (in greyscale) for the SEDIGISM coverage analysed in this paper. 
The peak intensity map was built after masking out voxels of the  $^{13}$CO datacube with intensities $< 2.5 \sigma_{\rm rms}$ (estimated locally for each line of sight). 
The clouds extracted with {\sc scimes} are overlaid as colours, 
where each cloud has a different (random) colour. }
\label{fig:lv_all}
\end{figure*}

In this paper we will use the complete contiguous dataset (i.e. the entire survey data except for the W43 field). This consists of 77 datacubes of roughly $2\degr\,\times1\degr$ (note that the latitude range is sometimes larger than $1\degr$), centred at all integer longitudes between $\ell=301$\degr\, and $\ell=17$\degr\, (i.e. spaced by 1\degr\, in longitude). This provides a $1\degr\,$ overlap in longitude between consecutive tiles, which ensures all sight lines (except for the first and last fields) are contained in two tiles. The velocity ranges from $-200$ to $+200$\,km\,s$^{-1}$ in all datacubes, and the pixel size is of 9.5\arcsec. Figure\,\ref{fig:lv_all} shows the full $\ell v$ map of the contiguous dataset from the SEDIGISM survey, that we use here.

\section{Molecular cloud extraction}
\label{sec:cloud_extraction}

\subsection{The method: SCIMES}

In order to decompose the $^{13}$CO emission from the SEDIGISM survey into discrete clouds, we use the {\sc scimes} algorithm (v.0.3.2)\footnote{https://github.com/Astroua/SCIMES}. The original algorithm is fully described in \citet[][]{Colombo2015}, and the improvements included in the version we use here are detailed in \citet[][]{Colombo2019}. In brief, {\sc scimes} brings a significant advancement with respect to other more commonly used cloud-extraction algorithms (e.g. Clumpfind by \citealt{Williams1994}, Gaussclumps by \citealt{Stutzki1990}, or Fellwalker by \citealt{Berry2015}), as it is a fully automated method that uses spectral clustering and graph theory to analyse the dendrogram of the emission, and decompose the hierarchical structure of the ISM into ``clusters'' of molecular gas emission (i.e. molecular clouds, considering the resolution of SEDIGISM). Unlike other cloud-extraction algorithms, {\sc scimes} relies on the natural transitions in the emission to define discrete structures, and it is robust against changes in the input parameters \citep[as demonstrated in][]{Colombo2015}.

The cloud extraction with {\sc scimes} was performed on each of the 77 tiles of $2\degr\,\times1\degr$. We ran {\sc scimes} on these relatively small cubes because it would be extremely computationally expensive (and memory intensive) to generate a single dendrogram from the full SEDIGISM dataset, and perform {\sc scimes}'s affinity matrix analysis, where each cluster is equivalent to an additional dimension in the clustering space. 

\subsubsection{Input parameters and files} 
\label{sec:input}

In order to optimise the performance of the {\sc scimes} clustering algorithm, we have performed a few preparation steps on the original DR1 data. Firstly, we enhanced the signal-to-noise ratio of the data set prior to running {\sc scimes} by smoothing the data in velocity. This was done by binning the data into 0.5\,km\,s$^{-1}$ channels. We then re-sampled these binned datacubes back into 0.25\,km\,s$^{-1}$ channels (using linear interpolation), simply so that the {\sc scimes} assignment masks (Sect.\,\ref{sec:dataproducts}) kept the same format as the original emission datacubes from DR1 (essential to have straight forward voxel-by-voxel match between the DR1 emission maps and the clouds' assignment masks). We have performed some tests on the science demonstration field \citep[][]{Schuller2017}, with binned and non-binned data, and this step allows us to remove high-frequency noise spikes, speeding up the dendrogram construction and the {\sc scimes} clustering, with minimal loss in the information retrieved.  

\begin{figure}
\centering
\includegraphics[width=0.45\textwidth]{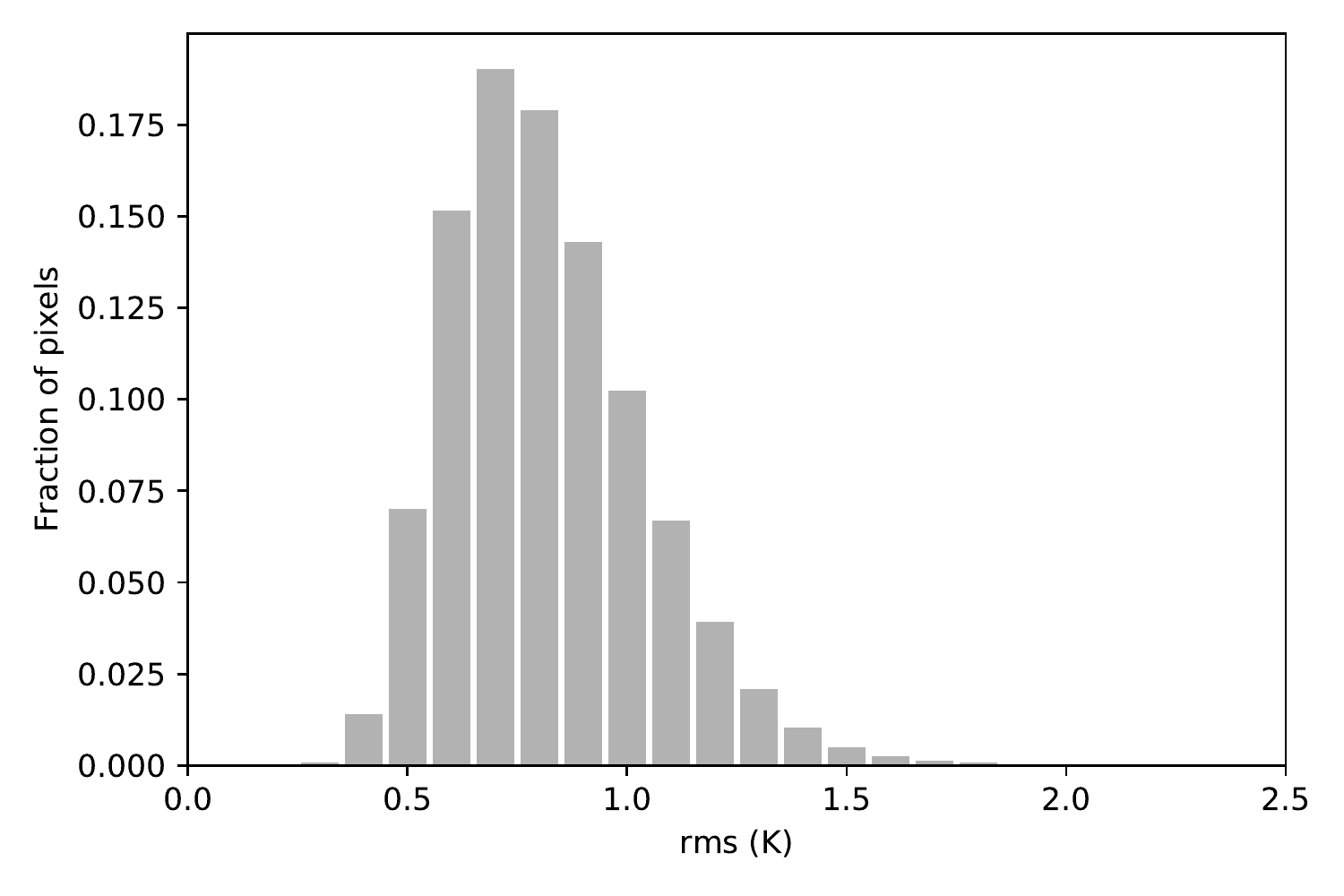}
\vspace{-0.5cm}
\caption{{Histogram of the rms noise level {of} the entire survey, from the velocity-smoothed datacubes that we use for the {\sc scimes} extraction, showing that it peaks at $\sim$0.7\,K, {with a median value of 0.78\,K.}}}
\label{fig:hist_rms}
\end{figure}

Secondly, given that the noise in the survey is not perfectly uniform (due to different observing weather conditions), it is also essential to mask the datacubes using the local noise level, in order to prevent high-noise regions from being used in the dendrogram tree, and incorrectly identified as clouds. 
For this purpose, we estimated the local noise level at each pixel (i.e. each line of sight) in the velocity-smoothed datacubes, by taking the first 50 channels (which are line-free, and on the high-frequency end, i.e. at negative systemic velocities), and computing the $1\sigma$ standard deviation. {Figure \ref{fig:hist_rms} shows the distribution of this local $1\sigma$-rms noise level for all the pixels in our velocity-smoothed dataset, showing that it peaks at $\sim$0.7\,K.}
We then create a mask of each datacube, by setting any 3D pixels (voxels) whose emission is lower that 2$\sigma$ of the local noise to zero. Note that since we already go down to 2$\sigma$ of the local noise, we do not perform any dilation of the masks after this step (which is a technique sometimes used to remove potential breaks in clouds in low signal-to-noise areas).

Using these masked datacubes, we computed the dendrogram tree of the 3D structures in the data \citep[using the {\sc astrodendro}\footnote{http://www.dendrograms.org} implementation, which is based on the original IDL procedures from][]{Rosolowsky2008}. The dendrogram is composed of three types of structures: {\it leaves}, which are at the top of the hierarchy and contain no substructure, i.e. they are associated with local peaks of emission; {\it branches}, which split into multiple substructures; and the {\it trunk}, which is at the bottom of the hierarchy (i.e. it has no parent structure), and comprises all {\it branches} and {\it leaves}. We built our dendrograms using the same input parameters as in the science demonstration field \citep[][]{Schuller2017}: we considered {a} noise level ($\sigma_{\rm rms}$) of 0.7\,K for all tiles ({corresponding to the peak of the noise distribution in Fig.\,\ref{fig:hist_rms}}), a 4$\sigma_{\rm rms}$ value as the minimum difference between two peaks for them to be considered as separate structures, and a lower threshold for detection of 2$\sigma_{\rm rms}$, to maximise the connections between different structures at contiguous lower intensity levels\footnote{These values are solely defined by the data quality, but tests using slight variations for the different parameters for the dendrogram construction were performed as part of our work on the science demonstration field \citep[][]{Schuller2017}. Those tests have shown that the {\sc scimes} clustering algorithm is robust against small differences on the parameters used to construct the dendrogram.}. Note that we specifically chose to use a single fixed value of $\sigma_{\rm rms}$ to build the dendrograms across the entire survey (rather than using a local signal-to-noise ratio approach) so that we could define our structures using a uniform {criterion throughout}. This not only makes it easier to replicate our results using other datasets, but it also ensures that the type of structures we extract are equivalent throughout the entire survey, and not dependent on the local noise conditions\footnote{Although the choice of a unique value for $\sigma_{\rm rms}$ does require an extra post-processing step to ensure that detected structures also have a high local signal-to-noise ratio (see Sec.\,\ref{sec:spurious}), doing the cloud extraction on a signal-to-noise map with the {\sc scimes} algorithm (which does not segment clouds at a fixed brightness threshold) means that we could end up with structures identified across the survey using different criteria, which is non-ideal.}. {The choice of a 4$\sigma_{\rm rms}$ for the significance of individual peaks, coupled with the fact that the dendrogram was built from datacubes that had been masked based on the pixel-based noise level, was so as to maximise the retained detailed information encoded in the survey, whilst minimising the inclusion of noise spikes.}
In addition, we set a minimum number of voxels for a structure to be considered as real to be 6 times the number of pixels per beam ($N_{\rm{ppbeam}} = 9$), so that structures are both resolved spatially (i.e. at least 3 beams), and in velocity (spanning at least 2 channels, which corresponds to our effective velocity resolution in the smoothed datacubes). Note, however, that the {\sc astrodendro} implementation that we use to build the dendrogram does not separate the spatial axes from the spectral axis. In practice, this means that this criterion will still allow some clouds to be retained whilst being unresolved in one of the axes. Those sources are dealt with in a post-processing step (see Sect.\,\ref{sec:spurious}). 

Once the dendrograms were constructed for each tile, we ran {\sc scimes} using both the ``volume'' and the ``flux'' (which in our case, refers simply to the surface brightness) as the clustering properties \citep[cf.][]{Colombo2015}. This extraction recovered a total of 20387 gas clusters from the 77 tiles, but most of these are duplicated due to the overlap between consecutive fields. In order to build the final catalogue (and respective assignment masks), we performed a cleaning up procedure to handle clouds in overlapping areas. This is described in the following section.

\subsubsection{Handling clouds in overlapping regions}

\begin{figure}
\centering
\includegraphics[width=0.45\textwidth]{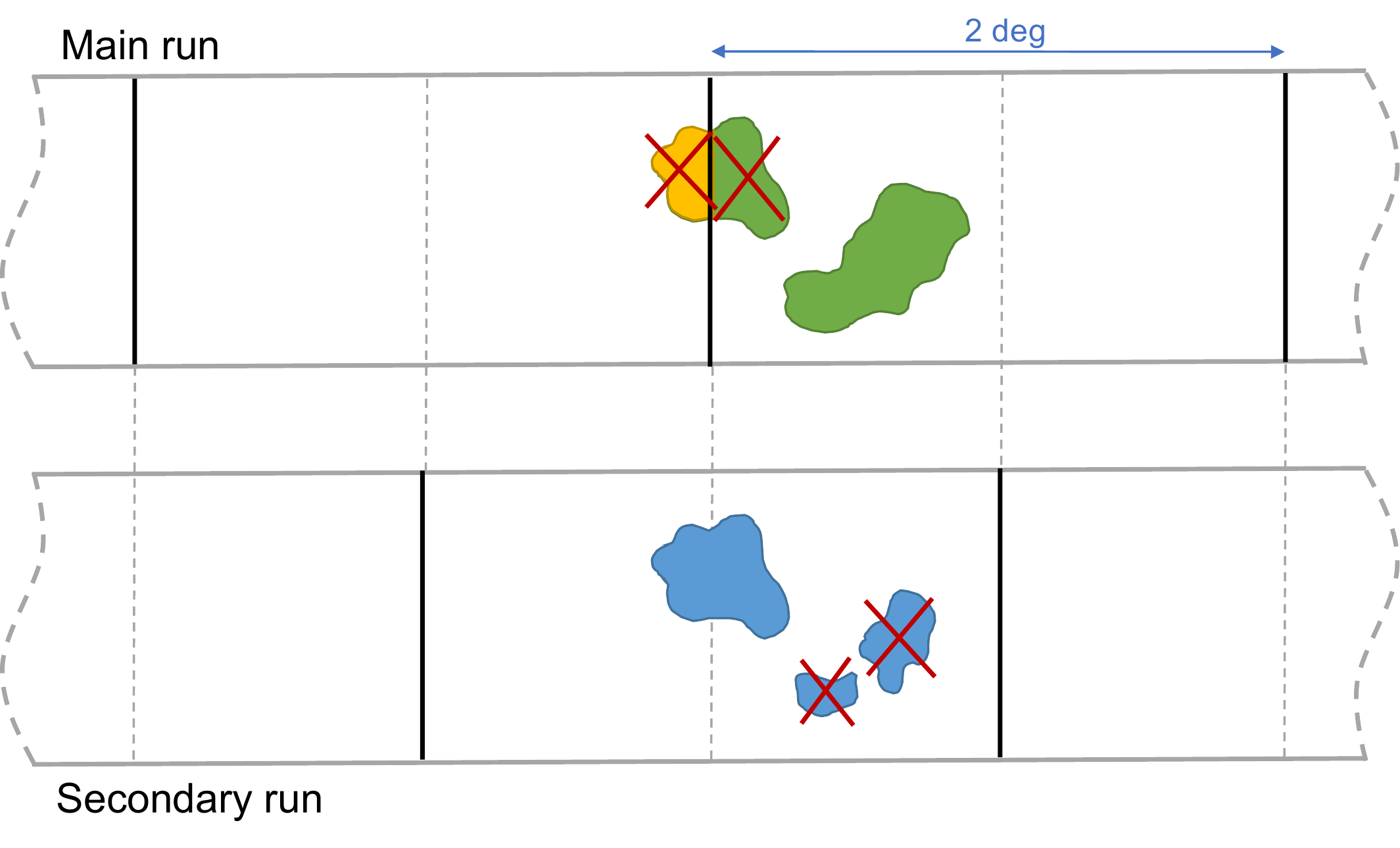}
\caption{Schematic sketch of the procedure used to decide which clouds to retain from the two overlapping runs, namely the removal of clouds that touch the edge of the tiles (which are then recovered in full in the complementary run), and the selection of the larger clouds for overlapping cases between the two runs.}
\label{fig:scheme_clouds}
\end{figure}

In order to handle the clouds that appear in overlapping areas, we have followed a procedure similar to that used by \citet[][]{Colombo2019}. This procedure is schematically described in Fig.\,\ref{fig:scheme_clouds}. In essence, we have split our dataset into a {\it main} run (which is composed of all tiles centred at odd longitudes), and a {\it secondary} run (which is composed of all tiles centred at even longitudes). We then exclude all objects that touch a tile edge on the longitude axis, since their contours are not closed, and they should be fully recovered in the complementary run. We only made an exception for objects that touch the first and last longitude edges of the contiguous coverage (i.e. at $\ell=18$\degr\, and $\ell=300$\degr), which are retained in the final catalogue with a tag that indicates that they are edge clouds. Similarly, we also retain clouds that touch the survey's upper and lower latitude edges, and tag them as being edge clouds. Finally, we proceed to checking the matches between the {\it main} and {\it secondary} runs. We start by including all objects that do not overlap between the two runs, and whenever two (or more) clouds overlap, we simply retain the larger object between the two runs.  After this procedure, we have compiled a total of 11638 unique molecular clouds.

\begin{figure*}
\centering
\includegraphics[width=0.9\textwidth]{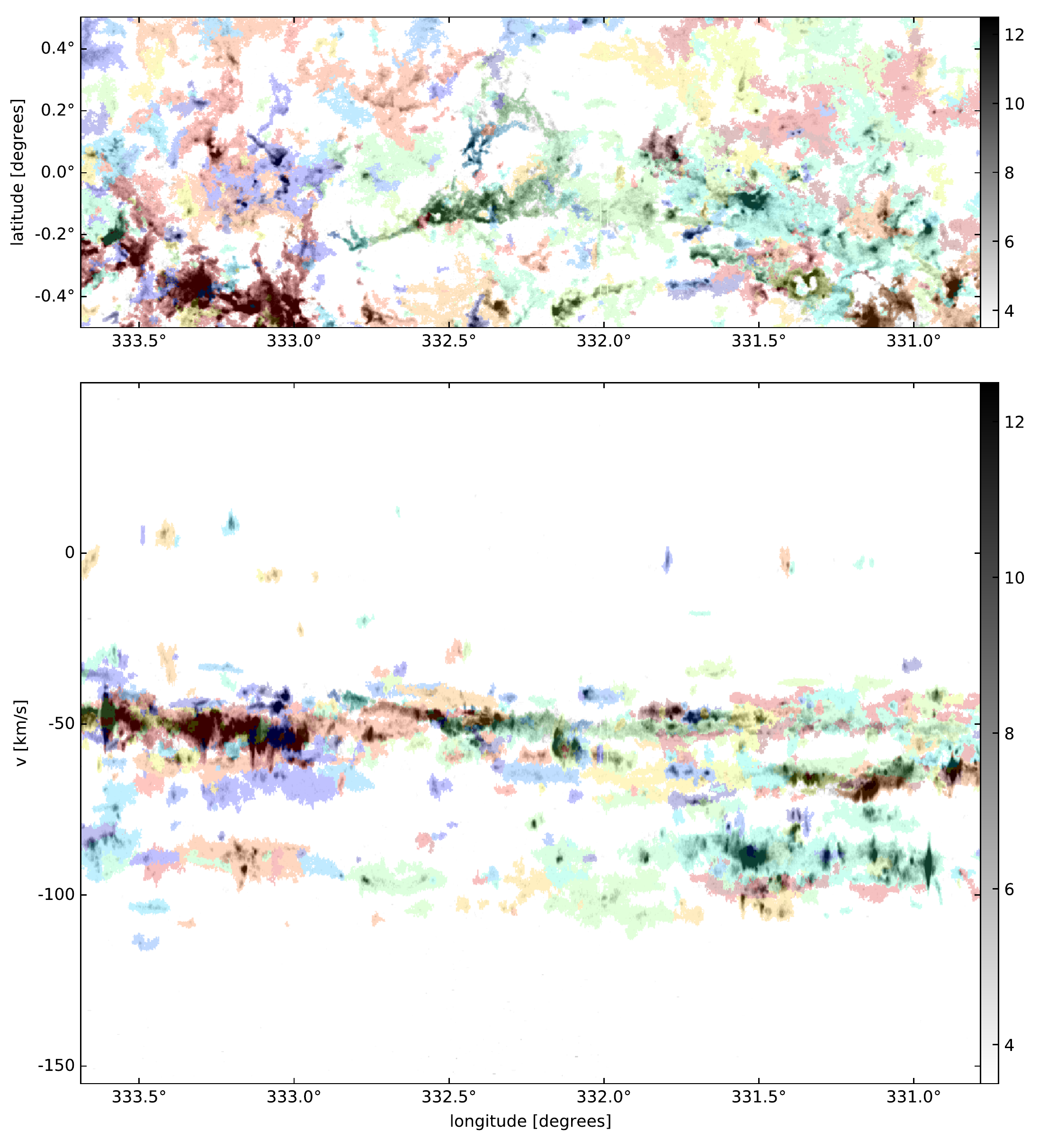}
\caption{Example of the {\sc scimes} cloud extraction results, on a small section of the SEDIGISM survey. The top panel shows the $\ell b$ map with the $^{13}$CO peak intensity in greyscale, and the SEDIGISM cloud masks overlaid as colours, where each cloud has a different (random) colour. The bottom panel shows the $\ell v$ map of the same field, with the same colour-scheme as the top panel.}
\label{fig:lb_lv_zoom}
\end{figure*}

\subsubsection{Removal of spurious sources}
\label{sec:spurious}

As mentioned in Sect.~\ref{sec:input}, despite our best efforts to avoid having  any noisy spikes in the dendrogram (by imposing a noise level threshold) or unresolved sources (by imposing a minimum number of voxels), some spurious sources still persist to the dendrogram construction and into our final catalogue. One of the reasons for this is the fact that we have applied an average noise $\sigma_{\rm rms}$ for the entire survey (so that the dendrogram for all fields was built upon a fixed physical value of emission intensity). This means that in areas where the local noise level is higher than this average $\sigma_{\rm rms}$, some noisy peaks would have been considered as robust emission peaks. Most of these sources are located near the noisier edges of the observed fields, and are relatively small (close to the beam size). We therefore applied the following selection criteria to remove spurious sources from the final catalogue: 1) any source touching an edge that has a projected (footprint) size of less than 5 beams\footnote{This size was determined by inspecting the datacubes. Unlike in the middle of the map where the noisy spikes are of the order of a beam size, the noisy spikes in the edges are typically much larger than a beam size due to gridding/convolution of the data whilst doing the data reduction. We also consider that even if some real sources were to be included in this criterion, those clouds would be both small and incomplete (since they touch an edge), and therefore their properties would be highly unusable.} \citep[where the angular size of the beam is taken to be $\Omega_{\rm{mb}} = \theta_{\rm{mb}}^{2}\,\pi/(4 ln(2)) \approx 888$\,arcsec$^{2}$, e.g.][]{Kauffmann2008}; 2) any source whose projected footprint size is less than two beam sizes; and 3) any source whose signal-to-noise ratio was less than 3.5 (estimated by taking the peak of emission and comparing it to the local noise level). Most spurious sources were successfully removed with this set of criteria, but some remained, in particular towards the noisier high-velocity end of the spectrum ({at positive velocities,} which can be clearly seen on the $\ell v$ plots of Fig.\,\ref{fig:lv_all}, and in Figs.\,\ref{fig:lb_lv_0} to \ref{fig:lb_lv_4}). Therefore, we applied another criterion to our removal procedure: 4) any source outside the Galactic centre region (i.e. outside a {+355}$ < \ell < 10$\degr\, range), and with a centroid velocity $v_{\rm lsr} > 160$\,km\,s$^{-1}$. The resulting catalogue contains 10663 molecular clouds (whose masks are shown as colours in Fig.\,\ref{fig:lv_all}). 

{By comparing the integrated intensities inside the cloud masks with the total integrated intensities along each sight line, we estimate that the extracted clouds contain $\sim70\%$ of the total integrated flux above $3\sigma_{\rm{w}}$ \citep[similar to][]{Barnes2016}, and $\sim50\%$ of the flux above $2\sigma_{\rm{w}}$, where $\sigma_{\rm{w}}$ is the standard deviation of the total integrated intensity map, defined as $\sigma_{\rm{w}} = \sqrt{N_{c}} \sigma_{\rm{rms}} \Delta v$, with $N_{c}$ being the total number of channels used for the integration, $\sigma_{\rm{rms}}$ the average noise level per channel (i.e. 0.7\,K), and $\Delta v$ the channel width (i.e. 0.25\,km\,s$^{-1}$). This suggests there is a non-negligible amount of molecular gas in a relatively diffuse inter-cloud medium. In addition, from the datacubes with the cloud masks, we find that of all $\ell b$ pixels with clouds, we have $\sim$82$\%$ of sight lines with a single cloud assignment, meaning that only $\sim$18$\%$ of the lines of sight have multiple clouds ($\sim$16$\%$ with two clouds, $\sim$2$\%$ with three clouds, and $<$1$\%$ with more than three clouds). }

\subsection{Data products: Cloud masks and catalogues}
\label{sec:dataproducts}

From our {\sc scimes} extraction, we have produced two main data products: a catalogue with the properties of all the molecular clouds; and the respective assignment datacubes in the same format as the input 3D datacubes of emission. These data products are made publicly available alongside the data release of the survey\footnote{\url{http://sedigism.mpifr-bonn.mpg.de}}. 

In the assignment datacubes, each voxel holds the unique ID number of the cloud it has been assigned to by {\sc scimes}, and the voxels with no assigned cloud take the value $-1$. These assignment datacubes are particularly useful for performing further studies on specific clouds, as they can be used to assign voxels to clouds, and therefore pull out the entire 3D structure of clouds from the original emission datacubes. Figure\,\ref{fig:lb_lv_zoom} shows an example of the results from the cloud extraction towards a small portion of the survey, with the $^{13}$CO peak intensity map in greyscale, and the cloud masks overlaid as colours. In Appendix\,\ref{app:Catalogue_columns}, we show the same images for the entire survey coverage (from Fig.\,\ref{fig:lb_lv_0} to Fig.\,\ref{fig:lb_lv_4}).

All the properties held in the catalogue of molecular clouds produced whilst running {\sc scimes} are listed in Table\,\ref{tab:catalogue_tabs} (in App.\,\ref{app:Catalogue_columns}). In essence, the catalogue contains two sets of properties: the directly measured quantities, and the physical properties derived from these after a distance has been assigned (see Sect.\,\ref{sec:distances}). Note that all the quantities we present in the catalogue were estimated using the default ``bijection'' paradigm, which is the most appropriate for characterising substructures within the nested dendrogram tree \citep[][]{Rosolowsky2008}. Amongst the directly measured properties are the ID number, the cloud name, the clouds' centroid longitude ($\ell $), latitude ($b$), and velocity ($v$), the velocity dispersion ($\sigma_{v}$), the projected footprint area ($Area$) and the respective equivalent radius ($R$), the average integrated intensity ($<I_{\rm ^{13}CO}>$), and the peak intensity ($T_{\rm ^{13}CO}^{\rm peak}$). We also include a tag ($edge$) to indicate whether a cloud touches an edge of the survey coverage, in which case it is an incomplete object.

Given that some clouds will be close to the resolution element of our survey, a beam deconvolution on the sizes is needed. This will only affect the smaller objects, and has only very marginal effects on the statistical properties that we derive. Nevertheless, in the catalogue we also provide the equivalent radius deconvolved from the beam ($R^{d}$).

\begin{figure}
\centering
\includegraphics[width=0.45\textwidth]{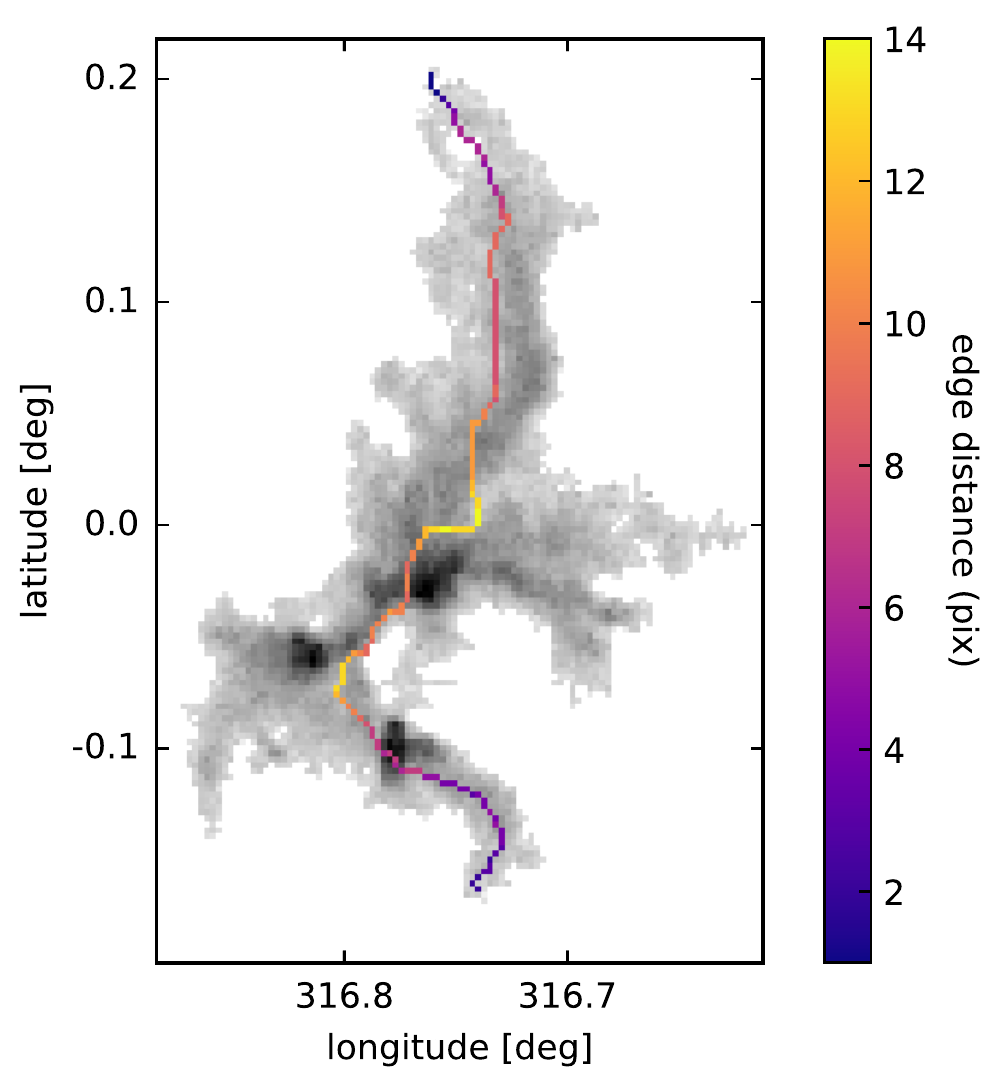}
\caption{Example of the medial axis for a molecular cloud in our sample (SDG316.766-0.020), which corresponds to an IRDC \citep[SDC316.786-0.044 from][]{peretto2009}, and the larger cloud often shortened to G316.75 \citep[e.g. studied in][]{Watkins2019}. The grey scale shows the $^{13}$CO\,(2-1) integrated intensity, estimated using the voxels within the cloud's mask as defined by {\sc{scimes}}, and the coloured pixels show the geometrical medial axis, colour-coded with the distance to the external cloud edge.}
\label{fig:med_axis}
\end{figure}

In addition to the properties already described, we also estimated some basic parameters to characterise the clouds' morphology. First, we estimated the projected semi-major and semi-minor axes from the second moment of the emission in 2D, weighted by the intensity ({\em major} and $minor$), along with the respective position angle ($PA$), and the aspect ratio ($AR_{\rm mom} = $ {\em major}$/minor$). However, this moment method is relatively limited in providing a good approximation of a cloud's morphology, and can easily underestimate the true aspect ratio. Therefore, we also determined the projected geometrical medial axis of the clouds, which is the longest running spine along the 2D-projected cloud's mask, which is farthest away from the external edges (any internal holes in the cloud's masks are filled before determining the medial axis). From that, we include in the catalogue also the medial axis length ($length_{\rm\,MA}$), as well as the medial axis width as being twice the average distance to the cloud edge ($width_{\rm\,MA}$), and the corresponding aspect ratio ($AR_{\rm\,MA} = length_{\rm\,MA}/width_{\rm\,MA}$). Figure\,\ref{fig:med_axis} shows an example of this medial axis for a cloud in our sample. Note that this is a purely geometrical medial axis (i.e. it is built on the assignment masks, with no information on the actual structure of the emission), and thus it is only a first approximation of the possible filamentary nature of clouds. A more accurate description of filamentary structures detected with ATLASGAL using the SEDIGISM survey data has been performed by \citet[][]{Mattern2018}, and shall be expanded to the entire SEDIGISM survey in future work.

The determination of the physical properties of the clouds requires a distance to be assigned. In Section\,\ref{sec:distances} we detail the procedures that we followed to determine distances to the SEDIGISM clouds. Once the distances have been assigned, we can compute the physical properties of clouds. In the catalogue, besides the measured sizes in angular scales, we also present the sizes in physical scales, i.e. already converted using the assigned distance.

We then estimated a few other physical properties, which required using an $X_{\rm{^{13}CO(2-1)}}$ conversion factor between the integrated intensities of $^{13}$CO (2-1) and the H$_{2}$ column densities. We adopted $X_{\rm{^{13}CO(2-1)}} = 1^{+1}_{-0.5} \times10^{21}$\,cm$^{-2}$\,(K\,km\,s$^{-1}$)$^{-1}$, as estimated in the SEDIGISM science demonstration field \citep{Schuller2017}, by comparing the SEDIGISM $^{13}$CO emission to the H$_{2}$ column densities as derived from the Hi-GAL survey data \citep{Molinari2010}\footnote{{The Hi-GAL column density maps for this calibration were built by fitting a pixel-by-pixel grey body curve to the spectral energy distribution from 160 to 500\,$\mu$m \citep{Elia2013}, assuming a dust to gas ratio of 1:100, and an opacity law with a fixed spectral index $\beta = 2,$ and $\kappa_{0} = 0.1$\,cm$^{2}$\,g$^{-1}$ at $\nu_{0}$ = 1200\,GHz \citep{hildebrand1983}.}}.
With this $X_{\rm{^{13}CO(2-1)}}$, and assuming a mean molecular weight $\mu_{H_2}$ of 2.8 \citep[][]{Kauffmann2008}, we derived the clouds' masses ($M$), average gas surface densities ($\Sigma$), and virial parameter ($\alpha_{\rm vir}$), defined as $\alpha_{\rm{vir}} = 5\sigma_{v}^{2}R / GM$ \citep{Bertoldi1992},
where $G$ is the gravitational constant, $\sigma_{v}$ the velocity dispersion, and $R$ is the equivalent radius. This formulation of $\alpha_{\rm vir}$ assumes a spherical geometry and a uniform density, and it only takes into account the balance between kinetic and gravitational energies. Thus, $\alpha_{\rm vir}$ is a very simplistic tool, and it should not be taken as a strict measurement of the gravitationally bound state of a cloud \citep[e.g.,][]{Bertoldi1992,kauffmann2013,Traficante2018a,Traficante2018b}. However, given its wide usage in the literature, we estimate it here to allow a direct comparison of our results with those of other surveys.

Finally, in the catalogue we provide the surface density and the virial parameters using both the measured $R$ (noted as $\Sigma$ and $\alpha_{\rm vir}$), and the deconvolved $R^{d}$ (noted as $\Sigma^{d}$ and $\alpha_{\rm vir}^{d}$). For the analysis presented in this paper we will use the deconvolved properties, although this choice has only a very marginal effect on the respective distributions, keeping the global trends virtually unchanged. Given the uncertainties on the distance estimates (which are of the order of $\sim30\%$) and on the X$_{\rm CO}$ factor (of a factor two), all these quantities have an uncertainty of at least a factor two.

In the catalogue, we also provide the Heliocentric and Galactocentric coordinates of each cloud, determined as explained in App.\,\ref{app:coordinates}.


\section{Distance determination}
\label{sec:distances}

\begin{figure}
\centering
\includegraphics[width=0.475\textwidth]{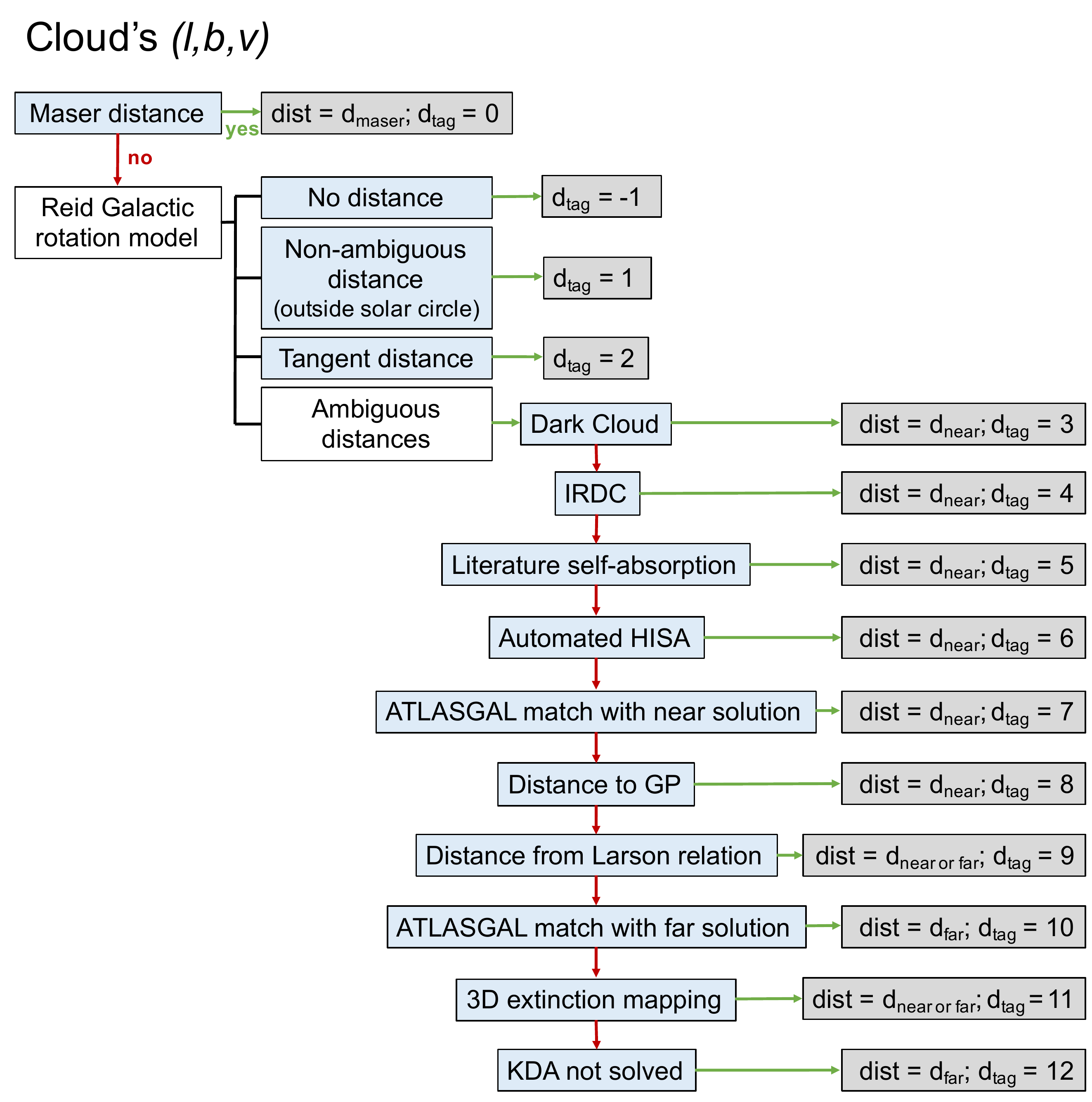}
\vspace{-0.3cm}
\caption{Flowchart showing the distance assignment procedure adopted for the SEDIGISM clouds. The blue boxes highlight the methods used, and the grey boxes show the corresponding assigned distance and tag. The green and red arrows show the directions taken if a specific method succeeds or fails in providing a distance solution, respectively.}
\label{fig:flowchart}
\end{figure}

In order to compute the physical properties of clouds, we require knowledge of the distances. However, for a large survey such as SEDIGISM, there are very few existing direct measurements of the distances towards molecular clouds, and we mostly need to rely on estimates based on the kinematic distances (i.e. by assuming a Galactic rotation model, see Sect.\,\ref{sec:kin_dist}), which rarely give a unique answer. Therefore, it is often required to search for ancillary indications to narrow down the distance assignment. In the following sections, we describe the computation of the kinematic distances, and how the problem of the kinematic distance ambiguities (KDA) were solved. The kinematic distance solutions, along with their uncertainties, and our final decisions are listed in the catalogue. For each cloud we include two distance tags: $d_{\rm sol}$ specifies the type of distance solution, and $d_{\rm flag}$ specifies the method used to reach the final distance assignment. The numbering of  $d_{\rm flag}$ reflects the order by which we check the different methods. Once a cloud gets a distance as per a given tag, we stop testing further methods.  
The flowchart depicting this decision process is shown in Fig.\,\ref{fig:flowchart}. These methods are all described in detail in Sect.\,\ref{sec:dist_amb}, and summarised in Table\,\ref{tab:dtag}. 

\begin{table}
\caption{Summary of the methods used to determine the distances of clouds, along with the number of clouds that had their distances assigned with each method.}
\label{tab:dtag}
\small
\renewcommand{\footnoterule}{}  
\begin{tabular}{l l c}
\hline 
\hline
$d_{\rm flag}$ & Description & Nb. clouds \\
\hline
$-$1 	& No distance information 			& 363 	\\
0	& Exact maser parallax distance		& 11 		\\
1	& No distance ambiguity 				& 551 	\\
2 	& Tangent distance					& 1080 	\\
3	& Dark Cloud (near distance)			& 77		\\
4	& IRDC (near distance)				& 751	\\
5	& Literature \hi SA (near distance)		& 91		\\
6	& Direct \hi SA measurement (near distance)  & 828	\\
7	& ATLASGAL source at near distance 	& 252	\\
8	& Solomon distance to GP (near distance)	& 34		\\
9	& {Size-linewidth} scatter (near or far distance) 	& {2263}	\\
10	& ATLASGAL source at far distance 		& {142}		\\
11	& Extinction (near or far distance)		& {3178}	\\
12	& Ambiguity not solved (defaulted to far) 	& {1042} 	\\
\hline
\end{tabular}
\end{table}

\subsection{Kinematic distances}
\label{sec:kin_dist}

To derive the kinematic distances of the clouds in our catalogue, we have used the Galactic rotation model of \citet{reid2016}, which has been constructed using maser parallax distance measurements. This model uses the revised values for $R_{0}$ and $V_{0}$ of 8.34\,kpc and 240\,km\,s$^{-1}$, respectively. Besides these rotation curve parameters, this model also uses a Bayesian approach that can consider the source's proximity to spiral arms, displacement from the Galactic mid-plane and proximity to parallax sources to estimate the most likely distance. Since molecular clouds are not always confined to the spiral arms or associated with star formation we have relaxed those constraints. 

Some clouds, however, have velocities that lie outside those allowed by the rotation model, and thus we are unable to assign them a distance. This is the case for 363 clouds, and they can be identified in the catalogue with the distance solution tag $d_{\rm sol} =$ NULL (and $d_{\rm flag} = -1$)\footnote{Note that in the catalogue we assign these clouds a distance of $-1$, which effectively means that we have estimated their physical properties as if they were at 1\,kpc distance, and that properties that have a linear dependency with distance will appear with negative sign.}. For the remaining clouds, if they lie outside the Solar circle, there is a unique kinematic distance solution. This is the case for 551 clouds, and these can be identified in the catalogue with the tag $d_{\rm sol} = $ NA, standing for \emph{No Ambiguity} (and $d_{\rm flag}=1$). When sources are located within the solar circle, there are two possible distance solutions, a \emph{near} and a \emph{far} one, which are equally spaced on either side of the tangent distance. Clouds that lie close to the tangent velocities \citep[i.e. within 5\,km\,s$^{-1}$, to accommodate for uncertainties due to streaming motions, e.g.][]{brand1993,wienen2015} were assigned the tangent distance, and given the tag $d_{\rm sol}$\,=\,T (and $d_{\rm flag}=2$).  This is the case for 1080 clouds. 

For sources with two possible distances, we performed an extensive cross-match with literature information, checked directly for \hi\ self-absorption (\hi SA) in each cloud, and checked whether the cloud properties would make them statistically more likely to be at a specific distance solution, in order to solve the distance ambiguity. Upon completion of this procedure, clouds that were assigned a near distance were tagged in the catalogue with $d_{\rm sol} = $ N (corresponding to a total of 3679 clouds), while far distance clouds have $d_{\rm sol} = $ F (which amount to 4979 clouds).  The full details on the procedure leading to our final distance decision are described in the following section. 

Note that, despite our extensive effort in assigning distances to clouds, there are regions within our Galaxy for which we know that our kinematic distances are not reliable. We have therefore included a flag in the catalogue, $d_{\rm{reliable}}$, which identifies clouds for which the distances are unreliable or nonexistent ($d_{\rm{reliable}} = 0$) and those that have a reliable distance estimate ($d_{\rm{reliable}} = 1$). In particular, we have given a $d_{\rm{reliable}} = 0$ for clouds with a $|v_{\rm lsr}| < 10$\,km\,s$^{-1}$ with a near distance assignment. For those, the kinematic distance is too uncertain, since the $v_{\rm lsr}$ of the clouds is dominated by local motions, and therefore the distance assigned from a global rotation model has a distance uncertainty on the order of the distance value itself. We also assigned a $d_{\rm{reliable}}=0$ to clouds for which we were not able to solve the distance ambiguity (i.e., clouds with a $d_{\rm flag}$ = 12, see Sect.\,\ref{sec:dist_other}). In addition, clouds towards the Galactic centre (and including most of the Galactic bar), i.e. within {+353}\degr\, < $\ell$ < 7\degr, also have a very uncertain distance estimate (and are given a $d_{\rm{reliable}} = 0$), as the Galactic rotation model used for our kinematic distance assignment is not tailored to reproduce the complex dynamics of the gas in the centre of the Galaxy. The only exception being the clouds for which we have a maser parallax distance (as that is an exact measurement, independent of kinematic considerations), which are retained with a $d_{\rm{reliable}} = 1$. The Galactic Centre will be studied in more detail in future work, and we will then revise the catalogued distances for those clouds accordingly.

\subsection{Solving the distance ambiguities}
\label{sec:dist_amb}

\subsubsection{Maser parallaxes, Dark Clouds, IRDCs, and \hi SA from literature}
\label{sec:maser}

We performed a cross-match of our entire catalogue with literature information for any known robust indication of the distance of our clouds. We started by cross-matching our clouds with a compilation of known maser parallax measurements \citep[][]{reid2009,reid2014,wu2014,Honma2012,Bobylev2013}. The matches were performed by checking if the position of the masers (in 3D) fell inside the mask of one of our SEDIGISM clouds. Sources with a known maser parallax measurement were assigned their maser parallax distance (instead of the kinematic distance). If there were more than one maser parallax measurement for a given cloud, then we take the average parallax distance. Clouds with a maser distance were given a $d_{\rm sol}$\,=\,M and $d_{\rm flag}$ = 0, and this was the case for 11 clouds. The small number of SEDIGISM clouds with a maser parallax is due to the fact that most of the maser parallax catalogues cover Quadrants 1, 2 or 3, hence only very few maser parallax distances have been measured for sources in our longitude range, and of those, about half lie outside our latitude range. 

We then did a cross match with other literature catalogues (including dark clouds, infrared dark clouds (IRDCs), and \hi SA), using the clouds' centroid Galactic coordinates and velocity. For catalogues in which the major axes, minor axes and position angles are given, the match was done by checking if the centroid position of the SEDIGISM cloud falls in the elliptical footprint of the catalogued source. For catalogues that give no position angle, or provide only the beam size or a radius, we use the effective radius and the match is done by checking if the centroid of the SEDIGISM cloud falls in the defined circular footprint. For catalogues which have velocity information, besides the spatial match, we require that the velocity difference between the SEDIGISM cloud and the catalogued sources must be less than 6\,km\,s$^{-1}$ \citep[assumed to be the typical cloud-cloud velocity dispersion, e.g.][]{Stark2006,Wilson2011}.

Using these criteria, we cross-matched our clouds with catalogued Dark Clouds with velocity information \citep[][]{Otrupcek2000}, as well as with IRDCs, some with and some without velocity information (\citealt{simon2006}, \citealt{jackson2008}, \citealt{du2008}, \citealt{peretto2009}, \citealt{chira2013}, \citealt{liu2013}). 
Their extinction makes Dark Clouds and IRDCs appear in silhouette against a bright background (in the visible and in the IR, respectively). Dark Clouds typically reach high optical depths very quickly, and thus are typically tracing nearby clouds that absorb the stellar light from the Galactic disc. The IRDCs probe a higher column density regime, which means we observe deeper into the molecular clouds. Nevertheless, the concept is the same, in that we are more likely to see a cloud in extinction, if there is enough IR background to absorb against, thus placing such clouds preferably at their near distance solution \citep[although this might not always be the case, e.g.][ found $\sim10\%$ of IRDCs to be located at the far distance]{Giannetti2015}. For our purpose, we have assumed any SEDIGISM sources which have a Dark Cloud, or an IRDC match to be at the near distance, and given a $d_{\rm flag}$ = 3 or 4, respectively. Note that, in cases where the cross-match with IRDCs was only spatial (i.e. in the absence of available velocity information), we only consider the match to be reliable if there is a single SEDIGISM cloud associated with each IRDC: if an IRDC is in the same line of sight as multiple SEDIGISM clouds, more information - such as velocity information or a more detailed morphological match - would be needed in order to produce a robust association.  

We also cross-matched our catalogue with known \hi SA (or H$_{2}$CO absorption) features from the literature within the SEDIGISM coverage \citep[][]{anderson2009a,Anderson2015,wienen2015,sewilo2004,pandian2008,busfield2006,urquhart2012_hiea}. \hi SA occurs when cold \hi\ gas in the foreground absorbs the warmer \hi\ emission from background gas at the same velocity \citep[e.g.,][]{Gibson2000}. Therefore, the existence of \hi SA at a given velocity is often used as an indication that the cold gas that is absorbing is at a near distance\footnote{Note that not all near-distance clouds are expected to show a strong \hi SA feature, given the simultaneous requirement of: 1) the existence of significant cold \hi\ gas at the same velocities as the molecular cloud traced by $^{13}$CO; 2) the existence of warm \hi\ gas in the background, at the same velocity as the cloud; and 3) the non-existence of intervening warm \hi\, gas between us and the cloud, that could fill in the cloud's intrinsic \hi SA. In addition, \hi{\sc i}\ regions can also produce a direct \hi\ absorption feature, even at the far distances, which can be mistakenly interpreted as a \hi SA feature. The  \hi SA  method for solving the KDA is therefore estimated to be $\sim$80\% reliable \citep[e.g.][]{anderson2009a}}  - as this makes it more likely to have background emission to absorb against, and that emission is less likely to be filled by other warmer \hi\ emission along the line of sight between the observer and the cold cloud \citep[e.g.,][]{Roman-Duval2009}. SEDIGISM sources with a known \hi SA feature from the literature were assumed to be at the near distances, and given a $d_{\rm flag}$ = 5. 

\begin{figure*}
\centering
\includegraphics[width=0.9\textwidth]{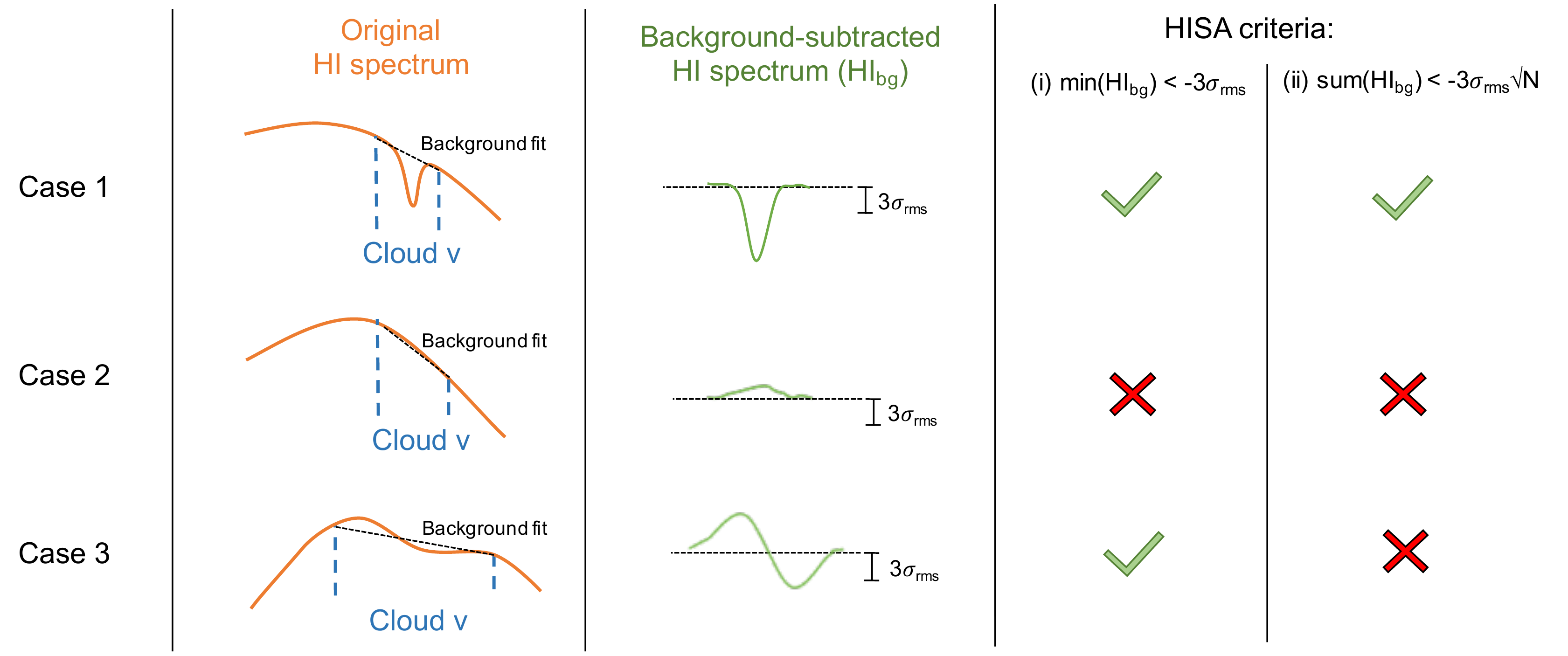}
\caption{Three sketch examples of the automated \hi SA method, showing the original \hi\ spectrum on the first column (in orange), with the cloud's velocity ranges denoted in blue, and the linear background fit done to the \hi\ spectrum in black dotted line. The second column shows the cloud's background-subtracted \hi\ spectra (in green), with the dotted dashed line representing the 0-emission level, and the vertical bar representing $3\sigma_{\rm rms}$ of the \hi\ emission. The two last columns show the criteria that we use to infer whether there is \hi SA in that particular sight line. Case 1 represents a line of sight with strong \hi SA, but the other two cases are not considered to have \hi SA. Case 3 shows an example of a false positive arising from criterion (i) alone, but which is mitigated by introducing criterion (ii).}
\label{fig:hisasketch}
\end{figure*}

\subsubsection{Direct and automated \hi SA determination}
\label{sec:hisa}

Given that many of the clouds in our catalogue do not have a counterpart with literature sources (given the improved sensitivity and resolution of the SEDIGISM survey), we have also checked for the presence of \hi SA directly for each individual cloud. We have done so in an automated way, making use of \hi\ 21\,cm ATCA and Parkes data from both the Southern Galactic
Plane Survey (SGPS; \citealt{McCG2005}), and the ATCA \hi\ Galactic Centre Survey \citep[]{McCG2012}. Both these datasets have a spatial resolution of 2\arcmin, a spectral resolution of 1\,km\,s$^{-1}$ and an average noise level, $\sigma_{\rm rms}^{\rm HI}$,  of $\sim 1$\,K. We combined the data from these surveys into a single datacube, using the {\sc convert}\footnote{http://starlink.eao.hawaii.edu/docs/sun55.htx/sun55.html} and {\sc kappa}\footnote{http://starlink.eao.hawaii.edu/docs/sun95.htx/sun95.html} packages from the Starlink software \citep[][]{Currie2014}, namely the {\sc wcsalign} and {\sc wcsmosaic} procedures. We then extracted sub-cubes covering the same spatial and velocity range of each of the SEDIGISM tiles, which we reprojected and resampled to the same pixel and channel sizes as the SEDIGISM data. Even though this procedure heavily oversamples the \hi\ data, it facilitates the automated check of the \hi\ emission in each of the SEDIGISM clouds, by directly using the assignment masks produced by {\sc scimes}.

Our automated \hi SA procedure works in the following way: First, it selects all the voxels that belong to each cloud, and creates a projected 2D image of the cloud in the plane of the sky. This allows us to identify all lines of sight which belong to that cloud. Then, for each sight line, it determines the ``background'' \hi\ emission, by taking the \hi\ emission one channel before, and one channel after the cloud's velocity range in that specific sight line, and fitting it with a linear function (see illustration on the first column of Fig.\,\ref{fig:hisasketch}). We then subtract the \hi\ emission inside the cloud from the fit of the background \hi\ emission, on a channel by channel basis. This ``subtracted'' datacube should have negative values whenever the \hi\ content of a cloud is self-absorbing against the background \hi\ emission (see the second column of Fig.\,\ref{fig:hisasketch}). Therefore we use these background-removed \hi\ datacubes as the basis for our decision on whether a given sight line in a cloud has a significant \hi SA or not (see last two columns of Fig.\,\ref{fig:hisasketch}). 

To determine if a specific line of sight has \hi SA, we impose two conditions:

\begin{enumerate}
\item{The minimum intensity of the background-removed \hi\ emission signal is lower than $-3\sigma_{\rm rms}^{\rm HI}$. This ensures that the self-absorption is significant, given the noise in the \hi\ data.} 
\item{The sum of the background-removed \hi\ signal is negative, and has an absolute value larger than 3 times the cumulative noise, given by $3\sqrt{N}\sigma_{\rm rms}^{\rm HI}$, where $N$ is the number of velocity channels across which the signal was summed up.}
\end{enumerate}

Step (ii) ensures that false positives are rejected. A false positive typically occurs when our simple background fit does not capture properly the variations of the \hi\ background emission (e.g. by under- or over-estimating the slope of the  \hi\ background emission), producing a signature similar to a p-Cygni profile, whose dip may be deeper than the \hi\ noise -- thus passing our criteria (i) (see Case 3 of Fig.\,\ref{fig:hisasketch}). However, while a true self-absorbed profile would have negative emission throughout the entire cloud velocity range, resulting in the sum of the background-removed \hi\ emission to be also negative (and significant), a false positive would have a sum that is within the noise of the \hi\ data. We therefore use this criterion to remove potential false positives.

We then consider a cloud to have strong \hi SA only if the number of sight-lines (i.e. 2D pixels) that satisfy condition (i) amount to at least one beam size in the \hi\ data, and that satisfy condition (ii) amount to at least one SEDIGISM beam size. The results from this automated \hi SA determination are compiled in the catalogue under the $tag\_hisa$ property, which is assigned a value of 1 for strong \hi SA, 0 if it is ambiguous (i.e. meeting only some of the criteria above), and $-1$ if there is no \hi SA. Clouds with a strong \hi SA from this method are taken to be at a near distance, and given a $d_{\rm flag}$\,=\,6. 

\subsubsection{ATLASGAL distances}
\label{sec:agal_dist}

The ATLASGAL survey \citep[][]{schuller2009,beuther2011} observed the dust continuum emission towards the inner Galactic plane at 870\,$\mu$m, and produced a catalogue of 10163 compact sources\footnote{Note on nomenclature: we will refer to the ATLASGAL compact sources as ``clumps'', as opposed to the larger scale SEDIGISM structures that we refer to as ``clouds''.} (CSC catalogue; \citealt{contreras2013,urquhart2014_csc}). In order to determine the distances to these clumps, there was a significant effort in assigning velocities to the continuum emission through a combination of extensive cross-match with molecular line data reported in the literature and dedicated follow-up observations (\citealt{wienen2012}, \citealt{csengeri2016_sio}, \citealt{wienen2018}, \citealt{urquhart2019}). This was then combined with the \citet{reid2016} Galactic rotation curve to calculate kinematic distances, and the distance ambiguities were resolved using the \hi SA method and using a friends-of-friends clustering algorithm to identify complexes. This successfully determined the distances to $\sim$8000 ATLASGAL clumps (see \citealt{urquhart2018_csc} for details).

Since all of the SEDIGISM survey is covered by ATLASGAL, we performed a cross match between all clouds in our sample, to the ATLASGAL clumps with known $v_{\rm{lsr}}$ from \citet{urquhart2018_csc}. This cross match was done by considering the centroid positions and velocities of the ATLASGAL clumps, and placing them in the respective voxel in our 3D datacubes. We then checked whether that voxel falls within the mask of a SEDIGISM cloud (i.e. a \emph{perfect match}), and otherwise estimate the distance to the nearest SEDIGISM cloud (in all 3 dimensions). We then consider ATLASGAL clumps that lie within one beam size of the edge of the nearest cloud, or within one $\sigma_{v}$ of the cloud, to be a \emph{partial match}. Out of the {5067} ATLASGAL sources within the SEDIGISM coverage, {4376} were matched as a \emph{perfect match} to a SEDIGISM cloud, and {448} as being a \emph{partial match}, leaving only {243} ATLASGAL clumps without a SEDIGISM counterpart. Most of these unmatched ATLASGAL clumps are either small clumps whose corresponding SEDIGISM emission did not satisfy our minimum size requirement, or they are in regions that form part of a smoother background that does not get assigned to a specific cloud (i.e. where the $^{13}$CO emission does not have a local peak rising above the 4-$\sigma_{\rm rms}$ requirement to be considered as independent peaks/leaves within the dendrogram). In total, these {4824} ATLASGAL clumps are contained within {1709} SEDIGISM clouds (i.e. $\sim 16\%$ of SEDIGISM clouds).

Given that the distances to the ATLASGAL sample were estimated with the individual $v_{\rm{lsr}}$ of clumps (rather than that of the parent cloud), we do not use their distances directly. Instead, we are only interested in the type of distance solution determined for each ATLASGAL clump (\emph{near} or  \emph{far}), in order to incorporate it in our distance assignment. In most cases, all ATLASGAL clumps within a given SEDIGISM cloud have a distance solution that agrees amongst them. However, there are a few cases where, within a SEDIGISM cloud, there are ATLASGAL clumps with both a \emph{near} and \emph{far} solution. In those cases, we define the ``global'' ATLASGAL solution as being \emph{near}, under the assumption that an indication for a near distance solution is more reliable than the absence of one (which is the most common reason for a far distance assignment). Note that, even though we had to do this step to provide a complete list of ``ATLASGAL distance solutions'' for our SEDIGISM sample, none of the clouds for which the ATLASGAL  distance solutions disagreed, actually took their final solution from ATLASGAL (instead they had their KDA lifted by other methods).

For SEDIGISM clouds with an ATLASGAL match, and for which the criteria in Sect.~\ref{sec:maser} and \ref{sec:hisa} did not have an indication for a near distance, we check the distance solution from ATLASGAL. If that solution is \emph{near}, then we adopt the near distance, and assign a $d_{\rm flag}$ = 7. If the ATLASGAL solution is \emph{far}, and there are no other indications of a \emph{near} distance solution (from methods 8 and 9, see Sect.\,\ref{sec:dist_other}), then we assign the far distance, and a $d_{\rm{flag}} = 10$.

\subsubsection{Other distance indicators}
\label{sec:dist_other}

In addition to the above methods, we also checked two often-used techniques that take the statistical distribution of the properties of molecular clouds into account. The first one is the method used by \citet[][]{Solomon1987}, which considers the physical distance of a cloud to the Galactic plane, should the cloud be assigned the far distance. If by taking the far distance the cloud is too far off the Galactic plane (i.e. > 140\,pc, which is the scale height of the gaseous Galactic disc, e.g. \citealt{Solomon1987,Tavakoli2012}), then the near distance is favoured, and the cloud is given a $d_{\rm flag} = 8$. Note that towards the far side of the Sagittarius and Scutum-Centaurus arm (around $\ell \sim 290\degr$), the Galactic mid-plane is known to be warped towards negative latitudes \citep[e.g.][]{Chen2019,Romero-Gomez2019}. This implies that on the far-distance side, in the latitude range of $300\degr$\,<\,$\ell$\,<\,$318\degr$, the Galactic plane descends below a latitude of $-0.5\degr$ \citep[e.g.][]{reid2016}, and therefore this area of the Galaxy is not well covered by our survey (since we cover a relatively narrow $b$ range). Nevertheless, since the Galactic warp only becomes significant at Galactocentric distances of 8\,kpc and beyond, any clouds in the longitude range of $300\degr$\,<\,$\ell$\,<\,$318\degr$ possibly following the warp are beyond the Solar circle, and should have unambiguous distances. Therefore, our criterion checking for the height above the Galactic plane is not affected by the existence of the Galactic warp.

The second method places each cloud on the {size-linewidth} relation ($\sigma_{v} - R$) \citep[e.g.]{Larson1981,Solomon1987}, for both near and far distance solutions, and checks which solution provides the smaller distance to the empirical relation. We use this method to favour a given distance solution \emph{only} if one solution is significantly closer to the empirical relation than the other solution (i.e. at least a factor 3 difference {in $log$-space}). {More details on this method can be found in App.\ref{app:size-linewidth}, and} clouds that used this criteria were given a $d_{\rm flag} = 9$. 

\begin{table*}
  \caption{Summary of different samples}
  \label{tab:samples}
  \begin{tabular}{l l c }
    	\hline \hline
	Sample Name & Description/conditions for selection & nb of sources \\
	\hline
	Full sample & Entire catalogue, with distances ($d_{\rm{flag}} \neq -1$) & 10300 \\
	Science sample & $d_{\rm{reliable}} = 1$, $Area > 3 \Omega_{\rm{beam}}$, edge = 0  & {6664} \\
	Distance limited sample & $d_{\rm{reliable}} = 1$, $Area > 3 \Omega_{\rm{beam}}$, edge = 0, 2.5\,kpc $<d<$ 5\,kpc & {1743}  \\
	Complete science sample & $d_{\rm{reliable}} = 1$, $Area > 3 \Omega_{\rm{beam}}$, edge = 0, $M > 2.6\times10^{3}$\,M$_{\odot}$, $R > 2.9$\,pc, $d < 14.5$\,kpc, $d_{\rm{flag}} \neq 2$ & {1680} \\
	\hline
  \end{tabular}
  \flushleft
\end{table*}

Finally, we also used a method based on datacubes of the visual extinction in K-band, as a function of distance \citep[][Marshall et al. in prep, Elia et al. in prep]{Marshall2006}. From those cubes, the structures of significant extinction can be identified along each line of sight, by taking the distances at which the extinction has a significant jump. We then compare the extinction distances with the near and far kinematic distance estimates, by taking into account a 30\% uncertainty on the extinction distance as well as the kinematic distance uncertainties. We solve the KDA by taking the kinematic distance which has an extinction counterpart, if one exists. Clouds that used this criteria were given a $d_{\rm flag} = 11$\footnote{This method does have a few limitations, one being that it becomes less reliable for far distances, mainly as the extinction cubes have a pixel of 5\arcmin, and therefore roughly 10 times larger than the SEDIGISM beam size. Small clouds assigned a distance using this method should therefore be used with caution.}.
In the future, this could potentially be expanded to also include Gaia-based 3D dust extinction maps \citep[e.g.][]{Lallement2019}, although at the moment these only probe distances up to 3\,kpc.

\subsection{Revisiting previous distance estimates}

In order to gauge how our distance estimates compare to the results from other surveys that covered the same area of the Galactic plane, we have compared the results from our distance solutions to those of the ATLASGAL survey, as a reference (given that ATLASGAL had already performed a detailed comparison with other surveys, e.g. \citealt{urquhart2014_atlas,urquhart2018_csc}). Note however, that as per Section\,\ref{sec:agal_dist}, only {1709} SEDIGISM clouds have an ATLASGAL counterpart (i.e. only $\sim 16\%$ of our sample), although this includes 95\% of all ATLASGAL clumps in our coverage (i.e. {4814} clumps). Of those, {1253} ATLASGAL clumps did not have an assigned distance, which we have now assigned\footnote{Note that most of these are towards the central Galaxy, for which the kinematic distances are less reliable. If we consider only the sample we use for science (as per Sect.\,\ref{sec:global_properties}), the number of ATLASGAL clumps that so far did not have a distance assigned and for which we are able to assign a reliable distance is {308}.}. For the ATLASGAL sources with a distance, the KDA solution between the two surveys agrees for 3080 ATLASGAL clumps. This leaves a total of 481 clumps (i.e. 13.5\% of the ATLASGAL clumps with distances) with a distance solution that was revised by us, in most cases from a far distance to a near distance, by one of the other methods listed in Section\,\ref{sec:kin_dist}. Most of these revisions were done using our HISA method (321 clumps, $d_{\rm flag}$ = 6), followed by 60 clumps revised using the IRDC matches ($d_{\rm flag}$ = 4), and 53 clumps with literature HISA ($d_{\rm flag}$ = 5). A further 23 clumps were revised using the maser parallax measurements ($d_{\rm flag}$ = 0), 7 clumps using the distance around the Larson relation ($d_{\rm flag}$ = 9), and 2 clumps using the Dark Cloud association ($d_{\rm flag}$ = 3). Finally, 2 clumps were re-assigned a near distance for being in the same complex as other ATLASGAL sources with a near distance ($d_{\rm flag}$ = 7), 1 clump was revised as having a non-ambiguous solution ($d_{\rm flag}$ = 1), and 12 clumps had their distances revised to a tangent distance ($d_{\rm flag}$ = 2), although for these cases the change from near or far solutions into the assumed tangent distance is within the uncertainties.

With the large survey coverage, and improved resolution and sensitivity of the SEDIGISM survey compared to other spectroscopic surveys covering the same Galactic longitudes (e.g. the MopraCO survey \citet{Burton2013}, the ThrUMMS \citet{barnes2015}, and the \citet{dame2001} survey), here we present the most extensive sample of molecular clouds towards the inner Galaxy yet, with 10663 clouds in total. With our comprehensive effort to combine different independent methods to determine the distance solutions for each SEDIGISM cloud, we have been able to assign distances to 10300 clouds, 7993 of which have well-characterised (reliable) distance assignments. 

\begin{figure*}
\centering
\includegraphics[width=\textwidth]{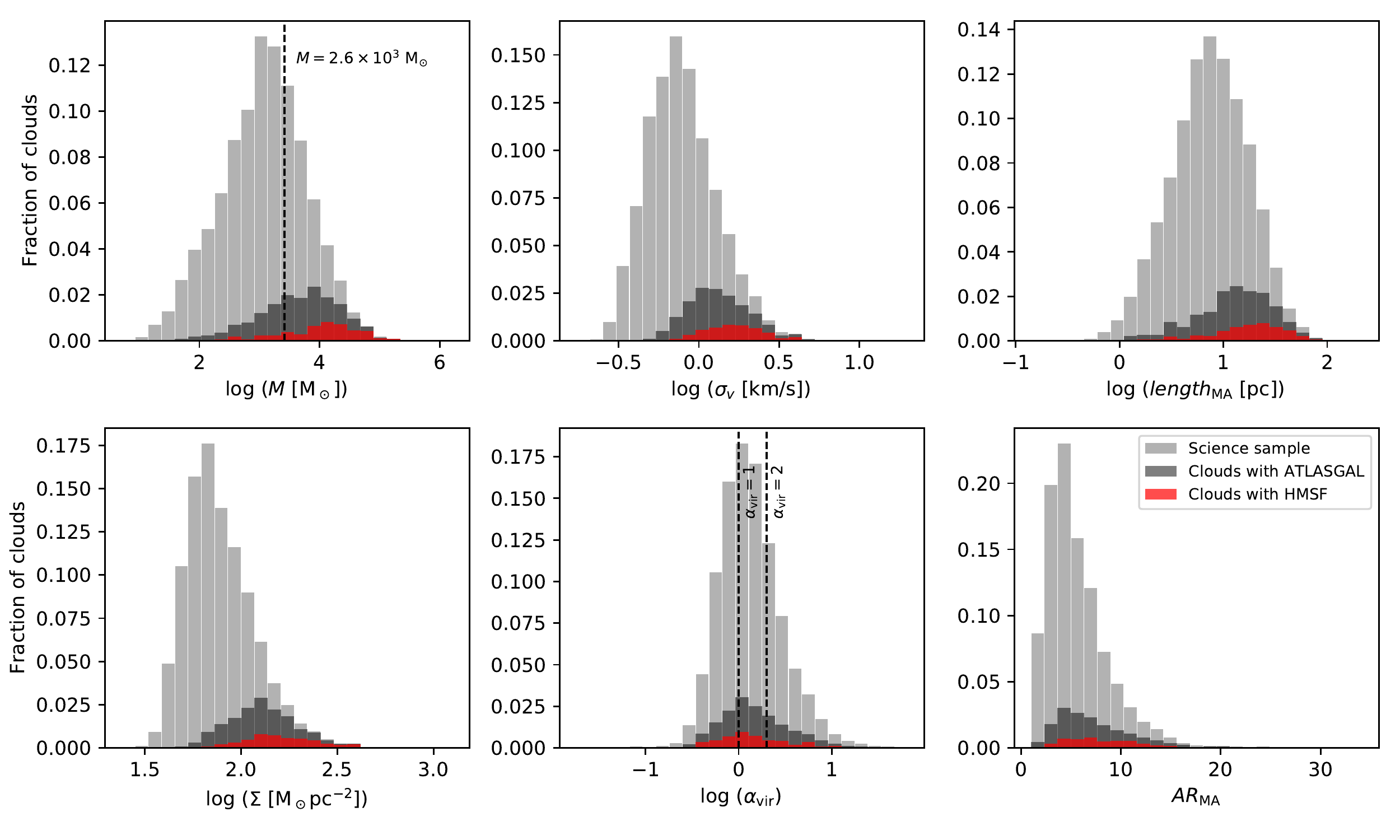}
\vspace{-0.5cm}
\caption{Histograms of global properties: Mass (top-left), velocity dispersion (top-centre), medial axis length (top-right), average surface density (bottom-left), virial parameter (bottom-centre), and aspect ratio from the medial axis (bottom-right). The histograms are for the science sample (light grey), clouds that have an ATLASGAL counterpart (dark grey), and clouds that have a HMSF signpost (red).  {The normalisation of all histograms was made with respect to the total number of clouds in the science sample.} The vertical dashed line on the mass histogram shows our mass completeness limit (see App.\,\ref{sec:completeness}), and the dashed lines on the virial parameter histogram represent an $\alpha_{\rm{vir}} = 1$ and 2.}
\label{fig:histograms}
\end{figure*}


\section{Global properties of the SEDIGISM sample}
\label{sec:global_properties}

\begin{table*}
  \caption{Statistics of some of the physical properties of the SEDIGISM clouds, namely the mass ($M$), velocity dispersion ($\sigma_v$), equivalent radius ($R_{\mathrm{eq}}$), medial axis length ($length_{\mathrm{MA}}$), medial axis aspect ratio ($AR_{\mathrm{MA}}$), surface density ($\Sigma$), and virial parameter ($\alpha_{\mathrm{vir}}$), for the entire science sample, and for a distance-limited sample (to minimise distance-biased results). Within these samples we also list the statistics for the subsets of clouds with an ATLASGAL counterpart or with a HMSF signpost. Q25 and Q75 represent the lower ($25\%$) and upper ($75\%$) quartiles of the distributions.}
  \label{tab:Statistics}
  \begin{tabular}{l c c c c c | c c c c c c}
    \hline \hline
    & \multicolumn{5}{l}{Science sample} &  & \multicolumn{5}{l}{Distance limited sample (2.5\,kpc $< d <$ 5.0\,kpc)} \\
    Sub-set & Median & Q25 & Q75 & Skewness & Kurtosis & & Median & Q25 & Q75 & Skewness & Kurtosis \\
    \hline
    $M$ [$\times10^3$ M$_\odot$] & & & & & \\
    Science         & 1.25 & 0.40 & 3.59 & 52.7 & 3521.1   & & 0.43 & 0.13 & 2.04 & 7.6 & 94.7 \\
    With ATLASGAL   & 5.13 & 1.69 & 13.83 & 24.1 & 690.6    & & 3.74 & 1.20 & 10.52 & 4.3 & 32.4 \\
    With HMSF       & 12.00 & 3.56 & 27.24 & 13.8 & 220.9	 & & 10.31 & 3.35 & 23.08 & 3.1 & 17.0 \\
    \hline
    $\sigma_v$ [km/s] & & & & & \\
    Science         & 0.76 & 0.55 & 1.07 & 6.9 & 162.0     & & 0.77 & 0.51 & 1.25 & 2.6 & 15.5 \\
    With ATLASGAL   & 1.29 & 0.97 & 1.81 & 7.5 & 126.7    & & 1.35 & 0.99 & 1.93 & 2.5 & 14.1\\
    With HMSF       & 1.66 & 1.25 & 2.20 & 7.7 & 95.4       & & 1.69 & 1.28 & 2.29 & 2.4 & 11.6 \\
    \hline
    $R_{\mathrm{eq}}$ [pc] & & & & & \\
    Science         & 2.31 & 1.39 & 3.64 & 3.4 & 37.2       & & 1.34 & 0.80 & 2.62 & 1.9 & 7.2 \\
    With ATLASGAL   & 3.56 & 2.16 & 5.65 & 1.5 & 6.8       & & 3.07 & 1.78 & 4.53 & 1.0 & 3.8 \\
    With HMSF       & 4.82 & 2.79 & 6.92 & 1.4 & 6.0       & & 4.10 & 2.66 & 5.81 & 0.6 & 3.1 \\
    \hline
    $length_{\mathrm{MA}}$ [pc] & & & & & \\
    Science         & 7.51 & 4.21 & 13.51 & 3.8 & 42.3    & & 4.82 & 2.54 & 10.83 & 2.1 & 8.8 \\
    With ATLASGAL   & 13.52 & 7.73 & 23.04 & 1.6 & 6.4     & & 12.61 &  6.32 & 21.07 & 1.2 & 4.5 \\
    With HMSF       & 18.88 & 10.73 & 29.74 & 1.2 & 4.5   & & 16.95 & 10.82 & 26.28 & 0.8 & 3.4 \\
    \hline
    $AR_{\mathrm{MA}}$ &&&& \\ 
    Science         & 4.9 & 3.4 & 7.1 & 1.6 & 7.7          & & 5.6 & 3.8 & 8.3 & 1.6 & 7.0 \\
    With ATLASGAL   & 6.6 & 4.5 & 9.6 & 1.4 & 6.3           & & 7.6 & 5.1 & 10.9 & 1.4 & 6.0 \\
    With HMSF       & 7.6 & 5.3 & 10.8 & 1.6 & 7.5          & & 9.0 & 6.2 & 11.7 & 1.6 & 7.7 \\
    \hline
    $\Sigma$ [M$_\odot$pc$^{-2}$] & & & & & \\
    Science         & 73.0 & 58.0 & 99.7 & 5.1 & 70.7     & & 75.4 & 57.6 & 112.7 & 3.0 & 19.3  \\
    With ATLASGAL   & 128.2 & 98.3 & 170.2 & 4.2 & 42.3    & & 139.9 & 103.9 & 190.8 & 2.2 & 12.1 \\
    With HMSF       & 158.1 & 120.4 & 221.1 & 3.7 & 27.6    & & 186.0 & 137.1 & 252.9 & 1.9 & 8.5 \\
    \hline
    $\alpha_{\mathrm{vir}}$ &&&&& \\
    Science         & 1.25 & 0.79 & 2.10 & 8.4 & 128.2      & & 1.85 & 1.23 & 3.08 & 4.3 & 33.0 \\
    With ATLASGAL   & 1.36 & 0.81 & 2.58 & 7.5 & 82.5       & & 1.79 & 1.05 & 2.98 & 2.9 & 16.0 \\
    With HMSF       & 1.28 & 0.76 & 2.62 & 7.0 & 61.3      & & 1.49 & 0.94 & 2.78 & 2.5 & 10.3 \\
    \hline
  \end{tabular}
\end{table*}

For our analysis of the statistical properties of the SEDIGISM molecular clouds, we have excluded any clouds whose projected footprint size is smaller than 3 beams (i.e. any clouds that are barely resolved). We also excluded clouds with an unreliable distance ($d_{\rm{reliable}} = 0$), and those that are incomplete because they touch a survey coverage edge ($edge = 1$). With these criteria, we select a total of {6664} clouds for our analysis, which we will refer to as our ``science sample''. In addition, we will refer to the science sample above the completeness limits (as per App.\,\ref{sec:completeness}) as our ``complete science sample'' (which also exclude clouds at a tangent distance - see Sect.\,\ref{sec:extreme} for more details). Table\,\ref{tab:samples} summarises the specific details of the several samples that we use in the paper.

\subsection{Distribution of individual properties}
\label{sec:distributions}

Figure\,\ref{fig:histograms} shows the distributions of a number of different properties, namely the total mass ($M$), the velocity dispersion ($\sigma_{v}$), the medial axis length ($length_{\rm{MA}}$), the average surface density ($\Sigma$), the virial parameter ($\alpha_{\rm{vir}}$), and the aspect ratio from the medial axis ($AR_{\rm{MA}}$). The histograms correspond to the full science sample (in light grey), from which we highlight the subset of clouds with an ATLASGAL counterpart (in dark grey), and from those, also clouds with a signpost of high-mass star formation (HMSF, in red), as per \citet{urquhart2014_atlas}. These signposts of HMSF include the existence of methanol masers (\citealt{urquhart2013_methanol,urquhart2015_methanol}, which used the masers from the Methanol Multibeam Survey, \citealt{caswell2010b,green2012_mmb}); H{\sc{ii}} regions (\citealt{urquhart2013_cornish}, which combined information from the CORNISH survey, \citealt{Hoare2012,Purcell2013}, and the GLIMPSE survey, \citealt{Benjamin2003}); or massive young stellar objects, YSOs (\citealt{urquhart2014_atlas}, which matched ATLASGAL sources with YSOs and H{\sc ii} regions identified by the Red MSX Source (RMS) survey, \citealt{Lumsden2013,urquhart2014_rms}). In total, we have 435 SEDIGISM clouds within the full sample (330 in the science sample, i.e. $\sim 4\%$ of clouds) that have signposts of active HMSF \citep[similar to the fraction of high-mass star forming clouds found by][]{Barnes2011}. We note, however, {that for this work, we did not cross-match our SEDIGISM clouds with HMSF tracers directly: } our sample of HMSF clouds is {purely} a subsample of the ATLASGAL sources, and so any HMSF signposts outside that are not accounted for. This will be explored in future work. We also computed the main statistics (i.e. the median, lower and upper quartiles, skewness and kurtosis) of these distributions, plus that of the equivalent radius ($R$), which we compile in Table\,\ref{tab:Statistics}. These distributions, however, could potentially be affected by our different completeness at different distances within our science sample. In order to check how this might affect the global results, we have also computed the histograms using a distance limited sample (with 2.5\,kpc $< d <$ 5.0\,kpc), shown in App.\,\ref{app:histograms} (Fig.\,\ref{fig:histograms_dlimited}). The statistics for the distance limited sample are also compiled in Table\,\ref{tab:Statistics}, showing that they follow broadly the same trends as the science sample.

Noticeably, the median values in Table\,\ref{tab:Statistics} and the histograms from Fig.\,\ref{fig:histograms} show that clouds with an ATLASGAL counterpart tend to be at the higher end of the distributions of mass, velocity dispersion, size, aspect ratio, and surface density, as compared to the science sample. This is even more so for clouds with a HMSF signpost (whose median values are again higher than those of the ATLASGAL sub-sample). The increase in the median values of those properties as we go from the science sample to the HMSF sub-sample range from a modest increase of a factor 2 (e.g. for the aspect ratio and velocity dispersion) up to an order of magnitude increase (for the mass). The only exception to this trend is the virial parameter, for which the median values (and the quartiles) are similar between all three subsets.

Interestingly, while the science sample typically has a distribution with a significant tail (i.e. with high kurtosis values), as we move from the full sample to the ATLASGAL sub-sample and then to clouds with a HMSF signpost, the shape of the distribution of all properties (except for the aspect ratio) becomes progressively flatter (smaller values of kurtosis) and symmetric (smaller values of skewness) -- with HMSF clouds occupying nearly the same parameter space as clouds with an ATLASGAL counterpart but without HMSF signpost. This is rather interesting as it suggests that there is no single ``global'' property of clouds that is sufficient to determine, on its own and unambiguously, their potential to host high-mass star formation, and perhaps a complex combination of several conditions is needed. It is worth noting that some global properties like magnetic fields are, of course, not considered here. In Section\,\ref{sec:extreme_hmsf} we will investigate if the ability to form high-mass stars might instead be influenced by the Galactic environment.

\subsection{Scaling relations}
\label{sec:scatter}

\begin{figure*}
\centering
\includegraphics[width=0.9\textwidth]{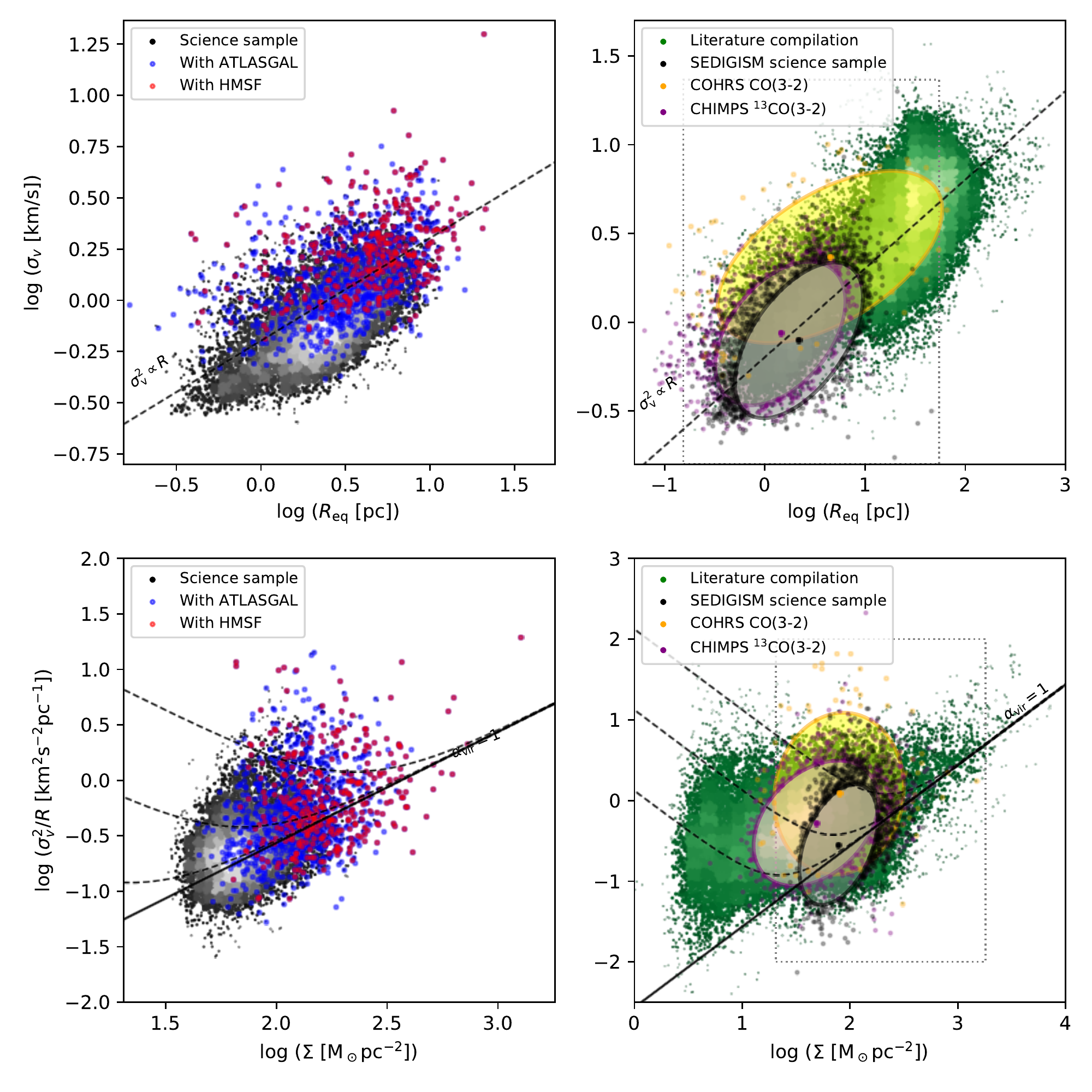}
\caption{Top row: size-linewidth relation ($\sigma_{v}$ versus $R_{\rm{eq}}$), where the dashed-line represents the Larson relation. Bottom row: scaling relation between $\sigma_{v}^{2} / R$ and gas surface density $\Sigma$, where the lines correspond to $\alpha_{\rm{vir}} = 1$: the solid line is without external pressure, and the dashed lines are when including external pressure (from top down, at a constant $P_{\rm{ext}} = $ 100, 10 and 1 M$_{\odot}$\,pc$^{-3}$\,km$^{2}$\,s$^{-2}$). The left panels show these relations for the SEDIGISM sample alone, where the grey scale represents the density of points for the entire science sample, the blue circles show the clouds with an ATLASGAL counterpart, and the red circles show the clouds that have a HMSF signpost. The panels on the right show, in green, the density of points from a compilation of literature catalogues which include both Galactic and extragalactic studies (see text for full list of references). Our SEDIGISM sample is represented by the black ellipse (from a PCA analysis, and where the ellipse contour contains $95\%$ of the data) and black points (which show the remaining $5\%$ of clouds). Similarly, we also show the {PCA} ellipses for the fiducial sample of the COHRS survey in orange \citep[][]{Colombo2019}, and the CHIMPS survey in purple \citep[][]{Rigby2019}, both of which are high-resolution surveys towards the 1$^{\rm{st}}$ Galactic quadrant - complementary to SEDIGISM. {For reference, the dashed grey boxes on the right panels show the plotting range of the corresponding left panel.}}
\label{fig:scatter}
\end{figure*}

Figure \ref{fig:scatter} shows two of the most common scaling relations in the literature: the size-linewidth relation in the top panels, where the dashed-line represents the Larson relation, $\sigma_{v}^{2} \propto R$  \citep{Larson1981,Heyer98}; and the Heyer relation, $\sigma_{v}^{2}/R \propto \Sigma$ \citep{Heyer2009}, in the lower panels, where the solid black line shows  $\alpha_{\rm{vir}} = 1$ as defined in Sect.\,\ref{sec:dataproducts}, and the dashed lines correspond to a $\alpha_{\rm{vir}} = 1$ when including the contribution of external pressure ($P_{\rm{ext}} = $ 1, 10 and 100 M$_{\odot}\,$pc$^{-3}$\,km$^{2}$\,s$^{-2}$). On the left panels, we show our SEDIGISM science sample in grey scale, and the subset of clouds with an ATLASGAL counterpart in blue, and those with a signpost of HMSF in red. From these, we can see that although our SEDIGISM clouds do show some correlation on both plots, neither of these follow the scaling relations proposed by previous works.

The right-hand side panels show a compilation of literature catalogues of molecular clouds in green colour scale, including both Galactic studies \citep[][]{Oka2001,Heyer2009,Roman-Duval2010,Rice2016,Barnes2016,miville-deschenes2017,Colombo2019,Rigby2019} and extragalactic studies \citep{Rosolowsky2005,Bolatto2008,Wong2011,Gratier2012,Wei2012,DonovanMeyer2013,Colombo2014,Leroy2015,Utomo2015,Faesi2016,Tosaki2017,Freeman2017,Pan2017,Schruba2017}. 
On those, we overplot the loci of the distribution of our science sample as the black ellipse, produced from a principal component analysis\footnote{The PCA analysis \citep[][]{Pearson1901_PCA} can be useful to identify the directions of maximal and minimal variance of data with large intrinsic scatter, thus equivalent to finding the direction and scatter of the underlying scaling relation (which are typically estimated using a linear regression fit). As we are simply interested in using the PCA as a representation of the loci of the distributions, we did not take into account the uncertainties in the measured quantities for this analysis.} \citep[PCA,][]{Pearson1901_PCA}, similar to \citet{Colombo2019}. The ellipse contours in the right panels of Fig.\,\ref{fig:scatter} correspond to a 2-sigma level, i.e. it contains $\sim 95\%$ of the data points, while the central point corresponds to the mean. The remaining $5\%$ of data points are overplotted as circles. 

We have also performed this PCA analysis for the cloud catalogues from the fiducial sample of the COHRS survey \citep[in $^{12}$CO (3-2),][]{Colombo2019}, and from the CHIMPS survey \citep[in $^{13}$CO (3-2),][]{Rigby2019}, which we plot in Fig.\,\ref{fig:scatter} as yellow and purple ellipses, respectively. Although both of these surveys have a slightly higher spatial resolution than SEDIGISM (17\arcsec\,versus 28\arcsec), they both cover the 1$^{\rm{st}}$ quadrant, making them highly complementary to the SEDIGISM survey. In fact, the native resolution of CHIMPS was smoothed to 27\arcsec\, for their source extraction and derivation of cloud properties that we use here, thus making it very similar to that of the SEDIGISM survey. For completeness, we summarise the directions of major variance from the PCA analysis for these three surveys in Table\,\ref{tab:pca_summary}, which can be compared to the expected slopes from the literature. Note, however, that even though the slopes from the PCA analysis can be suggestive of a correlation, in all the cases we performed the PCA here, the major and minor axis are similar (within a maximum of a factor 3 difference), which indicates that these are not tight correlations.

The clouds from the COHRS survey were extracted using the same method as us ({\sc scimes}) but, because it uses $^{12}$CO (3-2), it typically traces larger clouds, with larger velocity dispersions (partly due to the fact that $^{12}$CO traces more diffuse gas than $^{13}$CO, but also due to line broadening from optical depth effects, and from a coarser spectral resolution of 1\,km\,s$^{-1}$)\footnote{A comprehensive comparison of the COHRS cloud population with other surveys can be found in \citet[][]{Colombo2019}, namely their Fig.\,13, which can be used to compare with the relative position of the SEDIGISM cloud catalogue.}. The CHIMPS survey coverage overlaps with COHRS, but it uses the optically thinner $^{13}$CO (3-2). Even though the clouds from CHIMPS were extracted using Fellwalker \citep[][]{Berry2015}, which segments the emission into their individual peaks (hence not allowing for the grouping of several peaks into complexes), and their line tracer is not the same as ours (using a higher-energy transition of $^{13}$CO), the properties of the CHIMPS clouds agree remarkably well with those of our SEDIGISM sample. There is only a small shift in the sizes of the SEDIGISM clouds towards larger values (as we can see in the top-right panel of Fig.\,\ref{fig:scatter}) and, although the CHIMPS sample spans to lower average surface densities than the SEDIGISM sample (as we can see on the lower-right panel), both samples have clouds reaching similar values towards the high surface-density end. These differences can be easily understood as a consequence of: 1) the cloud segmentation used by the CHIMPS survey, breaks up the emission more, thus extracting smaller (and less dense) clouds, whilst the grouping of individual clumps into larger cloud complexes achieved by our usage of {\sc scimes} for the SEDIGISM segmentation will tend to incorporate such small diffuse clumps into larger complexes; and 2) the $^{13}$CO (3-2) transition used in CHIMPS has a higher critical density and will typically trace warmer gas than the brighter $^{13}$CO (2-1) transition of SEDIGISM, which will mean that the CHIMPS clouds will typically be able to trace less mass for a given brightness temperature.

\begin{table}
\caption{Slopes ($\alpha$ and $b$) recovered from a PCA analysis on the scaling relations, where $\sigma_v \propto R^{\alpha}$, and $(\sigma_v^2 / R) \propto \Sigma^{b}$. {The mean values of each pair or quantities (i.e. the centres of the ellipses in Fig.\,\ref{fig:scatter}), are noted with the upper-script $m$.}}
\label{tab:pca_summary}
\small
\renewcommand{\footnoterule}{}  
\begin{tabular}{l l l l l}
\hline 
\hline
Sample & $\alpha$ & [$\sigma_v^m$, $R^m$] & $b$ & [$(\sigma_v^2 / R)^m$ , $\Sigma^{m}$] \\
\hline
SEDIGISM & {0.52} &	{[2.19, 0.79]}	& {3.91} 	& 	{[78.3, 0.29]} \\
CHIMPS	 & 0.39  &	{[1.43, 0.86]} 	& 2.15 	& 	{[48.3, 0.52]} \\
COHRS	 & 0.27. &	{[4.48, 2.33]}	& 14.79	& 	{[79.1, 1.22]}  \\
\hline
Expected    & 0.5$^{a}$ &   & 1.0$^{b}$ &  \\
\hline
\end{tabular}
\\
$^{a}$ \citet{Larson1981}\\
$^{b}$ \citet{Heyer2009} 
\end{table}

Most interestingly, these plots show that the choice of tracer and the specific limitations of the surveys change our global view of the properties of molecular clouds. Looking at the $^{12}$CO emission from the COHRS survey, we could argue that these clouds are in a pressure-confined regime (i.e. lying above the $\alpha_{\rm{vir}} = 1$ line when external pressure is not included, but could be consistent with being virialised if a moderate external pressure is at play). However, looking at the same clouds with an optically thinner tracer (i.e. with CHIMPS) changes our perception of their energy balance, with clouds moving closer to a more gravitationally bound regime, or else requiring only a very weak external pressure to be virialised. This points out a rather important issue: although molecular clouds are highly hierarchical, they are part of a continuous medium that smoothly blends into the diffuse warm neutral medium, with no hard boundary. We know that the ISM is not composed of a discrete set of entities, and yet this discretisation is (and has been) a crucial step in our understanding of the cold molecular medium. What we use to define them thus changes what we actually trace. Simple measures of the energy balance of clouds at any one single level are incapable of providing a complete picture of the true physics that describe and regulate the evolution of clouds. Instead, we need to move into trying to put a sequence together for the general trend of the change in molecular cloud properties with tracer density (which could even perhaps be used as a proxy for time). Studies looking into the evolution of these global properties, within molecular clouds - i.e. as we move inside the internal hierarchy of clouds - are necessary for taking our understanding of the physics inside molecular clouds to the next level. This is one of the key advantages of using a dendrogram-based segmentation of the ISM, that we shall explore in future work. 


\section{Galactic distribution of the most extreme clouds}
\label{sec:extreme}

\begin{figure*}
\centering
\includegraphics[width=0.9\textwidth]{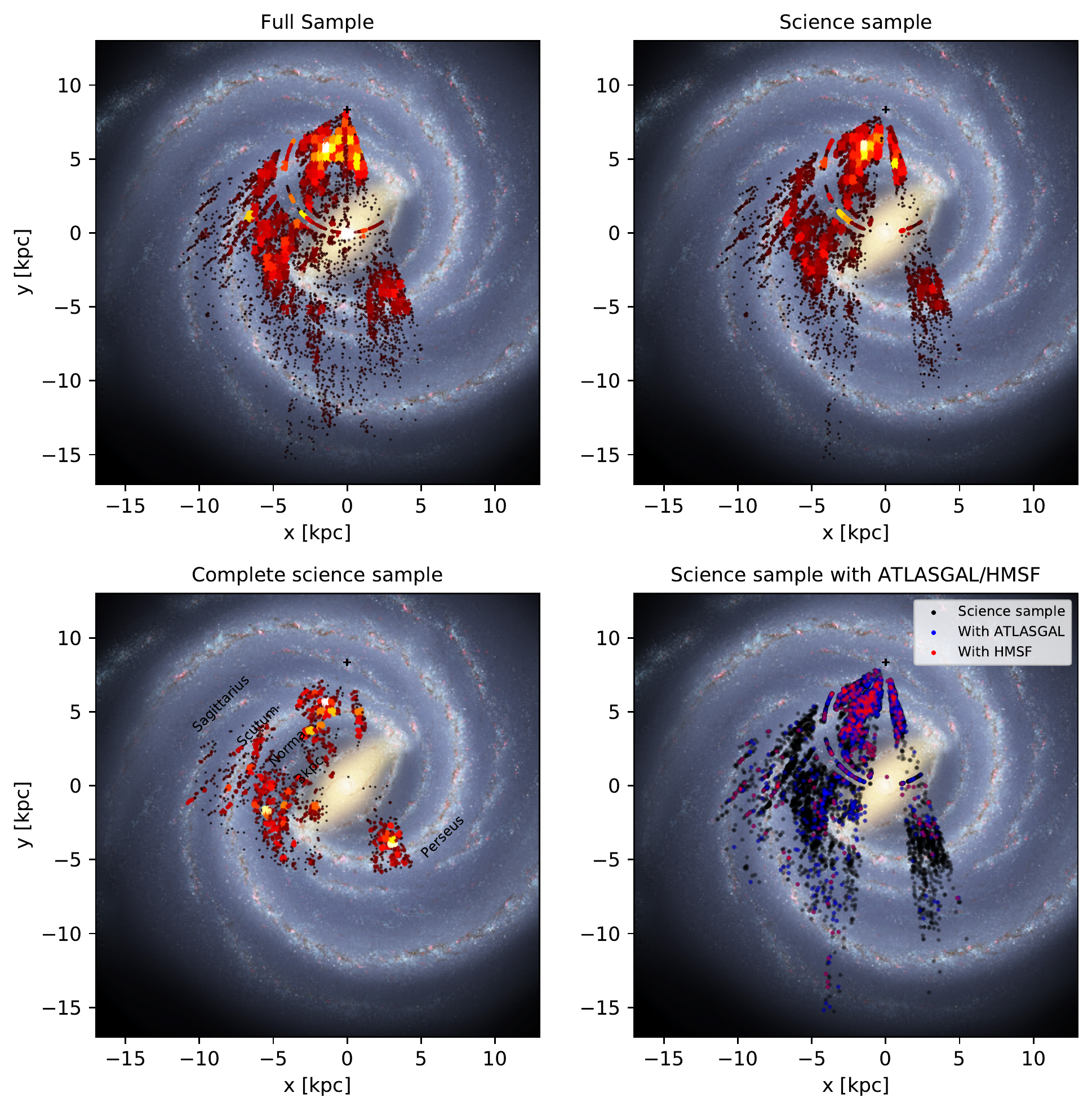}
\caption{Top down view of the Galaxy, with the deprojected position of SEDIGISM clouds overplotted on an artistic impression of the Milky Way (NASA/JPL-Caltech/R. Hurt (SSC/Caltech)). The position of the Sun is marked with a `+' in all panels. The top-left panel shows the density plot of the entire catalogue, and the top-right panel shows the science sample. The bottom-left panel shows the Galactic distribution of the clouds in the complete science sample (i.e. above our completeness limit, and excluding clouds with a tangent distance assignment). For these three panels, the colour scale and the size of the symbols is related to the local density of clouds (more crowded areas are shown in white, and with larger symbols). The bottom-right panel shows all the sources in the science sample colour-coded depending on whether they have an ATLASGAL counterpart (in blue), a HMSF signpost (in red), or neither (in black).}
\label{fig:top_down_agal}
\label{fig:top_down_1}
\end{figure*}

Using the longitude ($\ell $) and distance ($d$) of the clouds in our catalogue, we can estimate their Galactocentric coordinates, which we use to plot our clouds on a ``top-down'' view of the Galaxy. These are shown in Fig.\,\ref{fig:top_down_1}, overlaid on an artist's impression of the Milky Way (by NASA/JPL-Caltech/R. Hurt (SSC/Caltech)). The main known gaseous spiral arms are labeled in the bottom-left panel.  The top-left panel of Fig.\,\ref{fig:top_down_1} shows our full SEDIGISM catalogue with distances, the top-right panel shows the distribution of our science sample, and the bottom-right panel shows the science sample colour-coded depending on whether the clouds have an ATLASGAL counterpart (blue), or a HMSF signpost (red). {Using this top-down Galactic distribution of clouds in the science sample, we estimate a typical mass surface density of gas associated with clouds to be of the order of $1\times10^5$\,M$_{\odot}$kpc$^{-2}$ (and ranging from $\sim 4.4\times10^2$ to $1.3\times10^6$\,M$_{\odot}$kpc$^{-2}$). Note that the values for the average and minimum mass surface densities are only lower limits, as they are likely affected by our completeness limits.} On the bottom-left panel of Fig.\,\ref{fig:top_down_1} we show our complete science sample, i.e. clouds within the science sample that lie above our mass and radius completeness limit (as detailed in App.\,\ref{sec:completeness}), and are located within a heliocentric distance of 14.5\,kpc (the distance used to determine our completeness limit). The complete science sample also excludes clouds with a tangent distance. For those clouds, although the physical properties are reliable (since the near and far distances are relatively close together), their Galactic position falls into a single line at the tangent distance, which introduces some biases for the statistical tests we will be performing with this sample (see App.\,\ref{app:chisq_test} for more details). Our complete science sample consists of {1680} clouds.

We caution that showing clouds with this top-down perspective, although suggestive, can be misleading - indeed we know that the uncertainties on the distances can amount to $\sim 1$\,kpc, particularly when streaming motions around spiral arms can be important, and this can easily displace clouds across entire spiral arms. In addition, the exact position and strength of these arms is still quite uncertain \citep[e.g.][]{taylor1993,reid2014,Vallee2017}. In fact, the very existence of four strong spiral arms is still subject of debate, especially as studies in the Optical/near-IR \citep[e.g.][]{Drimmel2000,Siebert2011,Siebert2012,Gaia2018_KatzMW}, suggest that we only have two main stellar spiral arms - which could indicate that the four spiral arms that we see in the gas, are not as well defined as this figure depicts, and are perhaps more flocculent in nature. {This idea is also supported by our relatively low values of molecular gas mass surface densities, which place the Milky Way at the bottom of the distribution of the values retrieved for a sample of 15 nearby spiral galaxies \citet{Sun2018}, whose typical molecular gas mass surface densities are of the order of $10^{6}-10^{8}$\,M$_{\odot}$kpc$^{-2}$.}
Hence these top-down perspective plots are used here merely as a first look at the Galactic distribution of clouds. A more detailed study of arm/inter-arm dependency requires using a model of the spiral pattern, and is most accurately done in the $\ell bv$ space, which is beyond the scope of this paper. 

In order to look for effects that could depend on the Galactic environment, without the need to assume any specific spiral arm model, we have examined the spatial distribution of clouds with extreme properties (i.e. clouds that form the tails of a distribution), and compared those to the global Galactic distribution of clouds. The idea behind this exercise is a purely statistical one, which will test whether the most extreme clouds follow the same spatial distribution as the global population of clouds, or whether they show significant deviations from it. As an attempt to take this analysis a step further, we can make the loose assumption that the spiral arms should preferentially be represented by the crowded regions of the global population, while the inter-arm regions would be preferentially associated with the least crowded places. This assumption is purely qualitative (due to the uncertainties in the distances), and we make no attempt to effectively associate clouds with spiral arms or inter-arm regions.
For our purpose, we use the complete science sample as our global cloud population (bottom-left panel of Fig.\,\ref{fig:top_down_1}), from which we selected a number of sub-samples that comprise the most extreme clouds. This selection was made by taking the most extreme 100 clouds of each distribution (corresponding to the top or bottom {6}\%), and the specific selection criterion is indicated at the top of each panel in Figs.\,\ref{fig:top_down_extremes_1} and \ref{fig:top_down_extremes_2}. 

The comparison between the sub-samples and the global cloud population was done by performing the Pearson's $\chi^{2}$ statistical test, which tests whether the frequency distribution of certain events observed in a sample is consistent with a particular theoretical distribution. The full details of the $\chi^{2}$ statistical test that we performed are explained in App.\,\ref{app:chisq_test}. In brief, for our purpose, we used the 2D Galactic distribution of clouds in the complete science sample as our theoretical distribution. In practice, we built a normalised 2D histogram with the spatial distribution of clouds in the complete science sample (using their Galactocentric coordinates), using a spatial bin of $0.3\times0.3$\,kpc -- this map represents the probability of an observation falling in a specific spatial bin (see left panels of Figs.\,\ref{fig:chi-sq-1} to \ref{fig:chi-sq-3}). We then compute the $\chi^{2}$ statistics using the observed 2D distribution of each sub-sample (shown in the central panels of Figs.\,\ref{fig:chi-sq-1} to \ref{fig:chi-sq-3}), and the observed $\chi^{2}$-values are compared to the values obtained from a pure random draw of clouds from the theoretical distribution (i.e. effectively obtaining a $p$-value, which we call $p_{\rm{rnd}}$, see Fig.\,\ref{fig:chi-sq-test}). Given the statistical fluctuations, as well as the uncertainties in the distributions, and binning effects (neither of which are taken into account for this exercise), the exact $\chi^2$ values and $p_{\rm{rnd}}$ that we derive should not be taken at face value. Instead, they are more useful for a relative comparison of the sub-samples, as an indication for which sub-samples are most different to the global cloud population. The results from our $\chi^{2}$ statistical test are summarised in Table\,\ref{tab:Xsq_stats}. We describe all of the studied tails of distributions 
in the following Sects.\,\ref{sec:extreme_mass} to \ref{sec:extreme_hmsf}.

\subsection{The most massive molecular cloud complexes}
\label{sec:extreme_mass}

\begin{figure*}
\centering
\includegraphics[width=0.9\textwidth]{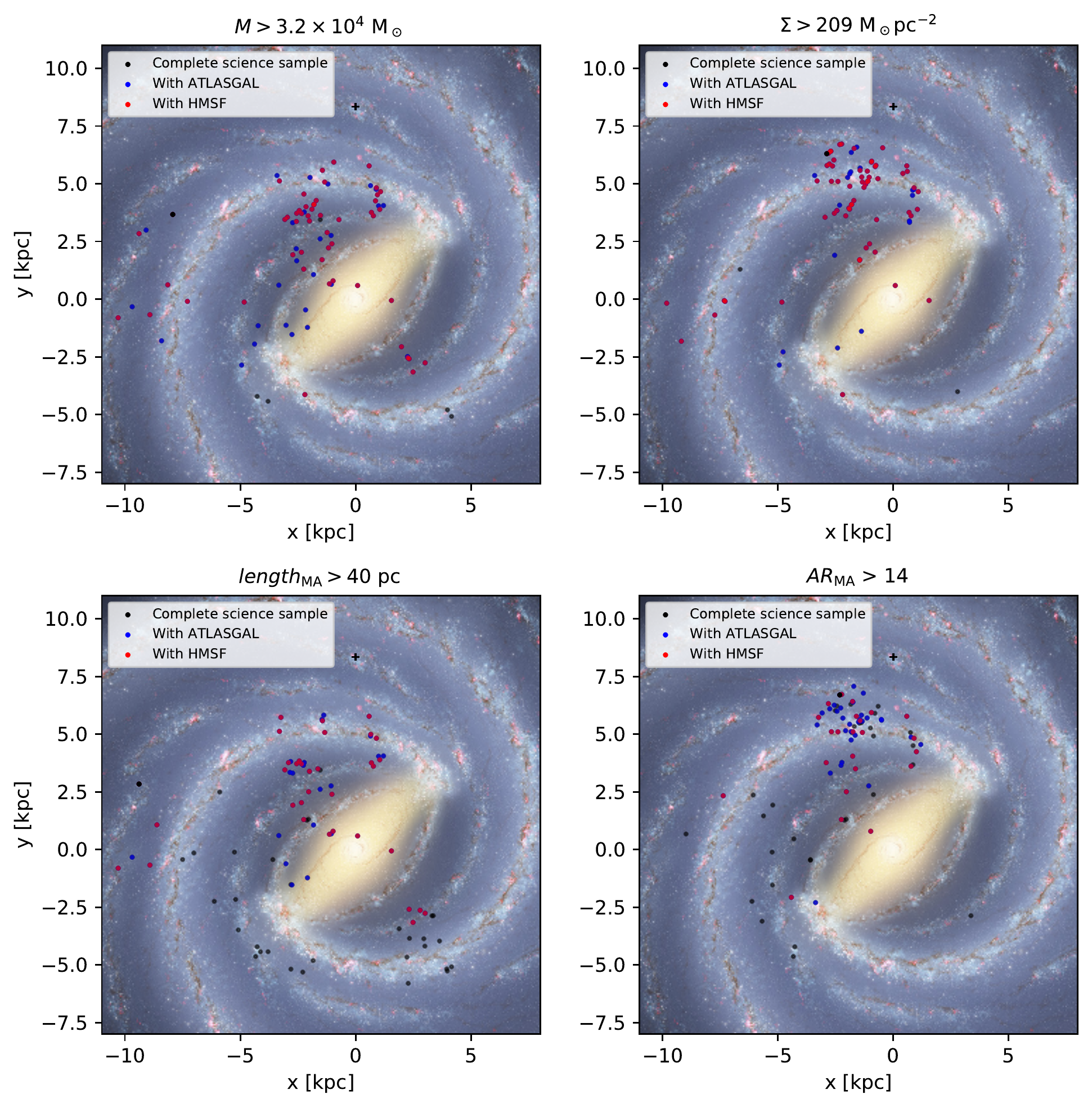}
\caption{Top down view of the Galaxy as in Fig.\,\ref{fig:top_down_1}, showing the SEDIGISM clouds of the complete science sample that are part of the top 10$\%$ of clouds in terms of Mass (top-left), surface density (top-right), medial axis length (bottom-left), and aspect ratio from the medial axis (bottom-right).  The specific condition that corresponds to this cut-off is indicated at the top of each panel. Clouds are colour-coded depending on whether they have an ATLASGAL counterpart (in blue), a HMSF signpost (in red), or neither (in black).}
\label{fig:top_down_extremes_1}
\end{figure*}

Some observations of nearby spiral galaxies (e.g. \citealt{Koda2009}, \citealt{Colombo2014}), as well as some galaxy-scale numerical models (e.g. \citealt{dobbs2008}, \citealt{Fujimoto2014}, \citealt{dc2016}, \citealt{Pettitt2018}) - both of which benefit from a more straightforward association of clouds to spiral arms - have suggested that the most massive clouds are preferentially located along spiral arms. 
This is widely accepted and understood in the context of spiral arms being able to concentrate more material, and thus able to form larger and more massive giant molecular clouds (GMCs). Similarly to the argument for encountering the most massive clouds in the arms, with a higher concentration of material in the spiral arms we ought to expect the highest surface density clouds to lie in the spiral arms as well. In this spirit, we have plotted the Galactic distribution of the 100 most massive clouds in our sample, and the 100 clouds with the highest surface density, in the top-left and top-right panels of Fig.\,\ref{fig:top_down_extremes_1}, respectively. 

Our $\chi^{2}$ tests comparing these two distributions to the global cloud population, give us $\chi^{2}$ values of {670 and 638} (which corresponds to a $p_{\rm{rnd}}$ of {0.05 and 0.16}), for the extreme mass and surface density clouds respectively. This suggests that the distribution of high-surface density clouds still follows the original distribution of clouds, implying that such clouds might be found in crowded areas (or spiral arms), simply from statistics. The distribution of the most massive clouds, however, is less consistent with a pure random draw of clouds from the parent distribution. If the disparities between the two distributions were caused by having more high-mass clouds in the spiral arms than what is statistically expected, then we should see an excess of high-mass clouds in the most crowded areas of the global distribution. However, considering the spatial distribution of these clouds on Fig.\,\ref{fig:top_down_extremes_1} (top panels), and the relative difference between the predicted and measured counts shown in Fig.\,\ref{fig:chi-sq-1} (top and middle rows), it is not obvious that this is the case, with clouds having both an excess and lack of counts in different crowded areas. The specific regions where the most high-mass clouds are found to be in excess or lacking, are not particularly striking in terms of their environment, leaving our interpretation inconclusive.

\subsection{The most elongated clouds}
\label{sec:extreme_morphology}

A subject of increasing interest in the SF community is the origin and properties of the most elongated clouds. While some numerical and observational studies suggest that extremely long filamentary clouds would be formed as the result of the Galactic shear in the inter-arm regions \citep[e.g.][]{Kim2002,Shetty06,Ragan2014,dc2016,dc2017}, other studies suggest that at least some of these might trace the ``spines'' of the spiral arms \citep[e.g.][]{Goodman2014,Wang2015,Zucker2015}. 

We have thus looked at the Galactic distribution of the 100 longest clouds in the SEDIGISM sample, as well as the 100 clouds with the largest aspect ratio. These are shown in the bottom panels of Fig.\,\ref{fig:top_down_extremes_1} (left and right respectively).  Our $\chi^{2}$ tests for these two distributions, give us a  $\chi^{2}$ value of {715 and 671} (corresponding to a $p_{\rm{rnd}}$ of {0.005 and 0.04}) for the extreme length and aspect ratio clouds respectively. This suggests that the Galactic distribution {of both these sub-samples are} different from the global cloud population {(although this is most evident for the sample of largest clouds)}. However, neither of them seem to show any clear preference for crowded or non-crowded areas (see also Fig.\,\ref{fig:chi-sq-1} bottom panel and Fig.\,\ref{fig:chi-sq-2} top panel).

This analysis has a few caveats, though. The first one is that there are more of these elongated clouds located at the near distance, than there are at the far distance. This could be linked to resolution limitations which will result in more distant filaments appearing less elongated. The second caveat is the fact that both of these quantities are purely the projected ones (the length and aspect ratio on the plane of the sky). If long filamentary clouds are indeed shaped by the shear from the Galactic differential rotation, we do not expect them to be randomly orientated. Therefore, this projection is likely to affect our ability to select the truly elongated structures, in specific parts of the Galaxy, being particularly critical in lines of sight where we expect the clouds' elongations to be roughly along our line of sight. The third caveat is the fact that even the longest molecular filaments in our Galaxy (such as the $\sim$100\,pc long Nessie filament, \citealt{Jackson2010}), do not appear in our segmentation as a single entity - they are instead composed of several (smaller) filamentary sections. Finally,  the relative lack of large (and massive) clouds nearby (with $d < 2.5$\,kpc), can also point at a possible bias from the cloud segmentation, in which we might still be more likely to break the most nearby clouds into smaller sub-structures (even though we use a clustering algorithm designed to minimise this effect). All of these effects could result in the underestimation of both the length and aspect ratio of clouds, and thus the 100 most extreme clouds we take for this analysis might not correspond to the most extreme cases in physical space.

\subsection{The most dynamically active clouds}
\label{sec:extreme_dynamic}

\begin{figure*}
\centering
\includegraphics[width=0.9\textwidth]{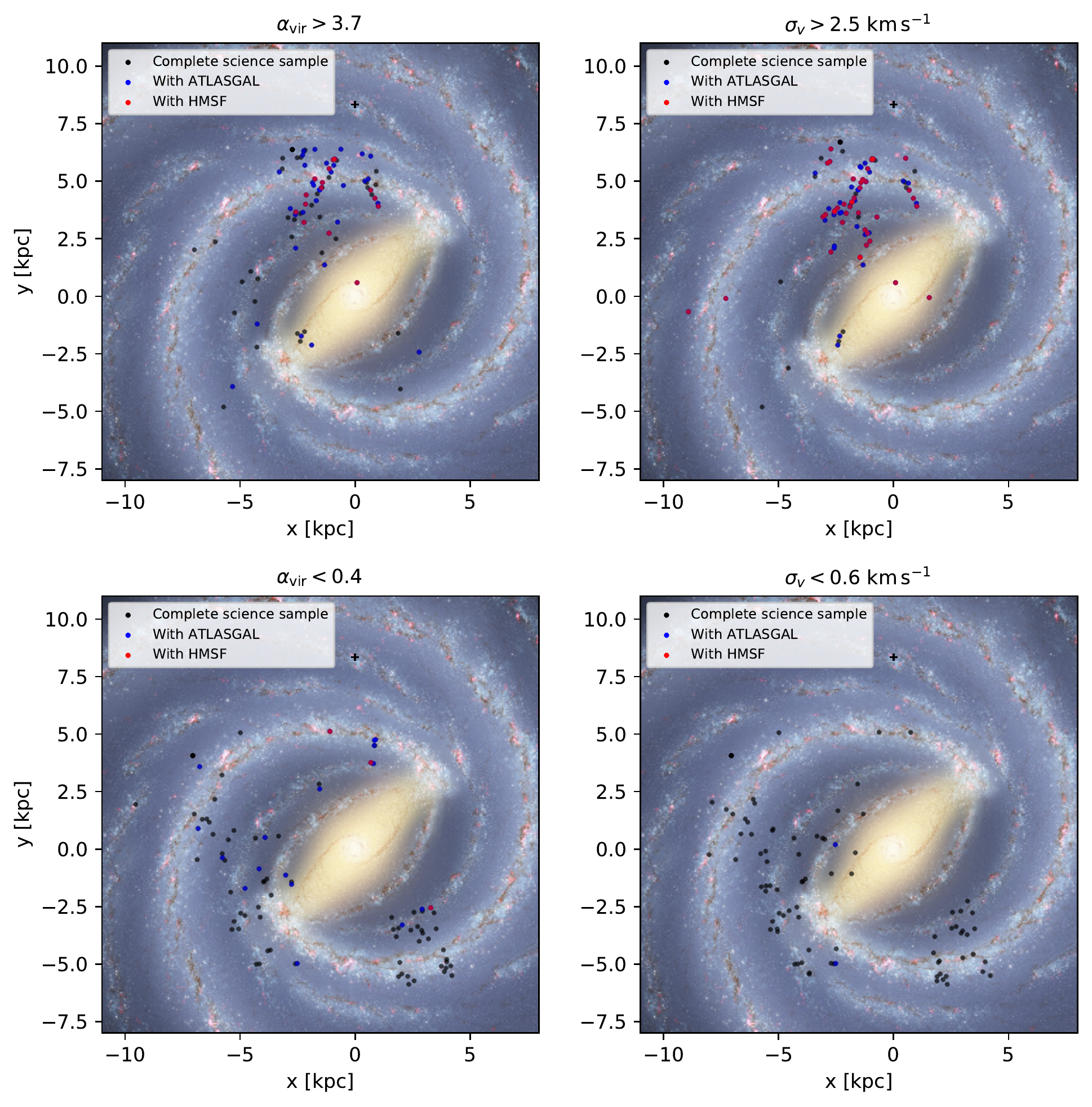}
\caption{Same as Fig.\,\ref{fig:top_down_extremes_1}, showing the SEDIGISM clouds of the complete science sample that are part of the top 10$\%$ of clouds with a high virial parameter (top-left), and high velocity dispersion (top-right). The lower panels show the bottom 10\% of clouds in the same properties: with a low virial parameter (bottom-left) and a low velocity dispersion (bottom-right).  The specific condition that corresponds to this cut-off is indicated at the top of each panel. Clouds are colour-coded depending on whether they have an ATLASGAL counterpart (in blue), a HMSF signpost (in red) or neither (black).}
\label{fig:top_down_extremes_2}
\end{figure*}

Given the complex global Galactic dynamics, we would expect to see, at least at first order, some link between the most dynamic places in the Galaxy, with the kinetic properties of clouds. In this sense, we have isolated the 100 clouds with the highest virial parameter, and highest velocity dispersion. Their Galactic distribution is shown in Fig.\,\ref{fig:top_down_extremes_2}, top panels. Our $\chi^{2}$ test for clouds with a large virial parameter gives us a $\chi^{2}$ value of {699} (corresponding to a $p_{\rm{rnd}}$ of {0.01}), indicating that they differ from a random statistical subset of the global cloud population, in terms of their Galactic placement (see also Fig.\,\ref{fig:chi-sq-2} middle panel). On the other hand, clouds with a large velocity dispersion have a higher $\chi^{2}$ value of {747} (which corresponds to a much smaller $p_{\rm{rnd}}$ of {0.001}), making this distribution less like the global cloud population. Most of the differences in the statistics of this sub-sample comes from an excess of high-velocity dispersion clouds relatively nearby (see top-right panel of Fig.\,\ref{fig:top_down_extremes_2}, and bottom panel of Fig.\,\ref{fig:chi-sq-2}), which then also propagates (although less severely) into clouds with high-virial parameters also being mostly nearby. We believe that these trends could be partly due to observational biases (see Fig.\,\ref{fig:low_dynamics}, and the discussion in App.\,\ref{sec:completeness}). 

Interestingly, these dynamically active clouds typically make up two types of populations. The first is most closely associated with crowded regions (potentially associated with the near Sagittarius, Scutum and Norma spiral arms), which is where we expect more frequent cloud-cloud interactions, in line with the results from numerical simulations of spiral galaxies \citep[e.g.][]{dc2017,Pettitt2018}. This population of clouds is also actively forming high-mass stars. The larger values of velocity dispersion and virial parameters could thus be also an indication of larger internal motions of clouds, perhaps partly driven by their active gravitational contraction, or by internal feedback from the forming stars, or both.

The second population of clouds are devoid of HMSF signposts, and some even lacking an ATLASGAL counterpart (i.e. less dense). Most of these are also at large distances, which could suffer from a completeness effect in the ATLASGAL and HMSF tracers. Alternatively, this second population could represent clouds relatively close to the Galactic bar, and/or in the streams of gas feeding the Galactic centre region -- all regions prone to experiencing a significant shear driven by the global Galactic dynamics. This dichotomy (of clouds in the two extremes of their SF history sharing the same integrated dynamical properties) highlights the caveats of performing a standard virial analysis and deriving any conclusions therefrom alone. 

\subsection{The most dynamically quiescent clouds}
\label{sec:extreme_low_dynamic}

On the opposite extreme of the dynamical status of molecular clouds, we have also explored the location of the clouds that are relatively quiet (which we refer to as the most ``dynamically quiescent'' clouds), which include clouds with a low virial parameter, or a low velocity dispersion. These types of clouds are often not subject of much attention (mostly as they typically lie close to survey limitations in terms of spectral resolution). Nevertheless, some recent numerical work by \citet[][]{Pettitt2018} has suggested that, in grand-design spiral galaxies, while clouds with high virial parameter are most often associated with spiral arms, clouds with low virial parameters have a weaker correspondence with the spiral arms, with many inter-arm clouds being remnants of large arm complexes or simply formed in-situ from small over-densities in filaments and arm spurs. 

To investigate these dynamically quiescent clouds in SEDIGISM, we have selected the 100 clouds in the complete science sample with the lowest virial parameter, and the lowest velocity dispersion. Their Galactic distribution is shown in the bottom panels of Fig.\,\ref{fig:top_down_extremes_2}. Our $\chi^{2}$ tests give us $\chi^{2}$ values of {675 and 669} (corresponding to a $p_{\rm{rnd}}$ of {0.04 and 0.05}) for the low virial parameter and low velocity dispersion respectively. This suggests that the Galactic distribution of the most dynamically quiescent clouds is only {mildly} different to that of the global cloud population.  
Their distribution in Fig.\,\ref{fig:top_down_extremes_2} (see also Fig.\,\ref{fig:chi-sq-3} top and middle row) suggests that they are not found in very crowded areas (possibly favouring inter-arm locations).  

Clouds with a low virial parameter are often interpreted to be gravitationally bound (i.e. where gravity dominates over turbulence). However, these clouds are not necessarily collapsing - indeed if they were, the collapse itself would increase the virial parameter again \citep[e.g.][]{kauffmann2013}. Our results show that these dynamically quiescent clouds are mostly devoid of HMSF or even high-column densities (which would result in an  ATLASGAL counterpart), perhaps indicating that their evolution is not regulated by their own gravity but by interaction with the Galactic potential, the large scale shear motions and perhaps also by large scale magnetic fields. 

We caution however, that even though a handful of dynamically quiescent clouds are relatively nearby, most of them are at $d > 8.0$\,kpc. In terms of absolute numbers, the science sample does contain nearby low-velocity-dispersion clouds, but most of those are below the size and/or mass threshold used to build the complete science sample. The usage of a completeness limit for the whole SEDIGISM sample (and especially one largely above the resolution element) was an attempt to remove any bias from the resolution and distance. However, our intrinsic observational limitations may still be responsible for at least part of this signature, as we can see that the average measured velocity dispersion of the complete science sample has a correlation with distance (see Fig.\,\ref{fig:low_dynamics}, and the respective discussion in App.\,\ref{sec:completeness}). Furthermore, at the far distances our sample may also not be complete in terms of the detection of an ATLASGAL counterpart or HMSF signposts, potentially biasing the interpretation above. 

Nevertheless, these type of clouds could potentially be interesting to follow up with the goal to investigate whether this tentative trend does hold up, with a more in-depth analysis, considering the survey limitations and a detailed modelling of the spiral pattern of the Galaxy.

\subsection{The high-mass star-forming clouds}
\label{sec:extreme_hmsf}

One of the questions we wanted to address here is whether the Galactic distribution of clouds that host ongoing high-mass star formation is uniform, or whether they are preferably located in spiral arms as our preliminary study of the SEDIGISM science verification field suggested \citep[][]{Schuller2017}. In particular, if high-mass star-forming clouds are tracing the arms, we are also interested in exploring whether that is purely due to a statistical sampling \citep[as suggested by e.g.][]{Elmegreen1986,Moore2012,Eden2013}; or whether there is an excess of high-mass star-forming regions in the crowded spiral arms, suggestive of SF triggering from the passage of a spiral wave (e.g.  \citealt{Lin1964}, \citealt{Roberts1969}, \citealt{Toomre1977}, \citealt{MartinezGarcia2009}).

Figure\,\ref{fig:top_down_agal} (bottom-right panel), shows the distribution of all clouds with a HMSF signpost in our science sample (in red). The $\chi^{2}$ statistical test, performed using only the clouds in the complete science sample (from which only 211 clouds have a HMSF signpost) gives a $\chi^{2}$ value of {735}, which translates into a $p_{\rm{rnd}}$ of 0.001. This indicates that the distribution of clouds with a HMSF signpost does not mimic the global distribution of clouds. Upon closer inspection of Fig.\,\ref{fig:top_down_agal} and \ref{fig:chi-sq-3}, it becomes clear, however, that most of the deviations from the global distribution of clouds do not arise from crowded or non-crowded areas, but rather shows a distance effect.  Indeed, most of the clouds with signs of on-going high-mass star formation are located relatively close to us. The extremely high density of points there (compared to elsewhere in the Galaxy), is likely to be a simple consequence of completeness in the HMSF signposts (namely H{\sc ii} regions and massive YSOs). 

\begin{figure}
\centering
\includegraphics[width=0.45\textwidth]{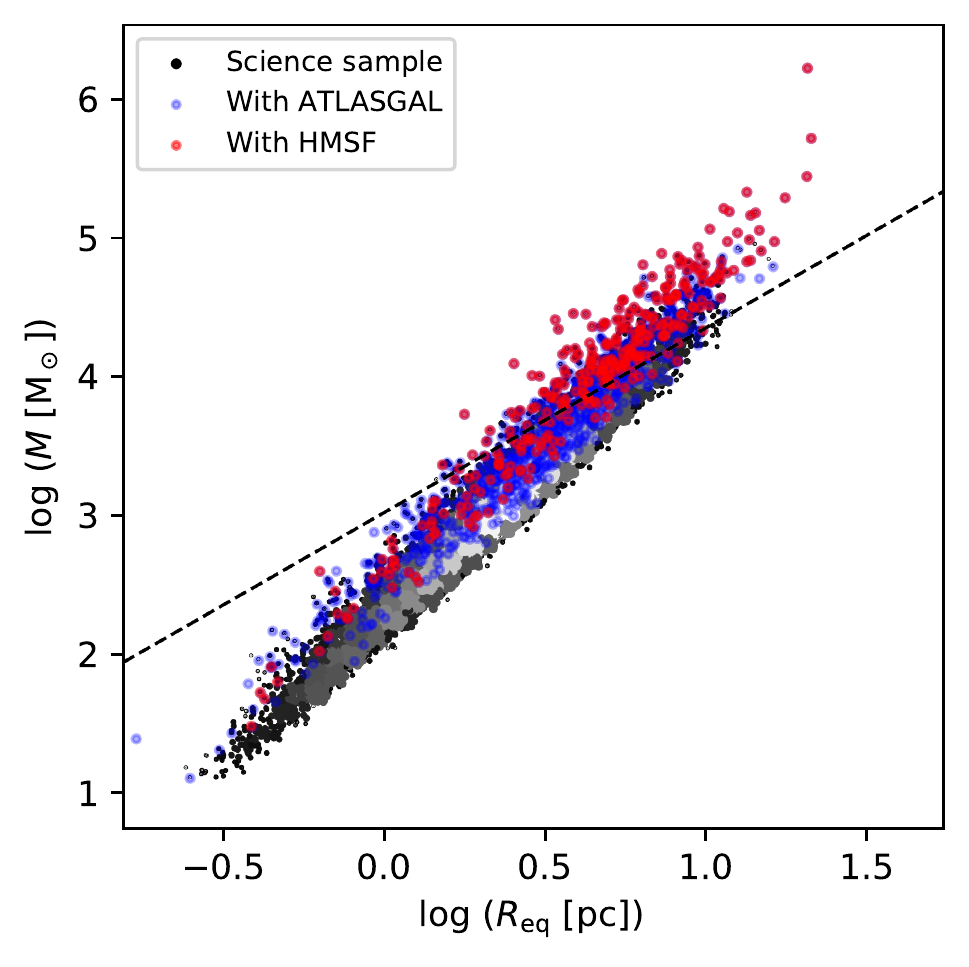}
\vspace{-0.3cm}
\caption{Mass-radius relation for the SEDIGISM clouds, where the grey scale represents the density of points for the entire science sample, the blue circles show the clouds with an ATLASGAL counterpart, and the red circles show the clouds that have a HMSF signpost. The dashed line shows the empirical relation from \citet{KauffmannPillai2010}, where clouds above this line are expected to form high-mass stars. {The plotted threshold is at $M [$M$\odot]=1053~(R [$pc$])^{1.33}$, which is scaled up from the original HMSF threshold from \citet{KauffmannPillai2010}, to account for the different opacity law used (see App.\,\ref{app:opacity}).}}
\label{fig:mass_radius}
\end{figure}

Interestingly, if we look at the higher-mass clouds or the higher-surface density clouds (Fig.\,\ref{fig:top_down_extremes_1} top panels), not all of these host high-mass star formation. This is true even if we just consider the most nearby clouds, where we should be less affected by completeness issues in terms of HMSF signposts. As we have seen in Sect.\,\ref{sec:global_properties}, there does not seem to be a unique global property of a molecular cloud that defines the ability of a cloud to form high-mass stars - and the same applies for the Galactic environment. Perhaps to isolate clouds with a potential to form massive stars, we need to use a combination of conditions that need to be satisfied, or even just the most extreme conditions within a cloud (rather than the integrated properties). Applying a single global threshold law \citep[such as a gas surface density threshold or mass-radius threshold, e.g.][]{Krumholz2008,KauffmannPillai2010, baldeschi2017} to define the potential to form massive stars, is probably not a single unique descriptor. Figure\,\ref{fig:mass_radius} highlights this issue, where we can see clouds with and without high mass star formation that have the same mass and radius. In this figure, we also show as a dashed line, the empirical relation for high-mass star formation inferred by \citet{KauffmannPillai2010}, {and confirmed by other works \citep[e.g.,][]{kauffmann2010a,kauffmann2010b,urquhart2018_csc}. Note that the plotted line is the \citet{KauffmannPillai2010} original threshold scaled up so as to be consistent with our adopted opacity law (see App.\,\ref{app:opacity} for more details).}

Although this empirical relation was determined for clumps, rather than for clouds (as we use here), the bulk of the parameter space that we probe is similar to that in \citet{KauffmannPillai2010}: their sizes range from <0.1\,pc to 10\,pc (compared to our range of 0.3\,pc to $\sim$30\,pc), and their masses range from 1\,M$_{\odot}$ to >\,$10^{4}$\,M$_{\odot}$ (compared to our range of 10\,M$_{\odot}$ to $>10^{5}$\,M$_{\odot}$).

If we use that relation directly with our clouds, we would miss some true positives ({107} out of 330 clouds with a HMSF signpost lie below the empirical threshold, i.e. missing {$\sim$33\%} of all clouds that we know are actively forming massive stars), as well as potentially provide a significant number of false positives ({455} out of a total of {678} clouds above the HMSF threshold do not have a detected HMSF signpost, i.e. {$\sim70$\%} of clouds above the empirical line for HMSF). Since completeness limits could play a role in the non-detection of the signposts for HMSF, we estimate that the number of false negatives (i.e. missed true positives) is a lower limit, while the number of false positives is an upper limit.  

{The detection of potential false positives was not ruled out by \citet{KauffmannPillai2010}: indeed they note that their threshold appears to capture a necessary condition for HMSF, but not a sufficient one.} 
Alternatively, it could also be that part of the clouds above the HMSF threshold line but for which we have no detected HMSF (i.e. the false positives), are in fact clouds that simply have not done so yet, because of the potential large latency periods prior to star formation. In that sense, the trends in properties going from the science sample to clouds with an ATLASGAL counterpart and then clouds with a HMSF signpost (from Sect.\,\ref{sec:global_properties}) could be an indication of the cloud evolution towards HMSF during this latency period (with clouds progressively building up their mass, becoming larger, denser, and more dynamically active - with larger velocity dispersions), even if this remains a stochastic process for each individual cloud \citep[e.g.][]{Barnes18}.

{More intriguing, however, are the missed true positives. These clouds lie below the empirical line supposedly representing the threshold below which HMSF would not occur, and yet they have tracers of ongoing HMSF. }
Nevertheless, it is possible that the material probed by \citet{KauffmannPillai2010} is intrinsically tracing higher density material than what we do, which could shift the exact position of the cloud sample with respect to the empirical line for HMSF, thus potentially making this relation inappropriate for usage with our sample.
{An indication that this might indeed be the case, is the fact that the subsample of SEDIGISM clouds with a HMSF signpost that we present here, is purely a subsample of the ATLASGAL clumps, which seem to confirm the \citet{KauffmannPillai2010} relation on clump scales \citep[e.g.][]{urquhart2018_csc}. This highlights a potential caveat of using such relations blindly, as perhaps they are not applicable on cloud scales, when the density profiles become shallower, and the more diffuse material contributes to increasing the sizes of the clouds, whilst providing only moderate increase to the enclosed mass. A hierarchical study of this transition within clouds would be required to understand where this relation might break.}


\section{Summary and conclusions}
\label{sec:summary}

The SEDIGISM survey has covered $\sim84$ square degrees of the inner Galaxy with $^{13}$CO (2-1). From the contiguous portion of the survey (i.e. excluding the W43 field), we extracted the entire molecular cloud population with a large dynamic range in spatial scales, using the Spectral Clustering for Interstellar Molecular Emission Segmentation ({\sc{scimes}}) algorithm. We determined the distances to the clouds, using the kinematic distances, and a number of methods to solve the distance ambiguities (including masers, IRDC, Dark Clouds, \hi SA, distance to the Larson's size-linewidth relation, distance to the Galactic plane, and extinction distances). The full catalogue that we release contains 10663 molecular clouds, 10300 of which with measurements of physical properties. 

In this paper, we have explored some of the global properties of clouds using a sub-sample of the full catalogue (i.e. our ``science sample''), consisting of {6664} well resolved sources and for which the distance estimates are reliable. In particular, we compare the scaling relations retrieved from SEDIGISM to those of other surveys, including Galactic and extragalactic work. We find that the locus of the SEDIGISM clouds is similar to that of other surveys, but that the specific scaling relations vary widely between surveys - even between those that cover the same area in the Galaxy, just with different tracers. The intrinsic scatter in these relations is very large, making all the correlations rather unconstrained. 

We also explored the properties of clouds with and without tracers of high-mass star formation, and we find that for most distributions (mass, size, surface density, velocity dispersion), the median values of the distributions is higher for clouds with a HMSF signpost, potentially indicative of an evolutionary sequence. However, the distributions become progressively flatter, with the clouds with HMSF spanning a wide range of values for all properties we looked at. These results suggest that there is no single global property of a cloud that is able to define their ability to form massive stars, and the usage of a simple threshold to isolate clouds forming high-mass stars is not complete (providing both false negatives and false positives). 

Finally, we have looked into potential links between the Galactic environment of clouds and their properties, by looking at the Galactic distribution of the most extreme clouds. For that purpose, we have isolated the most extreme 100 clouds in each distribution (i.e. clouds that make up the tails of the distributions), and compared their Galactic distribution to that of the cloud population above our completeness limits (i.e. our complete science sample), using a $\chi^{2}$ statistical test.  
This provides a means to determine whether extreme clouds follow a Galactic distribution that differs significantly from the global cloud population. We find that, for most properties, the Galactic distribution of the most extreme molecular clouds is  {is only marginally different to} that of the global cloud population. 
{The Galactic distribution of the largest clouds, the most turbulent clouds and the high-mass star-forming clouds are those that deviate most significantly} from the global cloud population. We also find that the least dynamically active clouds (with low velocity dispersion or low virial parameter) {are situated further afield,} mostly in the least populated areas, and therefore could hint at those being mostly in inter-arm regions. However, we find that part of these trends might be due  {to completeness limits (e.g. in case of the HMSF tracers), and} intrinsic survey limitations, which result in a trend of decreasing velocity dispersion with distance, hampering our ability to make any firm conclusions from this data alone. 

In future work, we shall follow up some of these tentative trends using distance-limited samples, with the incorporation of detailed models of the spiral arms, and with more complete cross-match with signposts of HMSF (e.g. by comparing with the Hi-GAL samples, and their $L/M$ ratio as an indicator for more embedded HMSF and their respective evolutionary stage) to mitigate some of the observational biases that are potentially at play in the work presented here.


\section*{Acknowledgements}

ADC acknowledges the support from the Royal Society University Research Fellowship (URF/R1/191609). ADC and AJR acknowledge the support from the UK STFC consolidated grant ST/N000706/1. DC acknowledges support by the Deutsche Forschungsgemeinschaft, DFG, through project number SFB956C. LB and RF acknowledge support from CONICYT grant Basal AFB-170002. HB acknowledges support from the European Research Council under the Horizon 2020 Framework Program via the ERC Consolidator Grant CSF-648505. HB furthermore thanks for financial help from the DFG via the SFB881 "The Milky Way System" (subproject B1). CLD acknowledges funding from the European Research Council for the FP7 ERC consolidator grant project ICYBOB, grant number 818940. M.W. acknowledges funding from the European Union’s Horizon 2020 research and innovation programme under the Marie Skłodowska-Curie grant agreement No 796461. S.B. and N.S. acknowledge support from the Agence National de Recherche (ANR/France) and the Deutsche Forschungsgemeinschaft (DFG/Germany) through the project GENESIS (ANR-16-CE92-0035-01/DFG1591/2-1).The Starlink software \citep{Currie2014} is currently supported by the East Asian Observatory. This publication is based on data acquired with the Atacama Pathfinder Experiment (APEX) under programmes 092.F-9315 and 193.C-0584. APEX is a collaboration among the Max-Planck-Institut fur Radioastronomie, the European Southern Observatory, and the Onsala Space Observatory.
 

\section*{Data availability}

With this paper, we release the complete catalogue of SEDIGISM molecular clouds as extracted using the {\sc{scimes}} code, alongside the masks of each molecular cloud in the catalogue as fits files, and additional ancillary tables in \url{http://sedigism.mpifr-bonn.mpg.de}.
 




\bibliographystyle{mnras}
\bibliography{fb,urquhart2016} 


\vspace{0.5cm}
{\it \small
\noindent $^{1}$ School of Physics \& Astronomy, Cardiff University, Queen's building, The parade, Cardiff, CF24 3AA, U.K. \\
$^{2}$ Max-Planck-Institut f\"ur Radioastronomie (MPIfR), Auf dem H\"ugel 69, 53121 Bonn, Germany. \\
$^{3}$ School of Physical Sciences, University of Kent, Ingram Building, Canterbury, Kent CT2\,7NH, U.K.\\
$^{4}$ Department of Astronomy, University of Florida, 211 Bryant Space Sciences Center, Gainesville, FL, USA \\     
$^{5}$ Laboratoire d'Astrophysique de Marseille, Aix Marseille Universit\'e, CNRS, UMR 7326, F-13388 Marseille, France \\
$^{6}$  West Virginia University, Department of Physics \& Astronomy, P. O. Box 6315, Morgantown, WV 26506, USA \\
$^{7}$  Space Science Institute, 4765 Walnut St. Suite B, Boulder, CO 80301, USA \\
$^{8}$ School of Science and Technology, University of New England, NSW 2351, Australia \\
$^{9}$ Observatorio Astrofisico di Arcetri, Largo Enrico Fermi 5, I-50125 Firenze, Italy	\\
$^{10}$ Max-Planck-Institut f\"ur Astronomie, K\"onigstuhl 17, D-69117 Heidelberg, Germany \\
$^{11}$ Laboratoire d'astrophysique de Bordeaux, Univ. Bordeaux, CNRS, B18N, allée Geoffroy Saint-Hilaire, 33615 Pessac, France. \\
$^{12}$ Departamento de Astronom\'ia, Universidad de Chile, Casilla 36-D, Santiago, Chile \\
$^{13}$ Department of Physics \& Astronomy, University of Exeter, Stocker Road, Exeter, EX4 4QL, United Kingdom \\
$^{14}$ Astrophysics Research Institute, Liverpool John Moores University, 146 Brownlow Hill, Liverpool, L3 5RF, United Kingdom \\
$^{15}$ Haystack Observatory, Massachusetts Institute of Technology, 99 Millstone Road, Westford, MA 01886, USA \\
$^{16}$ Korea Astronomy \& Space Science Institute, 776 Daedeokdae-ro, 34055 Daejeon, Republic of Korea \\
$^{17}$ Department of Physics, Faculty of Science, Hokkaido University, Sapporo 060-0810, Japan \\
$^{18}$ Istituto di Astrofisica e Planetologia Spaziali, INAF, via Fosso del Cavaliere 100, I-00133 Roma, Italy\\
$^{19}$ I. Physikalisches Institut, Universit\"at zu K\"oln, Z\"ulpicher Str. 77, D-50937 K\"oln, Germany \\
$^{20}$ European Southern Observatory, Alonso de Cordova 3107, Casilla 19001, Santiago 19, Chile \\            
$^{21}$ Astronomy Department, University of Wisconsin, 475 North Charter St, Madison, WI 53706, USA \\
$^{22}$ Dept. of Space, Earth and Environment, Chalmers University of Technology Onsala Space Observatory, 439 92 Onsala, Sweden \\
$^{23}$ INAF - Osservatorio Astronomico di Cagliari, Via della Scienza 5, 09047 Selargius (CA), Italy \\
$^{24}$ Univ. Grenoble Alpes, CNRS, IPAG, 38000 Grenoble, France \\
$^{25}$ School of Engineering, Macquarie University, NSW 2109, Australia \\
$^{26}$ Kavli Institute for Astronomy and Astrophysics, Peking University, 5 Yiheyuan Road, Haidian District, Beijing 100871, People's Republic of China \\
}


\appendix

\section{Data-products and catalogues}
\label{app:Catalogue_columns}

\begin{figure*}
\centering
\includegraphics[width=\textwidth]{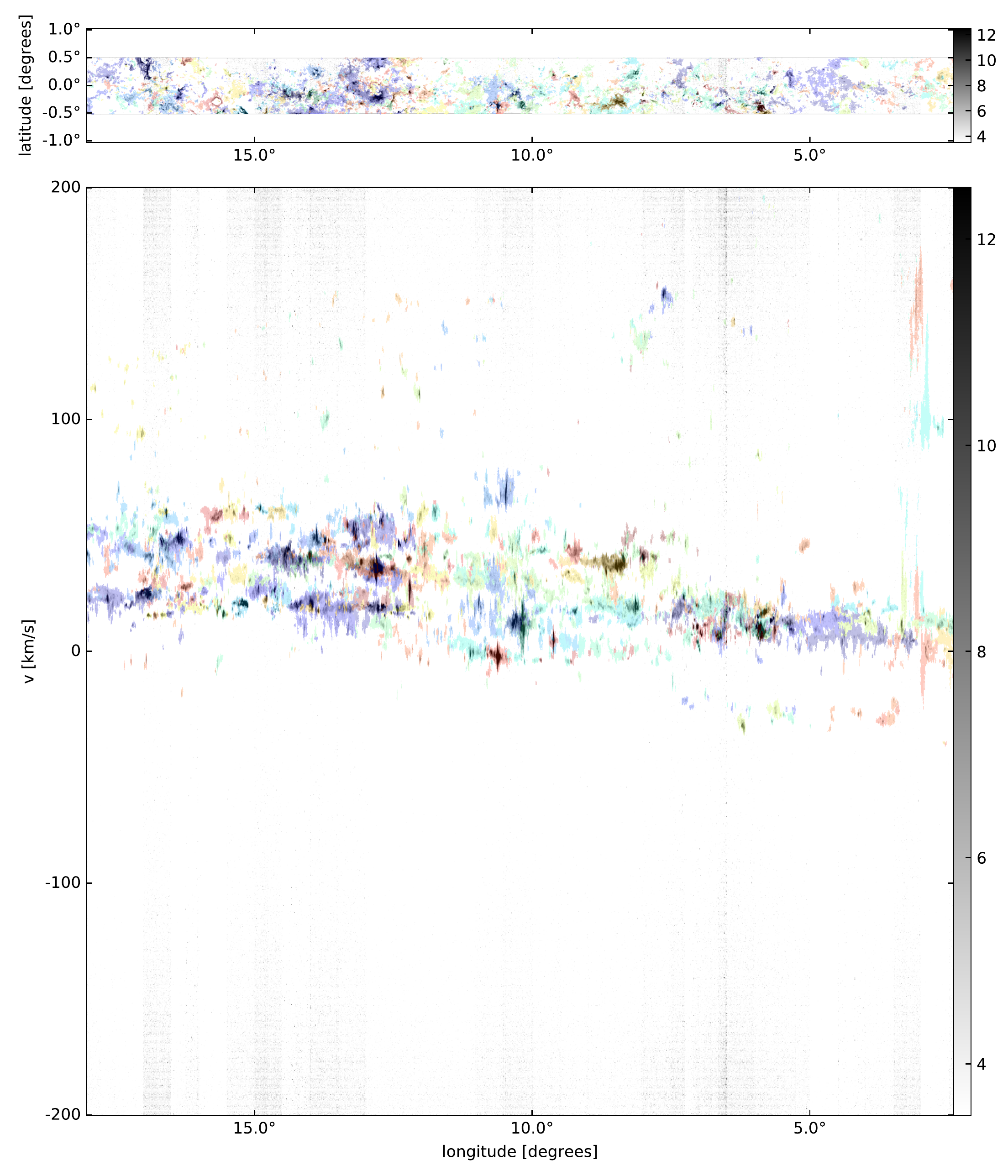}
\caption{$\ell b$ and $\ell v$ plots of the SEDIGISM survey, with the $^{13}$CO peak intensity as the background greyscale, and the SEDIGISM clouds overlaid as colours (each cloud has a different colour, and the colouring scheme is random, but consistent between $\ell b$ and $\ell v$ plots).}
\label{fig:lb_lv_0}
\end{figure*}

\begin{figure*}
\centering
\includegraphics[width=\textwidth]{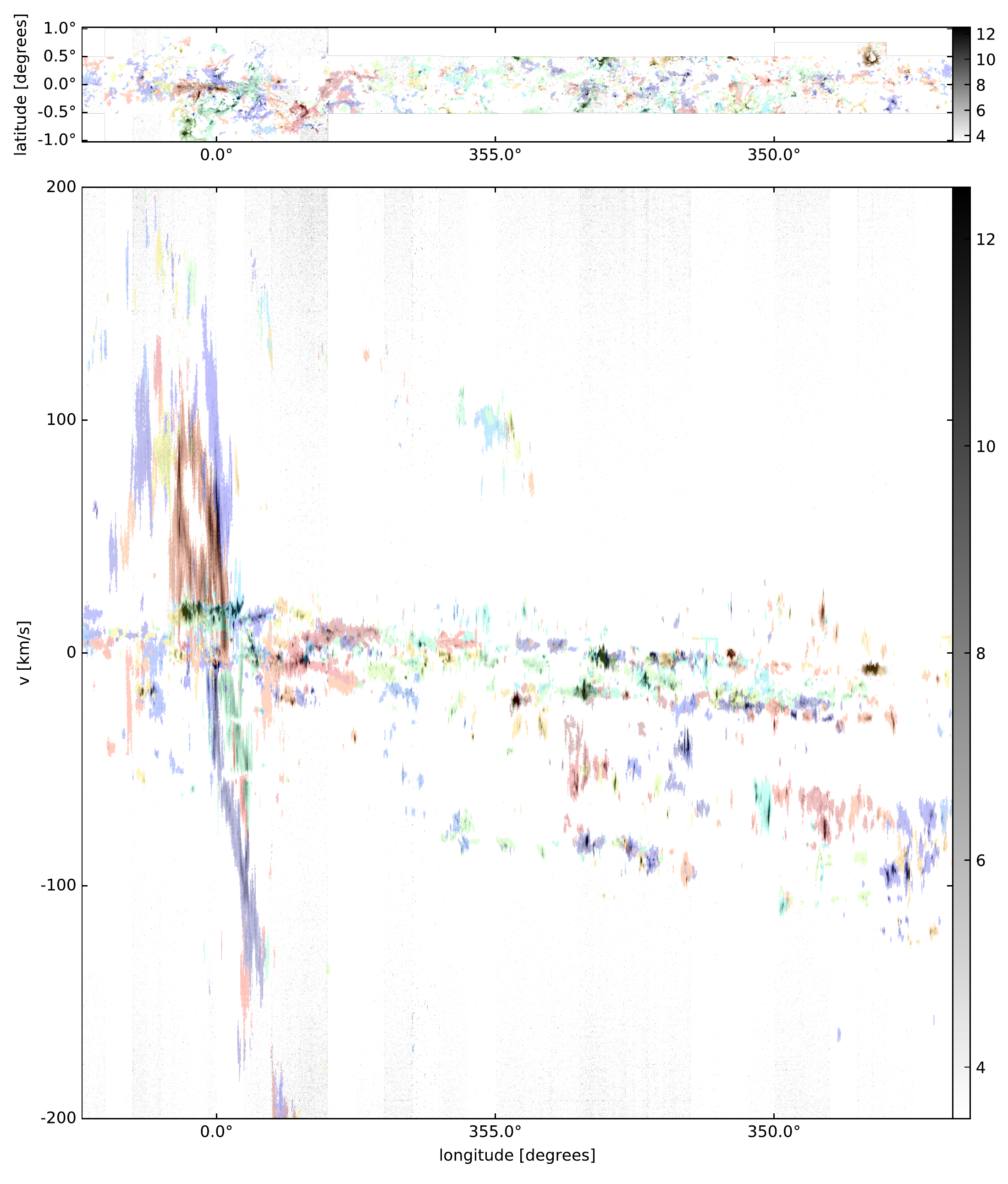}
\caption{Fig.\ref{fig:lb_lv_0} continued.}
\label{fig:lb_lv_1}
\end{figure*}

\begin{figure*}
\centering
\includegraphics[width=\textwidth]{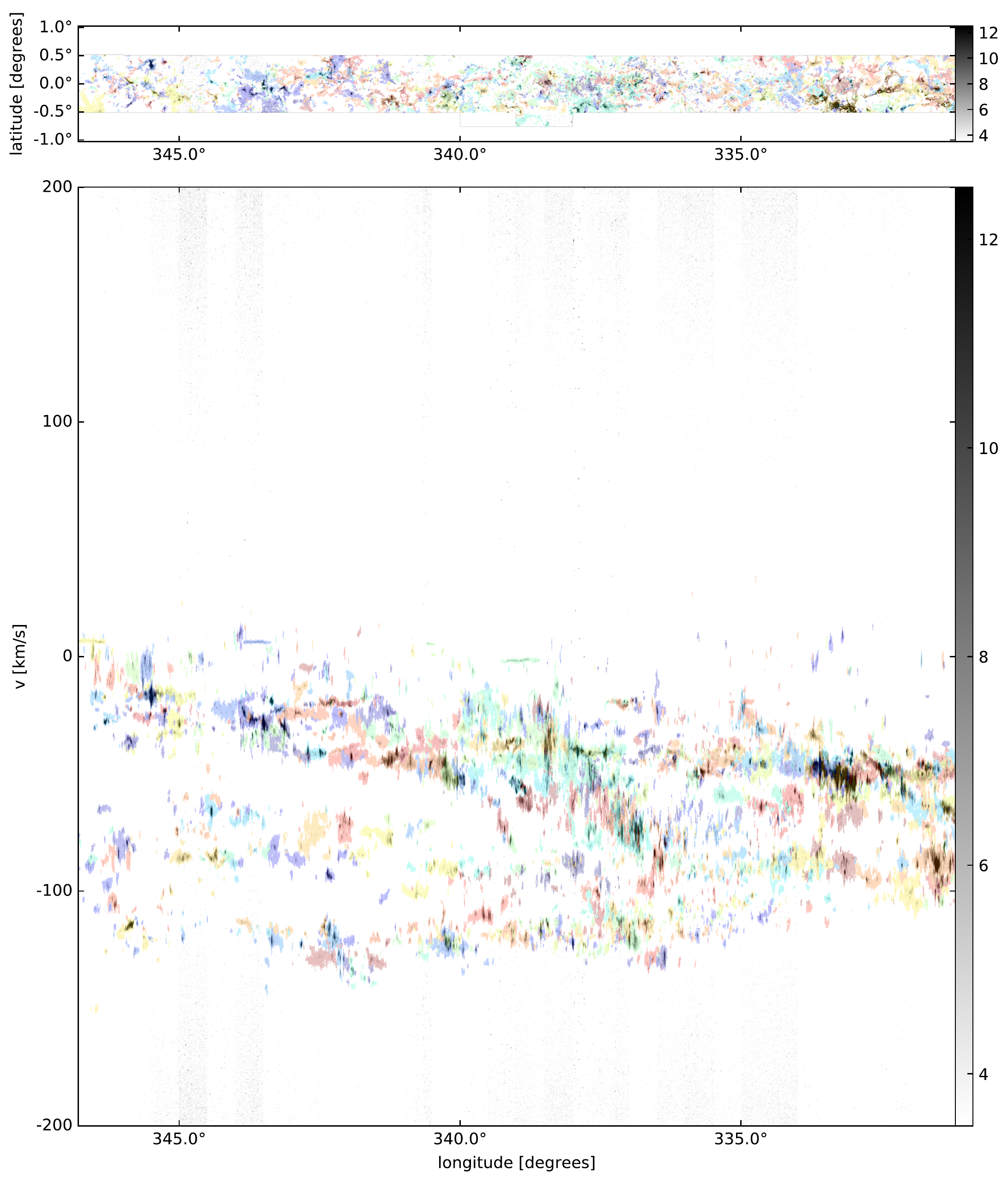}
\caption{Fig.\ref{fig:lb_lv_0} continued.}
\label{fig:lb_lv_2}
\end{figure*}

\begin{figure*}
\centering
\includegraphics[width=\textwidth]{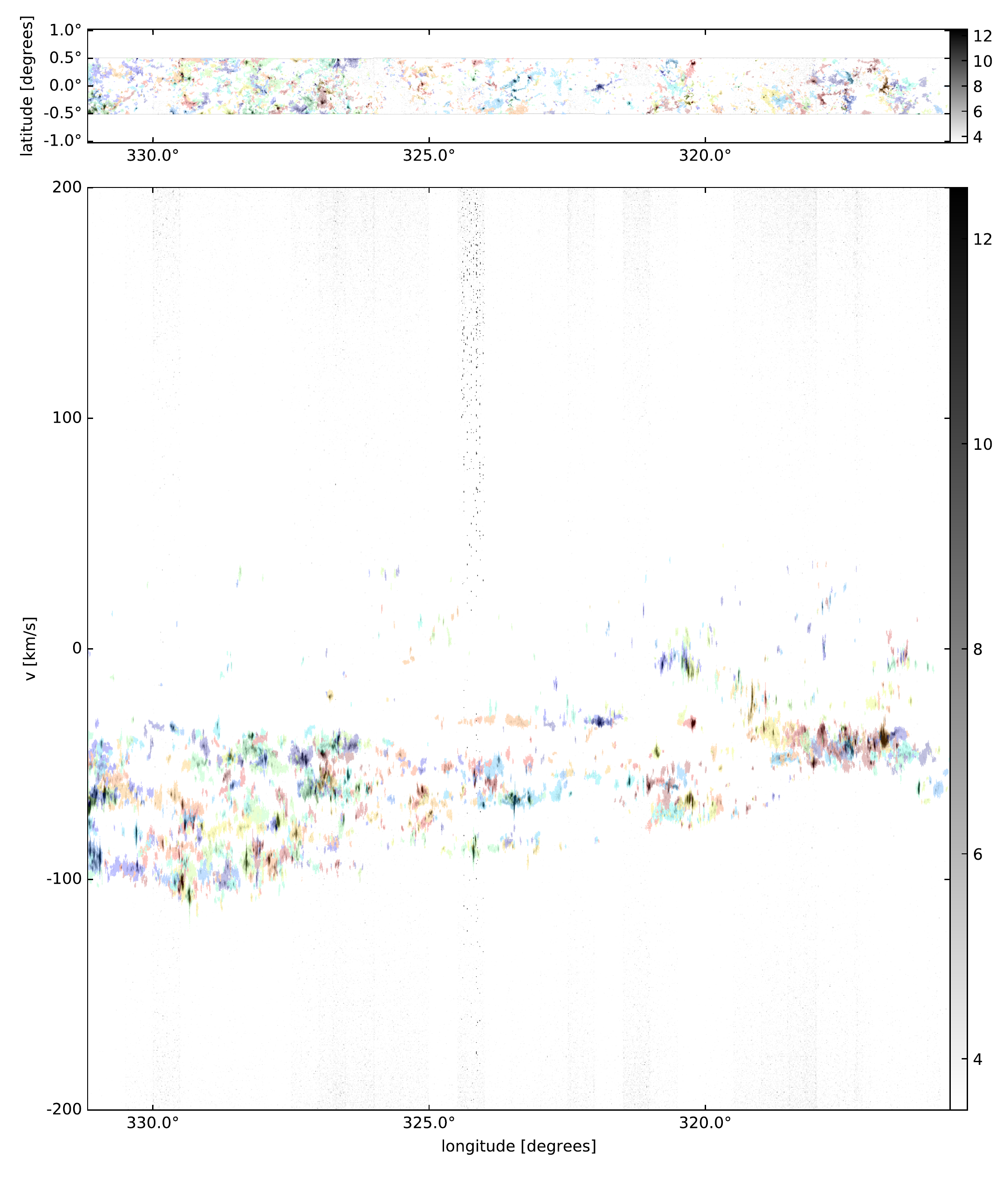}
\caption{Fig.\ref{fig:lb_lv_0} continued.}
\label{fig:lb_lv_3}
\end{figure*}

\begin{figure*}
\centering
\includegraphics[width=\textwidth]{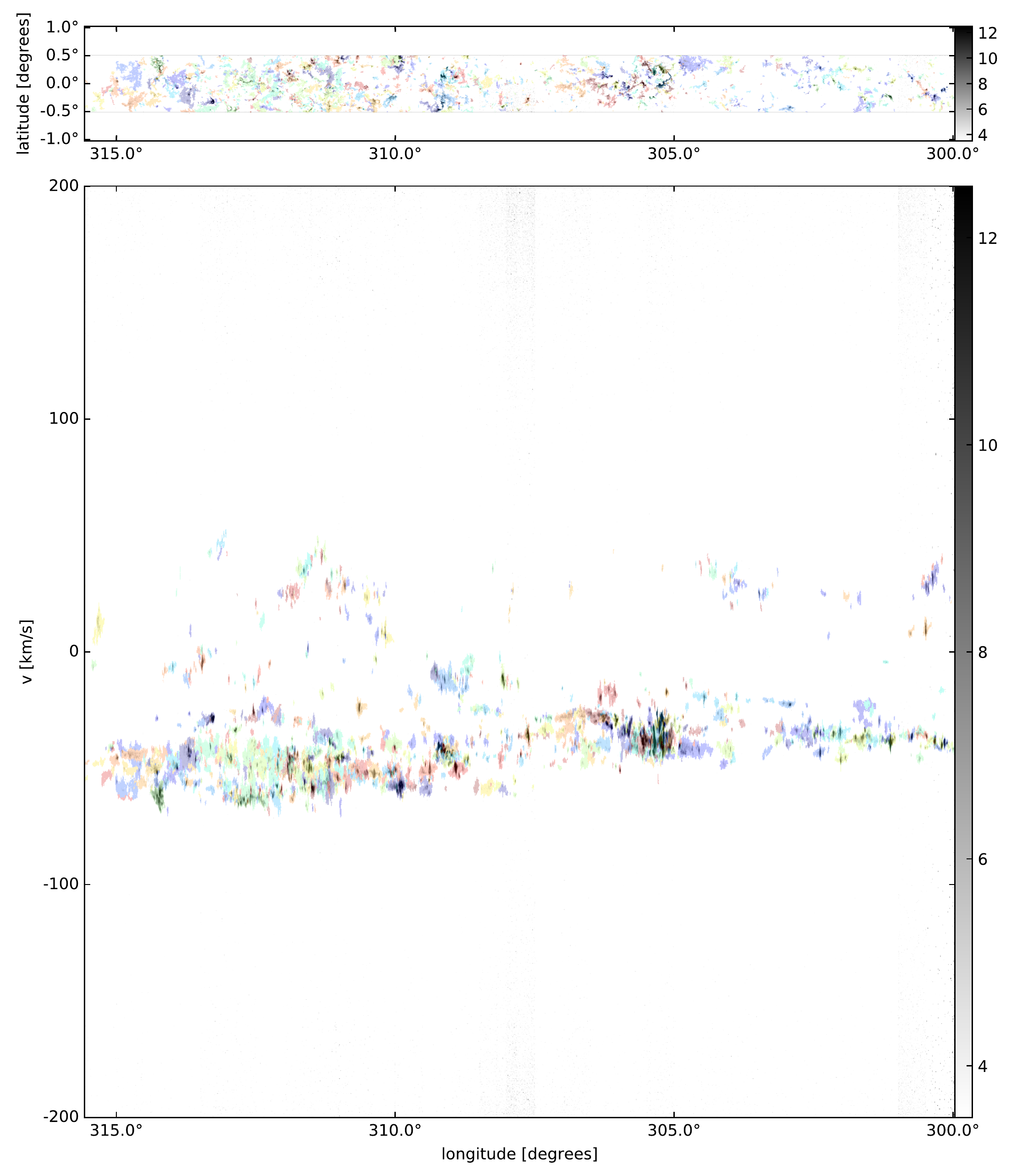}
\caption{Fig.\ref{fig:lb_lv_0} continued.}
\label{fig:lb_lv_4}
\end{figure*}

With this paper, we release the complete catalogue of SEDIGISM molecular clouds as extracted using the {\sc{scimes}} code, alongside the masks of each molecular cloud in the catalogue as fits files, in \url{http://sedigism.mpifr-bonn.mpg.de}. In Fig.\,\ref{fig:lb_lv_0} to \ref{fig:lb_lv_4} we show the sequence of $\ell b$ and $\ell v$ plots of the survey, with the $^{13}$CO peak intensity as the background greyscale, and the SEDIGISM cloud masks overlaid as colours. The full details of the extraction are given in Sect.\,\ref{sec:dataproducts}, and Table \,\ref{tab:catalogue_tabs} has a description of all the properties recorded in the released catalogue.

\begin{table*}
\caption{Description of the SEDIGISM catalogue contents}
\label{tab:catalogue_tabs}
\small
\renewcommand{\footnoterule}{}  
\begin{tabular}{l l }
\hline 
\hline
Catalogue column & Description  \\
\hline
cloud\_id & Unique cloud ID number \\ 
cloud\_name & Cloud name as per the SEDIGISM naming scheme, i.e. SDG followed by the Galactic coordinates of the cloud\\ 
lon\_deg & Galactic Longitude of the cloud's centroid, $\ell $ (deg) \\ 
lat\_deg & Galactic Latitude of the cloud's centroid, $b$ (deg)\\ 
vlsr\_kms & Systemic velocity, $v_{\rm{lsr}}$ (km\,s$^{-1}$)\\ 
sigv\_kms & Velocity dispersion, $\sigma_{v}$ (km\,s$^{-1}$)\\ 
area\_as & Exact footprint area (arcsec$^{2}$)\\
radius\_eq\_as & Equivalent radius, estimated using the footprint area, $R$ (arcsec)\\ 
major\_as & Semi-major axis, {\em major} (arcsec) \\ 
minor\_as & Semi-minor axis, $minor$ (arcsec)\\ 
pa\_deg & Position angle of the major axis, with $0\degr$ being along the $x$/$\ell $ axis, $PA$ (degrees) \\ 
pca\_axis\_ratio & Aspect ratio from the moments, $AR_{\rm mom}$ (i.e. major\_as/minor\_as)\\ 
medaxis\_length\_as & Projected geometrical medial axis length, $length_{\rm{MA}}$ (as) \\ 
medaxis\_width\_as & Projected geometrical medial axis width, $width_{\rm{MA}}$ (as) \\ 
medaxis\_ratio & Aspect ratio from the medial axis, $AR_{\rm{MA}}$ (i.e. medaxis\_length\_as /medaxis\_width\_as)\\ 
ave\_wco\_Kkms & Average $^{13}$CO (2-1) integrated intensity, $<I_{\rm ^{13}CO}>$ (K\,km\,s$^{-1}$)\\ 
peak\_Ico\_K & Peak $^{13}$CO (2-1) intensity, $T_{\rm ^{13}CO}^{\rm peak}$ (K) \\ 
sn\_ratio & signal-to-noise ratio (SNR = peak intensity / local noise level) \\ 
n\_pixel & Number of 3D pixels (i.e. voxels) in the cloud\\ 
n\_leaves & Number of individual dendrogram leaves comprised in the cloud\\ 
orig\_file & Name of the field from which the cloud was originally extracted\\ 
edge & tag identifying whether a cloud touches an edge of the field (yes=1, no=0)\\ 
\hline
d\_near & Near kinematic distance (kpc)\\ 
d\_near\_err & Uncertainty on near distance (kpc) \\ 
d\_far & Far kinematic distance (kpc)\\ 
d\_far\_err & Uncertainty on far distance (kpc)\\ 
dist\_kpc & Final adopted distance, $d$ (kpc)\\ 
dist\_err\_kpc & Uncertainty on final distance (kpc) \\ 
d\_flag & Flag describing the method by which the final distance was decided, $d_{\rm{flag}}$ (as per Table\,\ref{tab:dtag})\\ 
d\_solution & Flag describing the type of distance solution, $d_{sol}$ \\
		 & (NA = Not Ambiguous, T = tangent, N = Near, F = Far, M = Maser) \\
d\_reliable & Flag to indicate sources with a reliable distance, $d_{reliable}$ (1 = reliable, 0 = non-reliable -- as per Sect.\,\ref{sec:kin_dist})\\
tag\_hisa & Flag with the result from our automated \hi SA determination (1 = strong \hi SA; 0 = ambiguous; -1 = no \hi SA). \\ 
nb\_AGAL\_matches\_total & Total number of ATLAGAL matches \\ 
nb\_AGAL\_matches\_perfect & Number of ATLASGAL perfect matches\\ 
nb\_AGAL\_matches\_partial & Number of ATLASGAL partial matches\\ 
nb\_AGAL\_nodistance & Number of ATLASGAL matches with no distance assigned\\ 
HMSF & Tag identifying whether a cloud has a HMSF tracer (1 = yes, 0 = no)\\ 
\hline
area\_pc2 & Exact footprint area (pc$^{2}$)\\ 
radius\_eq\_pc & Equivalent radius, estimated using the footprint area, $R$ (pc)\\ 
major\_pc & Semi-major axis, {\em major} (pc) \\ 
minor\_pc & Semi-minor axis, $minor$ (pc)\\ 
medaxis\_length\_pc & Projected geometrical medial axis length, $length_{\rm{MA}}$ (pc) \\ 
medaxis\_width\_pc & Projected geometrical medial axis width, $width_{\rm{MA}}$ (pc) \\ 
\hline
Mass & Cloud mass,  $M$ ($M_{\odot}$)\\ 
Column\_density\_cm2 & Cloud's average column density, $N$ (cm$^{-2}$)\\ 
Surf\_density\_Mpc2 & Cloud's average gas surface density, $\Sigma$ ($M_{\odot}$\,pc$^{-2}$) \\ 
alpha\_vir & Virial parameter, $\alpha_{vir}$ \\ 
radius\_dec\_pc & Deconvolved equivalent radius, $R^{d}$ (pc) \\ 
Surf\_density\_dec\_Mpc2 & Surface density, calculated using the deconvolved radius, $\Sigma^{d}$ ($M_{\odot}$\,pc$^{-2}$)\\ 
alpha\_vir\_dec & Virial parameter, calculating using the deconvolved radius, $\alpha_{vir}^{d}$ \\
\hline
x\_sun\_kpc & x in Heliocentric coordinates, $x_{\odot}$ (kpc) \\ 
y\_sun\_kpc & y in Heliocentric coordinates, $y_{\odot}$ (kpc) \\ 
z\_sun\_kpc & z in Heliocentric coordinates, $z_{\odot}$ (kpc) \\ 
x\_gal\_kpc & x in Galactocentric coordinates, $x_{\rm{gal}}$ (kpc)\\ 
y\_gal\_kpc & x in Galactocentric coordinates, $y_{\rm{gal}}$ (kpc)\\ 
z\_gal\_kpc & x in Galactocentric coordinates, $z_{\rm{gal}}$ (kpc)\\ 
R\_gal & Galactocentric distance, $R_{\rm{gal}}$ (kpc) \\
\hline
\end{tabular}
\end{table*}

Besides these two main data-products, we also provide a few other ancillary materials online, which include dictionaries with the medial axis, and extra tables with more detailed information on the SEDIGISM-ATLASGAL matches, as well as the SEDIGISM matches with the other literature catalogues used for our distance assignment procedure. The full details on the format and content of this extra material are provided alongside those, as {\sc readme} files.

\section{Heliocentric and Galactocentric coordinates}
\label{app:coordinates}

In order to calculate the de-projected position of each cloud in the Galaxy, we have converted the ($\ell $, $b$, $d$) triad into a Heliocentric coordinate system ($x_{\odot}$,$y_{\odot}$,$z_{\odot}$), where the $x$-axis is defined along the line that connects the sun to the Galactic centre (GC), pointing towards the GC, and the $z$-axis points north out of the plane \citep[similar to][]{ellsworth2013}. Since the latitude ($b$) across the SEDIGISM coverage is always below $1\degr$, the contribution of the latitude is negligible in the determination of the $x$ and $y$ coordinates, and we have thus simplified the equations from \citet[][]{ellsworth2013} as:
\begin{equation}
\begin{split}
& x_{\odot}= d\,cos(l)\\
& y_{\odot}= d\,sin(l)\\
& z_{\odot}= d\,sin(b)
\end{split}
\end{equation}

We then also estimate the coordinates in a Galactocentric reference frame ($x_{\rm{gal}}$,$y_{\rm{gal}}$,$z_{\rm{gal}}$), centred in the GC, and in which the $y$-axis is now the line connecting the GC to the Sun, pointing outwards \citep[note that this is rotated by $90\degr$ with respect to the reference frame used in][]{ellsworth2013}. For $z_{\rm{gal}}$, we need to include the correction for the fact that the Sun does not lie exactly in the Galactic plane but is slight above \citep[e.g.][]{ellsworth2013}, by introducing a rotation angle $\theta = sin^{-1} (z_{0}/R_{0})$, where $z_{0}$ = 0.025\,kpc is the vertical displacement of the Sun above the Galactic midplane, and $R_{0}=8.34$\,kpc is the distance of the Sun to the Galactic centre \citep{reid2016}. As for the Heliocentric coordinates, we ignore the negligible contributions from the small latitude $b$ across the SEDIGISM coverage, as well as from $z_{0}$, on the calculations of $x_{\rm{gal}}$ and $y_{\rm{gal}}$. As such, our simplified equations for the determination of the Galactocentric coordinates are: 
\begin{equation}
\begin{split}
& x_{\rm{gal}}= d\,sin(l)\\
& y_{\rm{gal}}= R_{0} - d\,cos(l)\\
& z_{\rm{gal}}= R_{0}\,sin(\theta) - d\,cos(l)\,sin(\theta)
\end{split}
\end{equation}

The Galactocentric distance ($R_{\rm{gal}}$), can then be determined as $R_{\rm{gal}}= \sqrt{x_{\rm{gal}}^{2} + y_{\rm{gal}}^{2}}$.

\section{Sensitivity/completeness limit of the SEDIGISM cloud catalogue}
\label{sec:completeness}

\begin{figure*}
\centering
\includegraphics[width=0.8\textwidth]{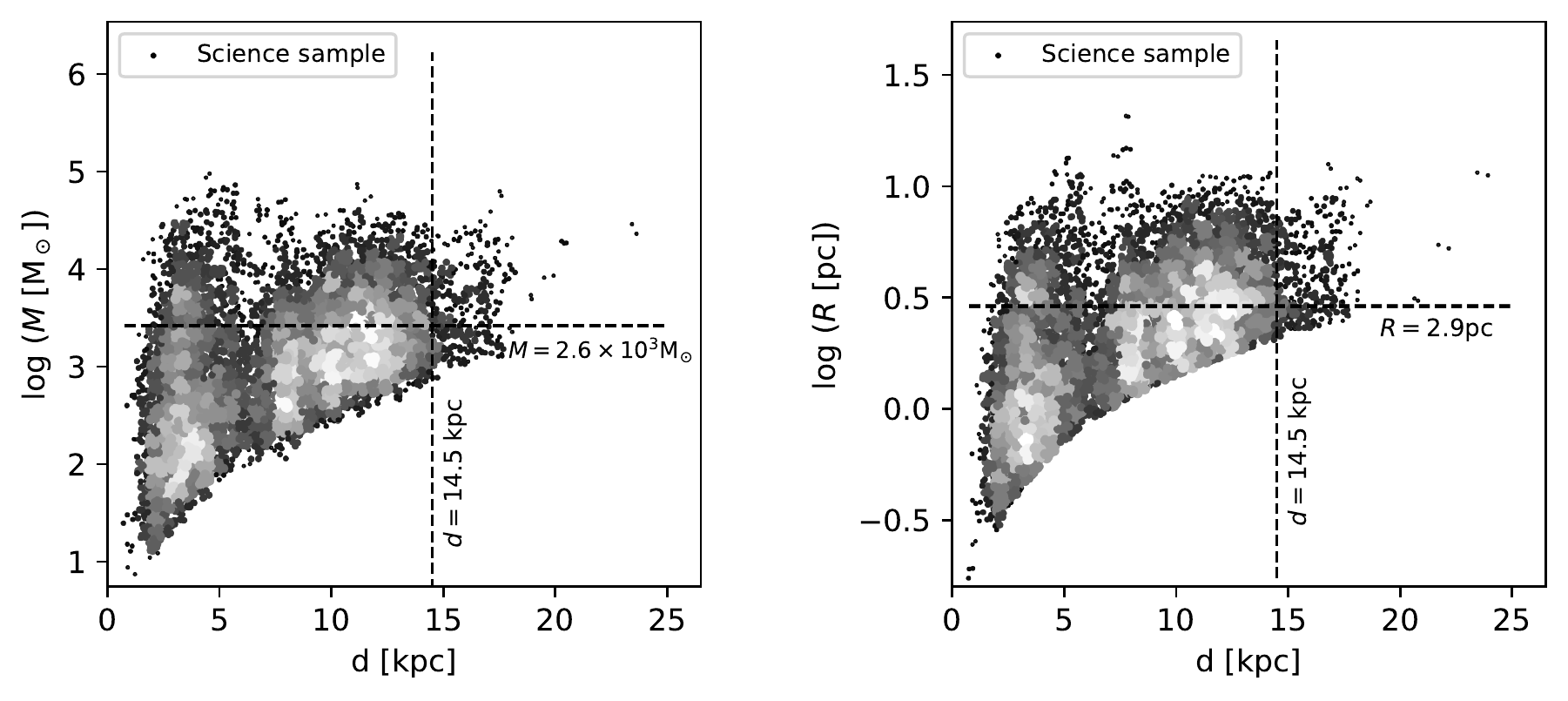}
\caption{{Scatter-density} plots showing the distances to the clouds in the science sample (in grey scale), with the completeness limits adopted for the complete science sample (in terms of mass on the left, and equivalent radius on the right) plotted as horizontal dashed lines. The vertical line delineates the distance at which the completeness limits were estimated from.}
\label{fig:d_limit}
\end{figure*}

\begin{figure*}
\centering
\includegraphics[width=\textwidth]{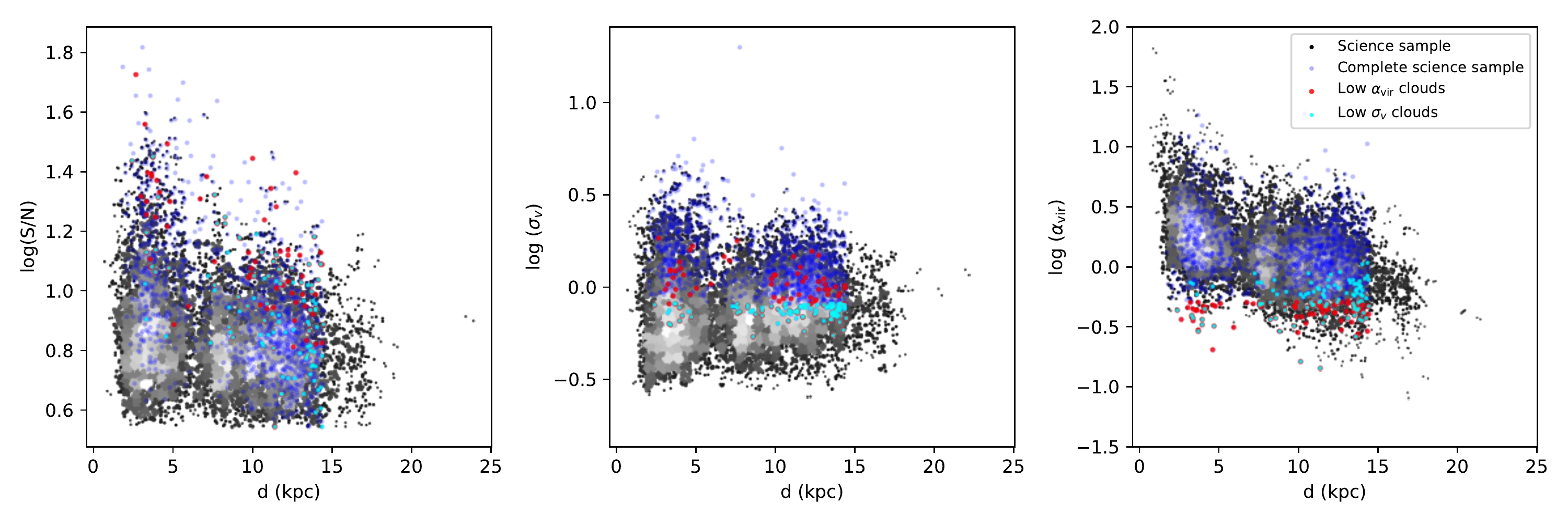}
\vspace{-0.5cm}
\caption{{Scatter-density} plots showing potential observational biases in some of the derived properties. In particular, we show the signal-to-noise ratio ($S/N$) as a function of distance on the left panel, the same for the velocity dispersion ($\sigma_{v}$) on the middle panel, and virial parameter ($\alpha_{vir}$) on the right panel. The grey scale shows all the clouds in the science sample, and the dark-blue points are the clouds within the complete science sample. The red and turquoise points are the sub-samples of 100 clouds with low velocity dispersion, and with low virial parameter, respectively, which are the two distributions that could be potentially most affected by observational biases.}
\label{fig:low_dynamics}
\end{figure*}

We have estimated a proxy for the completeness limit of the SEDIGISM dataset, based on the sensitivity and resolution of the data (i.e. estimating a robust detection limit), following the approach used by \citet[][]{heyer2001} and \citet[][]{Colombo2019}. To that purpose, we start by estimating a ``luminosity'' completeness limit, $L^{c}$, at a 5$\sigma$ confidence level, defined as:
\begin{equation}
L^{c} = L^{\rm{min}} + 5\sigma_{L}
\end{equation}

\noindent where $L^{\rm{min}}$ is the minimum luminosity we detect, defined as
\begin{equation}
L^{\rm{min}} {\rm{\,[K\,km\,s^{-1}\,pc^2]}} = N_{vox} T_{th} \Delta v \Omega_{p} d^{2},
\end{equation}

\noindent and
\begin{equation}
\sigma_{L} {\rm{\,[K\,km\,s^{-1}\,pc^2]}} = \sigma_{rms} \sqrt{N_{vox}} \Delta v \Omega_{p} d^{2},
\end{equation}

\noindent where $N_{vox}$ is the minimum number of 3D pixels (voxels) in the cloud, defined as $N_{vox} = N_{p} N_{c}$, where $N_{p}$ is the minimum number of (spatial) pixels that the cloud has to cover, and $N_{c}$ the minimum number of (spectral) channels. For this purpose, we want to be as conservative as possible, so that we minimise the possible biases, particularly when considering clouds that are at lower signal-to-noise levels. Therefore, we take $N_{p} = 6 N_{\rm{ppbeam}}$, where $N_{\rm{ppbeam}} = 9$ is the number of pixels per beam (i.e. corresponding to clouds whose footprint size is twice larger than the smallest clouds allowed into the science sample), and $N_{c} = 4$ (with 2 channels being our spectral resolution element). In essence, this corresponds to clouds that are $\sim4$ times larger (in terms of 3D pixels) than those allowed to go through to the dendrogram construction. $T_{th}$ is the sensitivity threshold as a main-beam antenna temperature, which we take as being $6\,\sigma_{rms}$, with $\sigma_{rms} = 0.7$\,K being the average noise level used for the dendrogram. This $T_{th}$ effectively corresponds to the first level leaves allowed into the dendrogram, i.e. starting at a level of $2\,\sigma_{rms}$, and with a leaf height of $4\,\sigma_{rms}$ above that. $\Delta v$ is the channel width (i.e. 0.25\,km\,s$^{-1}$), $\Omega_{p}$ is the pixel size ($\Omega_{p}=9.5\arcsec \times 9.5\arcsec \approx 2.12\times10^{-9}$\,sr), and finally, $d$ is the distance to the cloud. Given that the vast majority of the SEDIGISM clouds are within 14.5\,kpc we use that as our maximum distance for these calculations, although we note that we are still sensitive to clouds beyond those distances, with clouds lying up to kinematic distances of 23\,kpc. 

From this luminosity completeness limit, we derive a mass completeness limit as $M^{c}$\,$=$\,$\alpha_{^{13}\rm{CO(2-1)}} L^{c}$, using a conversion factor $\alpha_{^{13}\rm{CO(2-1)}}$\,$=$\,$22.43$\,M$_{\odot}$\,(K\,km\,s$^{-1}$)$^{-1}$\,pc$^{-2}$, estimated from our $X_{\rm{^{13}CO(2-1)}}$ (Sect.\,\ref{sec:dataproducts}), assuming a molecular weight per hydrogen molecule of 2.8 \citep[][]{Kauffmann2008}. We thus retrieve a mass completeness limit of $M^{c} = 2.6\times10^{3}$\,M$_{\odot}$ at 14.5\,kpc.

We also derive a size completeness limit, which is directly linked to the minimum source size that we can robustly recover. As for the luminosity completeness limit, we take that to be 6 beam sizes, which corresponds to a completeness radius $R^{c}=2.9$\,pc, at 14.5\,kpc distance. 

We consider these completeness/detection limit estimates to be rather conservative, since the majority of the sources in our catalogue lie well below a distance of 14.5\,kpc, but also because our full catalogue does contain sources that are as small as two beam sizes, making it possible to find clouds that lie well below our completeness limits at 14.5\,kpc and beyond.  This is clear from Fig.\,\ref{fig:d_limit} which shows the mass ($M$) and radius ($R$) of the science sample, as a function of distance ($d$), and where we overplot the respective completeness limits at 14.5\,kpc (as horizontal lines). For reference, if we simply take clouds up to a distance of 5\,kpc (e.g. as in our distance limited sample), then our completeness limits decrease to $M^{c} = 3.1\times10^{2}$\,M$_{\odot}$ and $R^{c}=1$\,pc.

By using the {\sc{scimes}} cloud extraction algorithm, which clusters the individual peaks of emission into clusters, we are less prone to have severe observational biases that often arise from the large range of distances probed and the extraction of objects close to the resolution elements of the survey (i.e. both spectrally and spatially). In addition, the rather conservative limits imposed to build the complete science sample should, in principle, also guarantee that the cloud properties that we extract are comparable across the entire survey. However, to test whether this was truly the case, we have investigated whether we could see evidence for any remaining observational biases in the complete science sample, that could potentially affect our results.

The left panel of Fig.\,\ref{fig:low_dynamics} shows the peak signal-to-noise ratio ($S/N$) as a function of distance, for the science sample (in grey) and for the complete science sample (in blue). We can see that while across the entire science sample the average $S/N$ is roughly constant, we have a larger amount of clouds with a high $S/N$ nearby. When we consider only the complete science sample (in dark blue), we effectively discard most of the nearby clouds with lower $S/N$, which then produces a trend of decreasing $S/N$ with increasing distance from the sun. This could become a problem, as the properties of clouds with lower $S/N$ are less well constrained than those with high $S/N$. A particularly problematic property is the velocity dispersion (and the respective virial parameter derived from it), because for further away clouds - since we do not do any bootstrap to extrapolate the emission into the noise - we might artificially recover a lower FWHM than we would have, had we been able to detect the full extent of those clouds at higher $S/N$. The middle and left panels of Fig.\,\ref{fig:low_dynamics}, highlight this issue, where we can see a clear trend of decreasing $\sigma_{v}$ and $\alpha_{\rm{vir}}$ as a function of distance. There is no obvious physical reason why we should expect these two properties to correlate with their distance to the sun, suggesting that there are still some remaining observational biases at play, despite our best efforts to neutralise them. This could therefore be responsible (at least in part) for the signatures seen in Sect.\,\ref{sec:extreme_dynamic} and \ref{sec:extreme_low_dynamic}. Surprisingly, though, Fig.\,\ref{fig:low_dynamics} also shows that the clouds with the low velocity dispersion (marked in turquoise) and those with low virial parameter (shown in red), are not the clouds with the lower $S/N$ values of the sample (which we ought to expect, if these trends were purely driven by cloud segmentation biases and survey limitations). It thus remains to be seen whether some of the signal seen in Sect.\,\ref{sec:extreme_dynamic} and \ref{sec:extreme_low_dynamic} could be physically driven.

\section{Solving KDA using the size-linewidth relation}
\label{app:size-linewidth}

\begin{figure}
\centering
\includegraphics[width=0.45\textwidth]{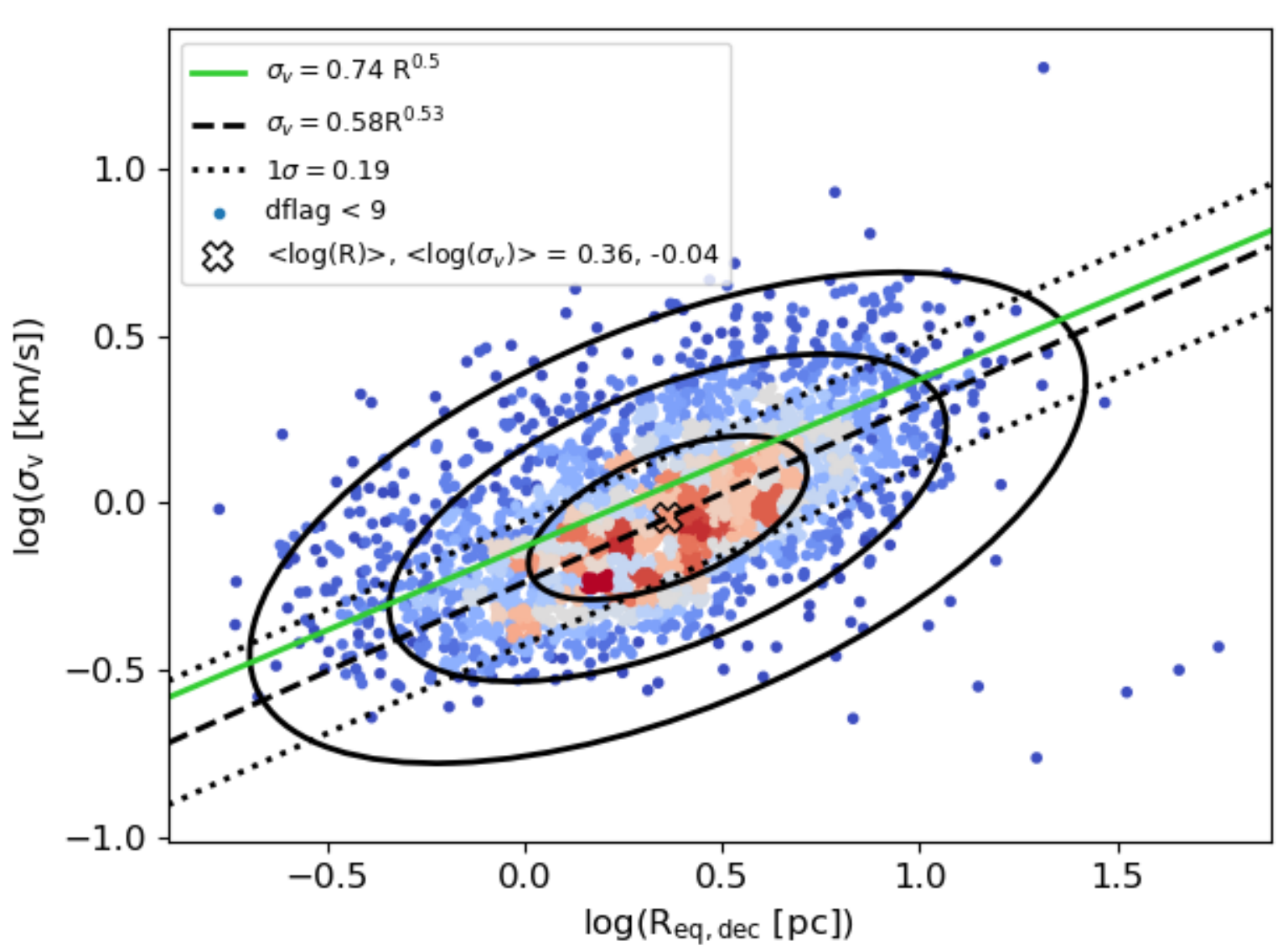}
\hfill
\includegraphics[width=0.45\textwidth]{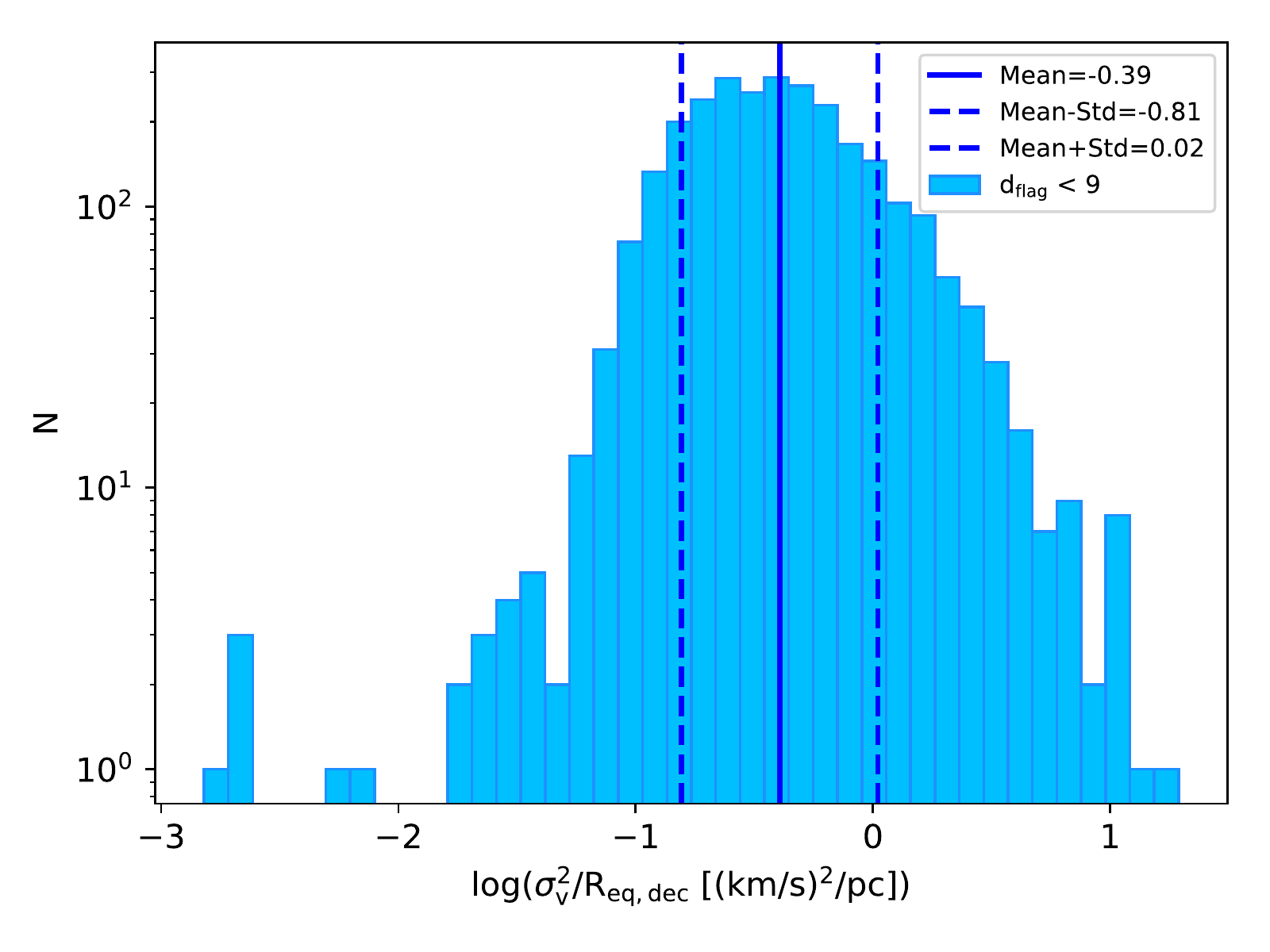}
\caption{Top: Scatter-density plot of the size-linewidth relation for our full sample of clouds with a solved KDA using methods 0-9. The ellipses show the 1, 2 and $3\sigma$ from PCA analysis on our sample, and the dashed line shows the respective slope (i.e. $\sigma_v \propto R^{0.53}$). This slope is consistent with the relation from \citet{Solomon1987}, shown as a green line, that lies well within $1\sigma$ of our relation. Bottom: Histogram of the size-linewidth relation for all clouds with a solved KDA using methods 0-9, assuming the original $\sigma_v \propto R^{0.5}$ relation. The blue solid and dashed lines show the mean and $1\sigma$ standard deviation respectively (in $log$-space), that we use to determine if a distance solution of a cloud is significantly more likely than the other.}
\label{fig:size-linewidth}
\end{figure}

{In an attempt to solve the kinematic distance ambiguity for clouds for which our methods 0-9 did not work, we explored the position of clouds in a size-linewidth plot, using both near and far distances. {We then} determined whether one of those solutions was more likely, based on their position relative to the bulk distribution of clouds. 

The original size-linewidth relation was first looked at by \citet{Larson1981}, but redefined by \citet{Solomon1987}, taking the form of $\sigma_v \propto R^{0.5}$. However, the exact positioning of this relation (with \citealt{Solomon1987} placing it at $\sigma_v^{2}/R = 0.55$) is sensitive to the specific way by which the radius and velocity dispersion are estimated (e.g. sensitive to the specific tracer, cloud extraction algorithm, etc.). Therefore, for our purpose, we calibrate the size-linewidth relation using our data, for clouds with a solved KDA using methods 0-9. The size-linewidth relation for our data is shown in Fig.\,\ref{fig:size-linewidth} (top panel), and has an exponent (from a PCA analysis), that is consistent to the exponent $\sigma_v \propto R^{0.5}$ found by \citet{Solomon1987}. The bottom panel of Fig.\,\ref{fig:size-linewidth} shows the histograms of the values of $\sigma_v^{2}/R$ for our data, showing that they follow a log-normal distribution. We use these mean and standard deviation of the $\sigma_v^{2}/R$ values (in log-space) to compare to the $\sigma_v^{2}/R$ values of the clouds using both the near and far distance solutions. We favour a given distance solution \emph{only} if that solution is significantly closer to the empirical relation than the other solution (i.e. at least a factor 3 difference {in $log$-space}), and \emph{only} if we do not have both solutions placing the clouds within $1\sigma$ of the underlying distribution (as in that case, both near and far solutions would be equally plausible).
}

\section{Global properties for distance-limited science sample}
\label{app:histograms}

\begin{figure*}
\centering
\includegraphics[width=\textwidth]{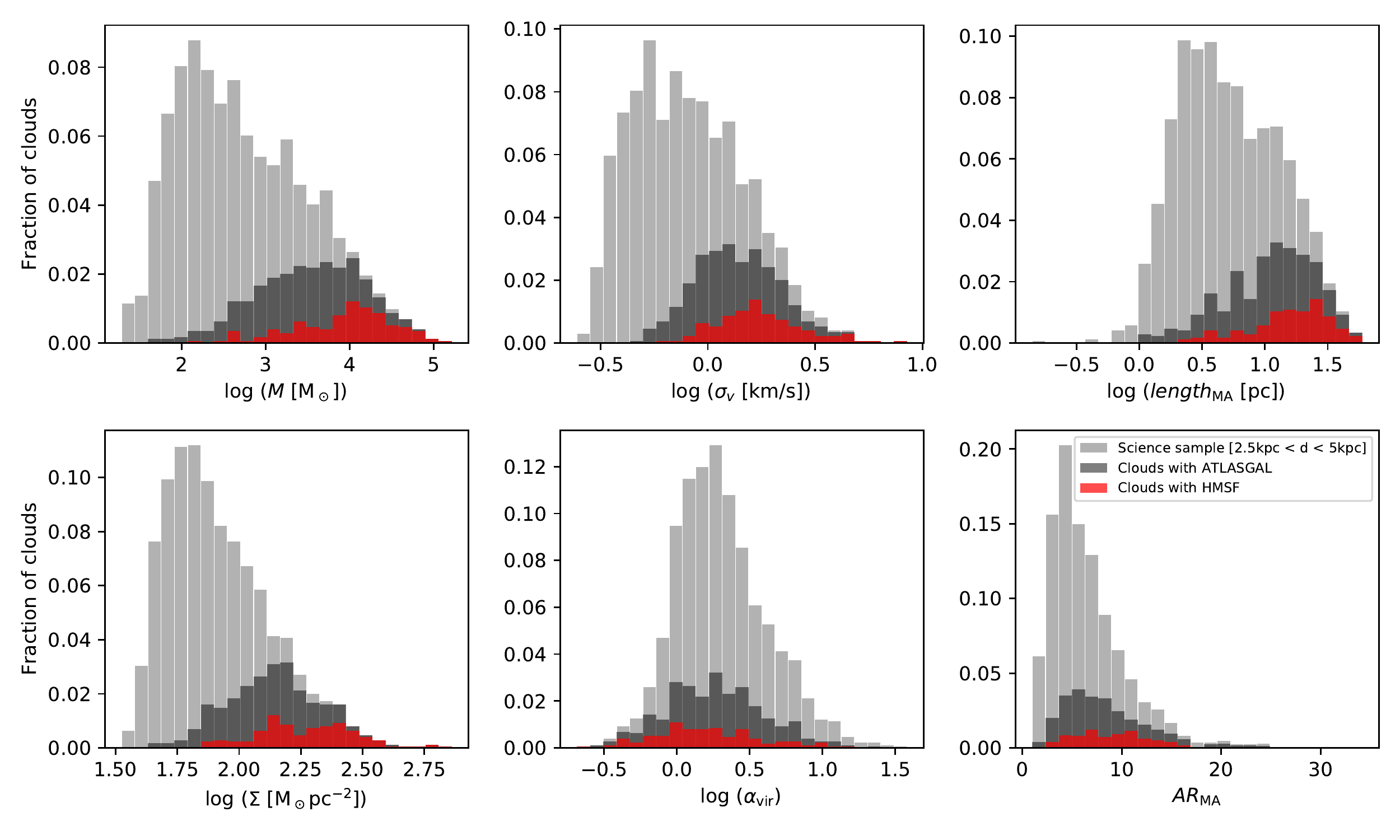}
\vspace{-0.5cm}
\caption{Same as Fig.\,\ref{fig:histograms}, showing the histograms of global properties but for a distance-limited sample (2.5\,kpc $< d <$ 5.0\,kpc), in order to minimise distance biases. The panels represent the distributions of: Mass (top-left), velocity dispersion (top-centre), medial axis length (top-right), average surface density (bottom-left), virial parameters (bottom-centre), and aspect ratio from the medial axis (bottom-right). The histograms shown are for the distance-limited science sample (light grey), along with clouds that have an ATLASGAL counterpart (dark grey), and clouds that have a HMSF signpost (red). {The normalisation of all histograms was made with respect to the total number of clouds in the distance-limited sample.}}
\label{fig:histograms_dlimited}
\end{figure*}

Figure\,\ref{fig:histograms_dlimited} shows the distributions of a number of different properties, namely the mass ($M$), the velocity dispersion ($\sigma_{v}$), the medial axis length ($length_{\rm{MA}}$), the average surface density ($\Sigma$), the virial parameter ($\alpha_{\rm{vir}}$), and the aspect ratio from the medial axis ($AR_{\rm{MA}}$), for a distance-limited sample, with 2.5\,kpc $< d <$ 5.0\,kpc, in order to minimise the selection effects due to distances/resolution. Note that at these distances, we are mostly looking at a spiral arm (the Scutum arm), and thus we tend to focus on the sample that has more ATLASGAL matches, missing some of the more diffuse clouds. Overall, the histograms are consistent with the full sample, and the shapes of the distributions follow the same trends - as quantified through the statistics for both the full and the distance-limited samples reported in Table\,\ref{tab:Statistics}. This would suggest that the results inferred from the global sample are significant.

\section{2D statistical test}
\label{app:chisq_test}

In order to determine if the spatial distribution of clouds in our sub-samples with extreme properties has any dependency on Galactic environment, we would require a proper modelling of the spiral arms and bar in PPV space, and even then, the uncertainties in the distances would always be a limitation to the interpretation of the results. Therefore, we have instead simply chosen to test whether the spatial distribution of clouds in our sub-samples (as per their de-projection onto the top-down view of the Galaxy) is statistically consistent with the global distribution of clouds. This makes no assumption on the Galactic structure, and is less affected by distance uncertainties - since the sub-sample is purely drawn out of the global population; both are affected in the exact same way. We have used the Pearson's $\chi^{2}$ statistical test, which tests whether the frequency distribution of certain events observed in a sample is consistent with a particular theoretical distribution. 

The specific calculation of the $\chi^{2}$ statistics is given by:

\begin{equation}
    \chi^{2} = \sum_{i=1}^{n} \frac{(O_i - E_i)^{2}}{E_i} = N \sum_{i=1}^{n} \frac{(O_i/N - p_i)^{2}}{p_i} 
\end{equation}
\\

\noindent where $n$ is the number of bins (or cells) considered, $O_{i}$ is the number of observed counts into bin $i$, $N$ is the total number of observations, $E_{i} = N p_{i}$ is the number of expected counts in bin $i$ where $p_{i}$ in the probability of an observation falling into bin $i$. 

\begin{figure*}
\centering
\includegraphics[width=\textwidth]{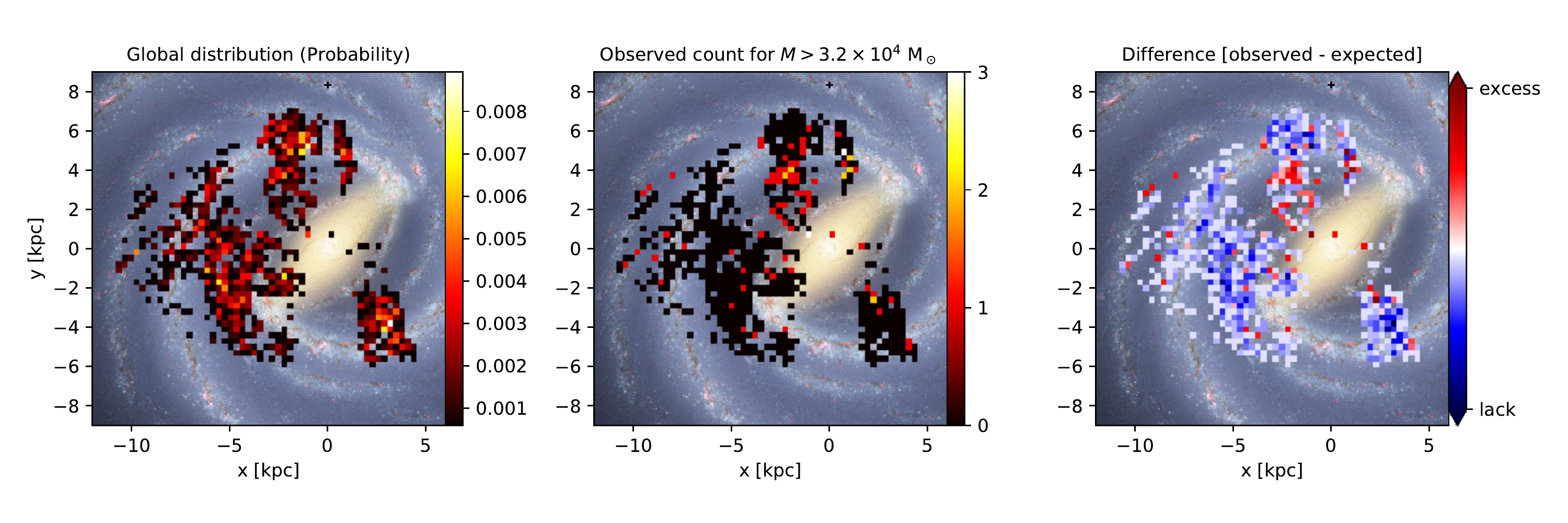}\\
\vspace{-0.5cm}
\includegraphics[width=\textwidth]{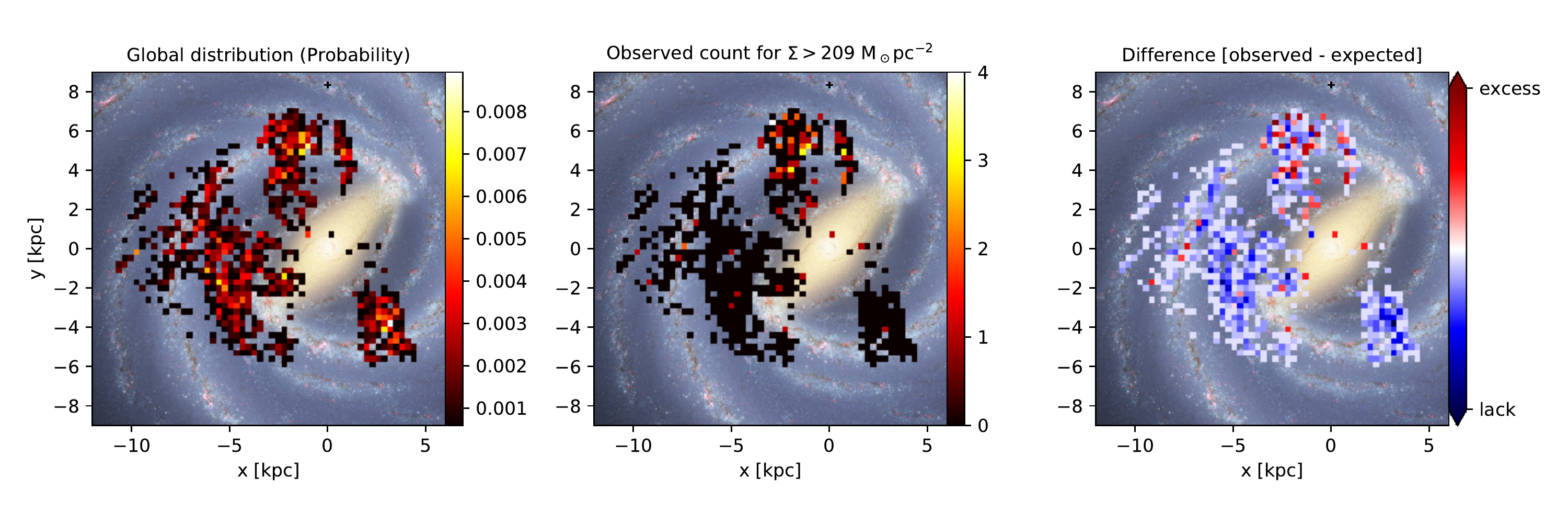}\\
\vspace{-0.5cm}
\includegraphics[width=\textwidth]{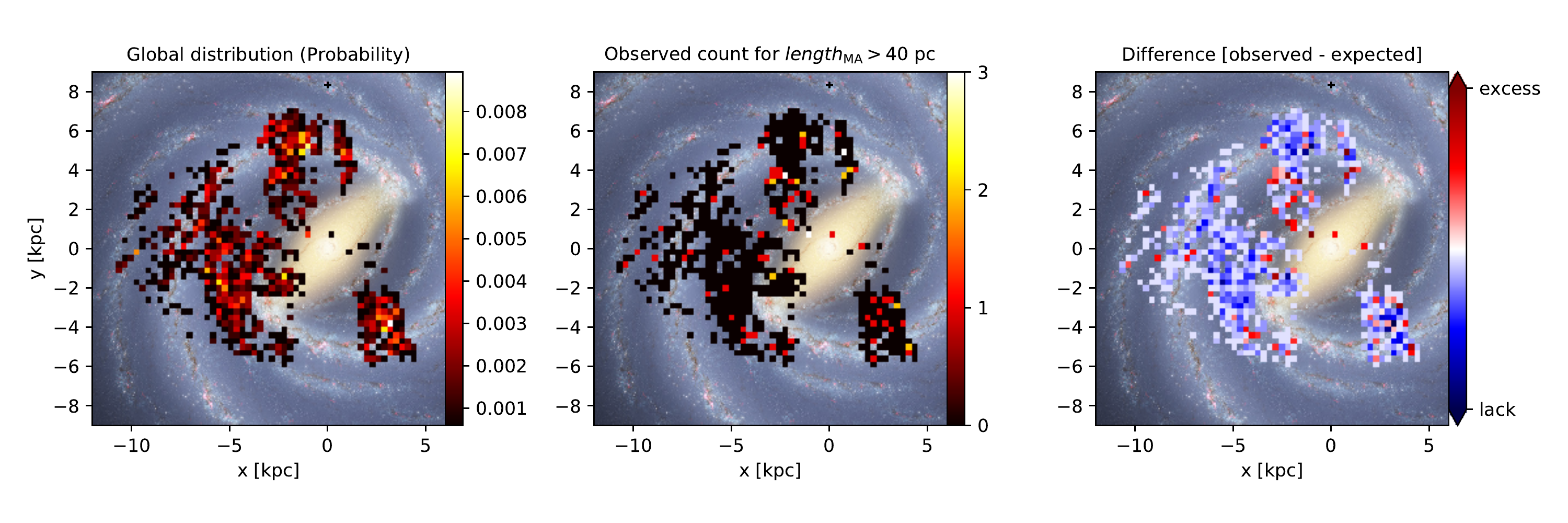}\\
\vspace{-0.5cm}
\caption{Left column: Normalised 2D histogram of the Galactic distribution of all clouds within the complete science sample, representing our ``theoretical probability'' for the $\chi^{2}$ test. Each 2D bin corresponds to a cell of 0.3\,kpc\,$\times$\,0.3\,kpc.  Middle: Observed number of clouds from a specific sub-sample in each of the 2D bins (defined as in the left panels). Each row shows a different tail of a distribution, using the 100 clouds with highest mass (top row), surface density (middle row), and length (bottom row). The specific condition to select these subsamples is specified on the top of each of these middle panels. Right column: Rescaled difference between the observed and expected distributions (i.e. between the middle and left panels), with red representing a relative excess, and blue a relative lack of counts with respect to the statistical prediction.}
\label{fig:chi-sq-1}
\end{figure*}

\begin{figure*}
\centering
\includegraphics[width=\textwidth]{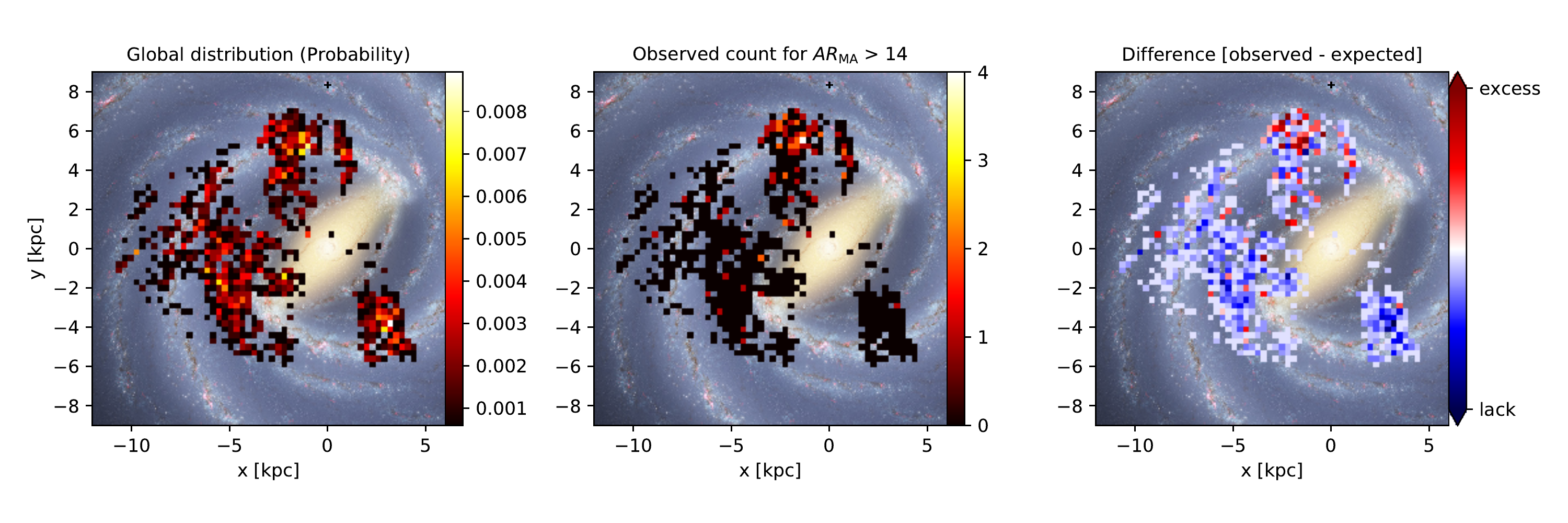}\\
\vspace{-0.5cm}
\includegraphics[width=\textwidth]{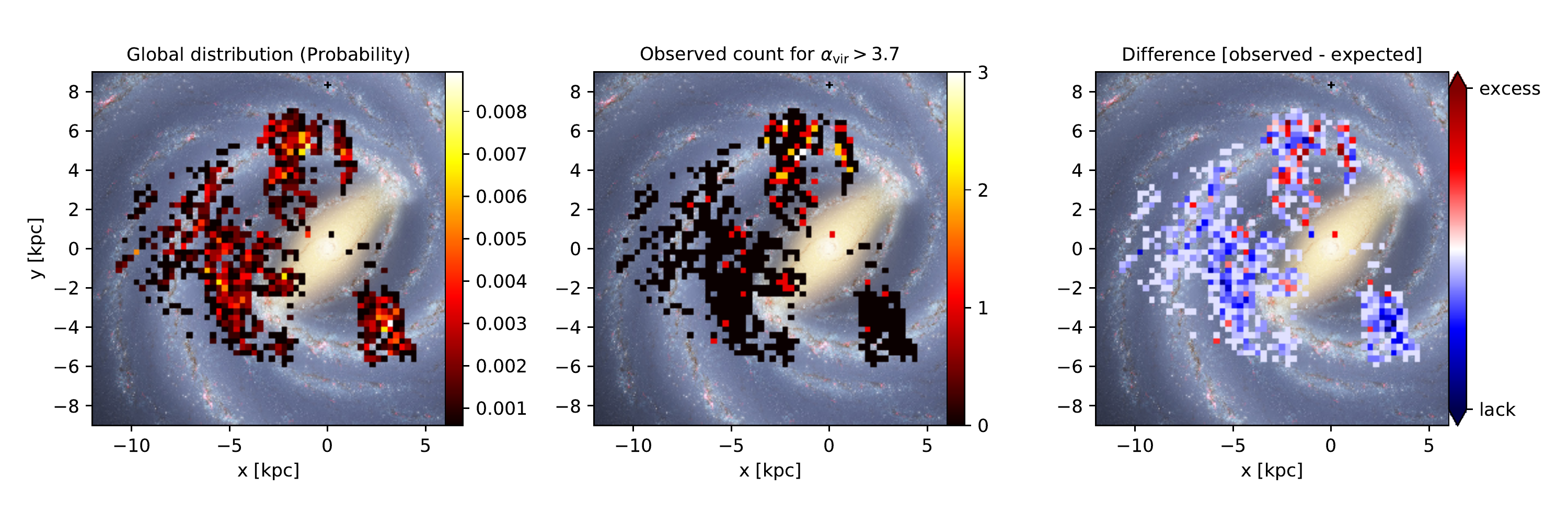}\\
\vspace{-0.5cm}
\includegraphics[width=\textwidth]{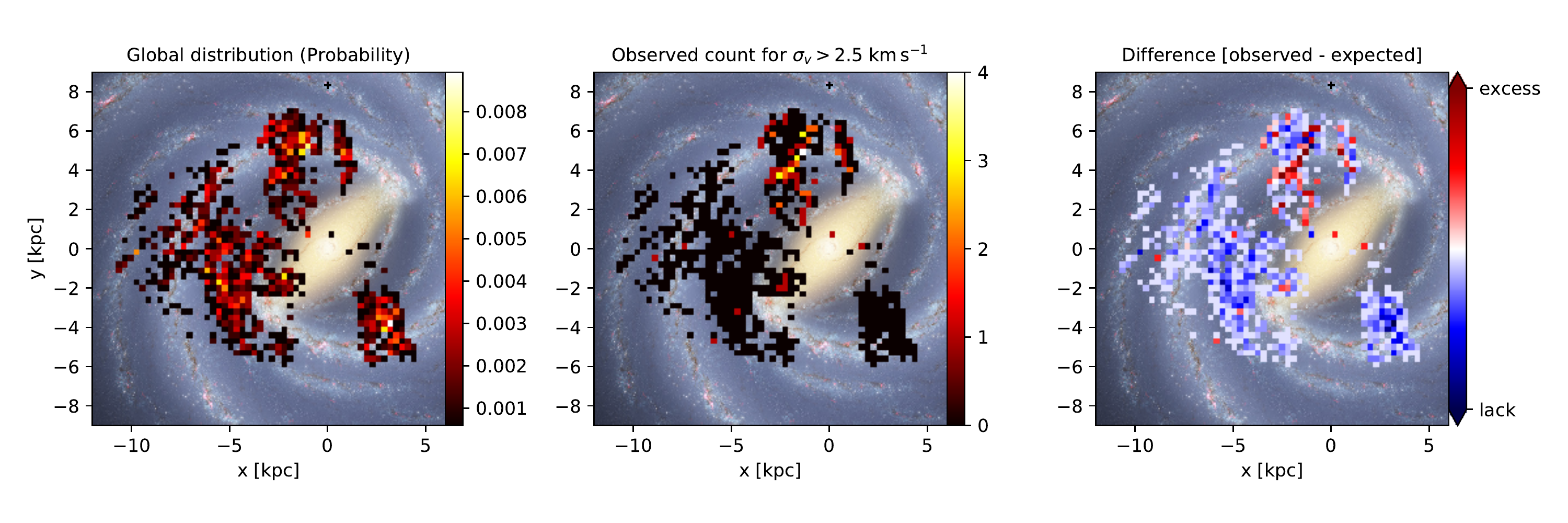}\\
\vspace{-0.5cm}
\caption{Same as Fig.\,\ref{fig:chi-sq-1}, but for the 100 clouds with the largest aspect ratio (top), highest virial parameter (middle), and highest velocity dispersion (bottom).}
\label{fig:chi-sq-2}
\end{figure*}

\begin{figure*}
\centering
\includegraphics[width=\textwidth]{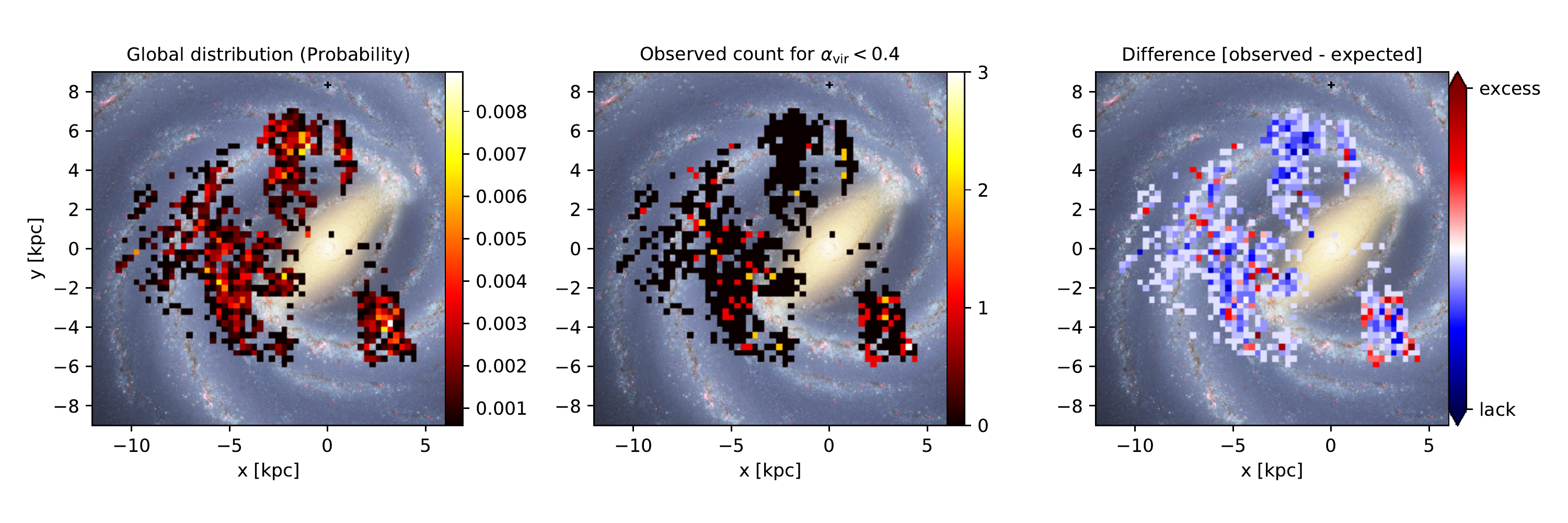}\\
\vspace{-0.5cm}
\includegraphics[width=\textwidth]{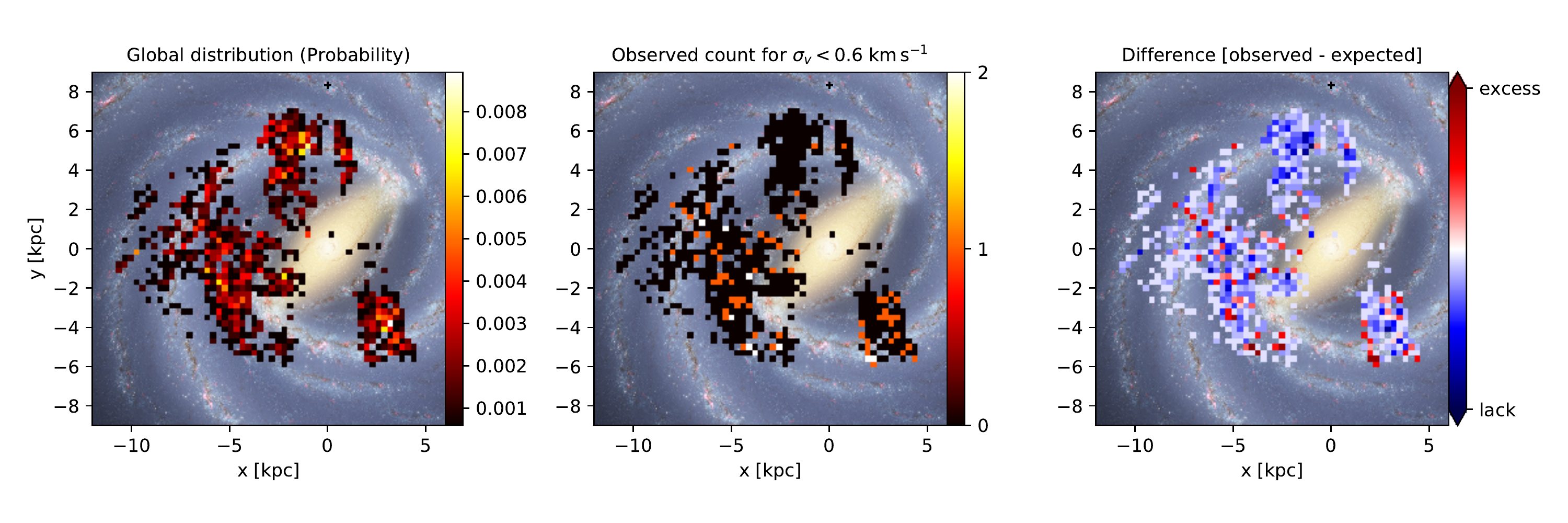}\\
\vspace{-0.5cm}
\includegraphics[width=\textwidth]{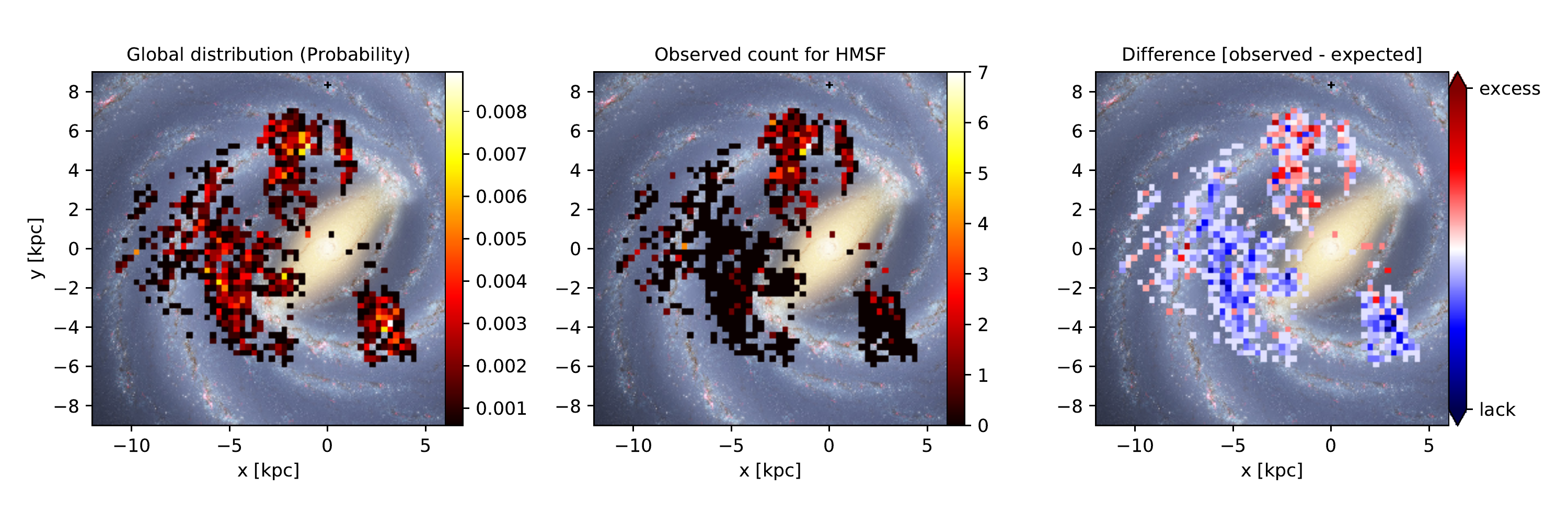}\\
\vspace{-0.5cm}
\caption{Same as Fig.\,\ref{fig:chi-sq-1} and \ref{fig:chi-sq-2}, for the 100 clouds with lowest virial parameter (top) and velocity dispersion (middle), and also for the 211 clouds within the complete science sample that have a HMSF signpost (bottom).}
\label{fig:chi-sq-3}
\end{figure*}

In our case, we assume that the global distribution of clouds in the complete science sample represents our probability function (i.e. our ``theoretical'' distribution), and we want to assert if the spatial distribution of the most extreme clouds simply follows (statistically) the same distribution of the entire sample, or whether it shows significant deviations. We thus constructed our $p_{i}$, by building a 2D probability density function (pdf), of the complete science sample (i.e. a normalised 2D histogram of the spatial distribution - in Galactocentric coordinates - of clouds within the complete science sample), using a spatial bin of $0.3\times0.3$\,kpc\footnote{The need to produce these regular spatial bins is the reason why we have excluded clouds at the tangent distance for this exercise. If using those, we would be effectively including bins that have sources regrouped from a larger spatial range than the bin size. In order to include tangent clouds, we would have to introduce a different weight to the bins at the tangent distances, to effectively account for the larger areas covered. This, however, is not straight forward to produce, since the binning of sources onto their tangent distance was made based on their line of sight velocity, which effectively means a variable spatial range, and also directed solely along the line of sight (rather than along any of the Galactocentric cartesian coordinates).} (see left panels of Figs.\,\ref{fig:chi-sq-1}, \ref{fig:chi-sq-2} and \ref{fig:chi-sq-3}).
Note that because our data are sparse, $p_{i}$ can be $0$ in many bins, but we only compute the $\chi^{2}$ statistics for the $n$ bins that have $p_{i} > 0$. Similarly to $p_{i}$, we then construct $O_{i}$ as the 2D pdf of the distribution of the $N$ clouds in our sub-sample, using the exact same bins for $p_{i}$ (see the panels in the central column of Figs.\,\ref{fig:chi-sq-1},  \ref{fig:chi-sq-2} and  \ref{fig:chi-sq-3}). $N$ in our case is always 100 clouds, i.e. 5\% of the complete science sample, except for the HMSF clouds which amount to a total of 211 clouds. The results from this $\chi^{2}$ statistics, for all of our tested sub-samples, are summarised in Table\,\ref{tab:Xsq_stats}.

\begin{figure*}
\centering
\includegraphics[width=0.9\textwidth]{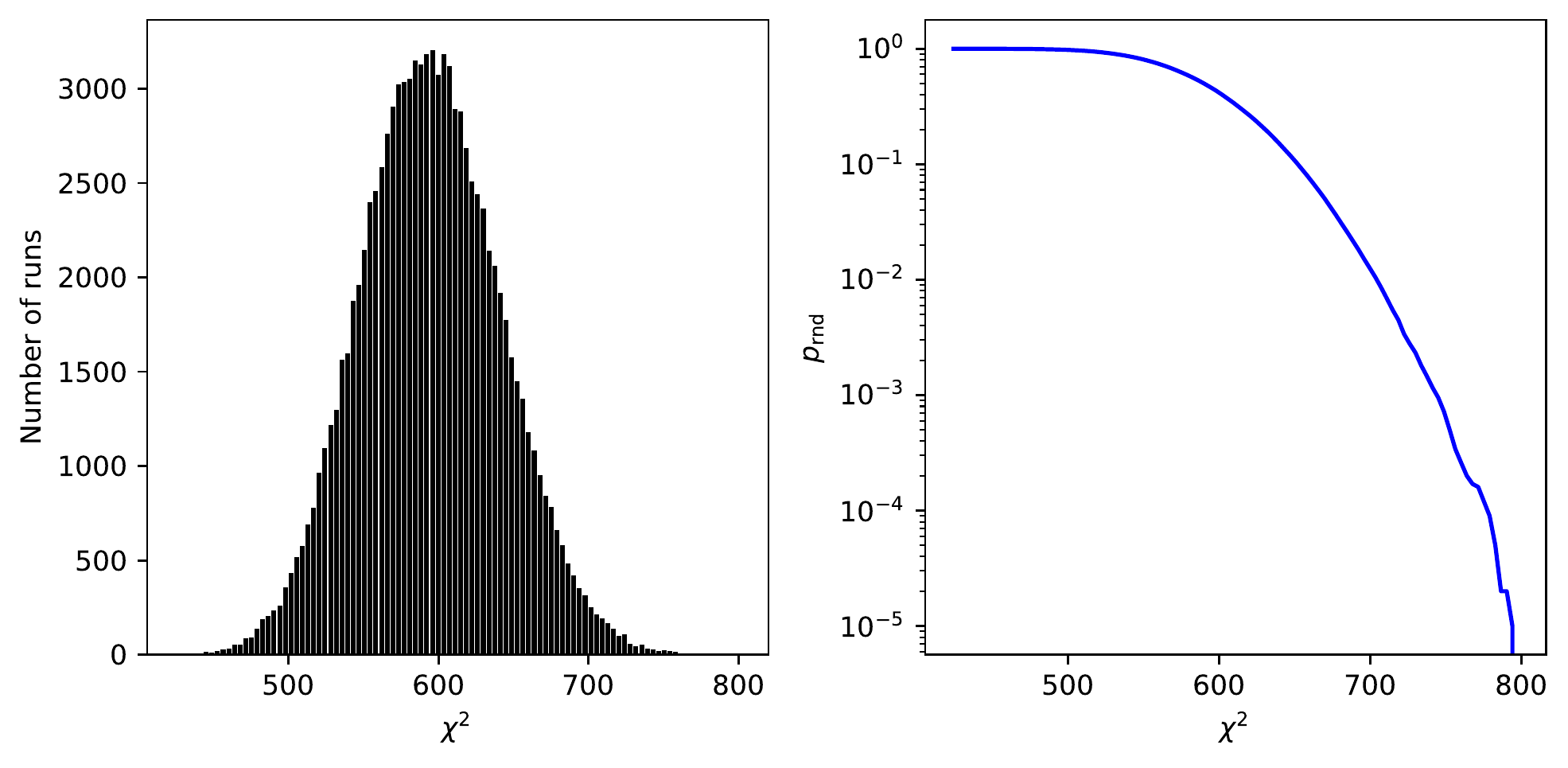}
\caption{Left panel: Distribution of $\chi^{2}$ values obtained as a result of performing our $\chi^{2}$ statistical test on 100\,000 random draws of 100 clouds from the complete science sample. Right panel: cumulative fraction of those runs (in log-scale) that had a $\chi^{2}$ value above a certain value, i.e. the probability, $p_{\rm{rnd}}$, of observing $\chi^{2}$-values above a specific value, purely from a random draw of clouds from our theoretical distribution.}
\label{fig:chi-sq-test}
\end{figure*}

In order to quantify the statistical significance of these results for our specific purpose, we have performed a test to determine the likelihood of obtaining a given $\chi^{2}$-value purely out of a random sampling of our theoretical distribution. To do so, we performed 100\,000 draws of $N = 100$ clouds randomly selected from the original sample of clouds (i.e. from the complete science sample), without replacement. For each of those draws, we construct the $O_{i}$ as the 2D pdf of the distribution of the $N$ clouds, and perform the $\chi^{2}$ test in the exact same way as for the extreme cloud samples. Figure\,\ref{fig:chi-sq-test} shows the distribution of $\chi^{2}$ values obtained as a result of these 100\,000 random draws (left panels). The right panel of Fig.\,\ref{fig:chi-sq-test} shows the cumulative fraction of runs with a $\chi^{2}$-value above a certain value. From this, we derive our probability, $p_{\rm{rnd}}$, that the observed $\chi^{2}$-value comes from a pure random sampling of the theoretical distribution.  For instance, there is a 1\% change of obtaining a $\chi^{2}$ above 705 from a pure random sampling of the theoretical distribution. Similarly, there is a 2\% chance that the $\chi^{2}$-value lies above 690, 5\% above 670, 10\% above 650, 20\% above 630, and 30\% above 615. Table\,\ref{tab:Xsq_stats} compiles the $p_{\rm{rnd}}$ for each of the extreme cloud samples that we studied.  

Although the exact $\chi^2$ values and $p_{\rm{rnd}}$ should not be taken at face value (given the statistical fluctuations, as well as the uncertainties in the distributions, and binning effects, neither of which are taken into account), they can be useful for a relative comparison of the sub-samples. Indeed, the lower the $\chi^2$ value, the higher the $p_{\rm{rnd}}$, and the closer the distribution of the sub-sample matches the global one. From these statistics we can start to quantify which distributions are less-like the original distribution of clouds, and identify which properties of clouds could be most affected by the Galactic environment. 

To support this interpretation, besides the pure statistical test that compares our sub-samples to the global cloud population, we also checked where (spatially) the disparities between the theoretical and observed distributions were coming from. In order to do that, we produced ``difference maps'' between the observed and expected distributions, after rescaling the observed distribution to have the same mean and standard deviation as the theoretical one (so that they share a common reference frame). These difference maps are shown on the right panels of Figs.\,\ref{fig:chi-sq-1}, \ref{fig:chi-sq-2} and  \ref{fig:chi-sq-3}. Note that because we performed a re-scaling of the observed distribution, the absolute values of the difference are not meaningful. Instead, these plots are only meant to illustrate, qualitatively, where the differences between observed and predicted distributions are, with regions that have a relative excess of counts shown in dark red, and regions with a relative lack of counts shown in dark blue.

\begin{table}
  \caption{Results from the 2D $\chi^{2}$ statistical tests, for all the sub-samples of extreme clouds, compared to the global spatial distribution of clouds in the complete science sample. $p_{\rm{rnd}}$ specifies the likelihood of obtaining the respective $\chi^2$-value from a pure random sampling of $N=100$ clouds from the theoretical distribution.}
  \label{tab:Xsq_stats}
  \begin{tabular}{l c c c}
  \hline
  \hline  
  Condition & & $\chi^{2}$ & $p_{\rm{rnd}}$ \\
  \hline
    $M  > 3.2\times10^{4}$ M$_\odot$   	    & &  {670}   & {0.05} \\
    $\Sigma > 209 $ M$_\odot$pc$^{-2}$  	& &  {638}   & {0.16} \\
    $length_{\mathrm{MA}} > 40$ pc      	& &  {715}   & {0.005} \\
    $AR_{\mathrm{MA}}$ > 14  	         	& &  {671}   & {0.04} \\
    $\alpha_{\mathrm{vir}} > 3.7 $      	& &  {699}   & {0.01} \\
    $\sigma_v > 2.5 $ km$\,$s$^{-1}$    	& &  {747}   & {0.001} \\
    \hline
    $\alpha_{\mathrm{vir}} < 0.4 $      	& &  {675}   & {0.04} \\
    $\sigma_v < 0.6 $ km$\,$s$^{-1}$    	& &  {669}   & {0.05} \\
    \hline
    HMSF                        			& &  {735}   & {0.001$^{(*)}$}\\
    \hline
    \end{tabular}
   
   $^{(*)}$ estimated for a random sampling of $N=211$ clouds.
\end{table}

\section{Opacity laws and their effect on the HMSF threshold}
\label{app:opacity}

{
The original empirical threshold for HMSF, of $M [$M$\odot]=870 (R [$pc$])^{1.33}$, from \citet{KauffmannPillai2010} was determined using a combination of dust extinction and dust emission measurements. When determining gas masses from dust emission, however, we are required to adopt an opacity law and specific dust opacities, both of which are still largely uncertain. Some works \citep[such as][]{Battersby2011}, adopt the \citet{Ossenkopf1994} specific opacity of $k_0 = 4$\,cm$^2$g$^{-1}$ at 505\,GHz, and an opacity law as $k_{\nu} = k_0 (\nu / 505 $GHz$)^{1.75}$. \citet{KauffmannPillai2010} also use the \citet{Ossenkopf1994} opacities, but include an additional correction of a factor 1.5, so that the mass estimates from dust extinction were consistent to those from dust emission at 1\,mm. In other words, the equivalent opacity law from \citet{KauffmannPillai2010} would be $k_{\nu} = 12.1$\,cm$^2$g$^{-1} (\nu / 1200 $GHz$)^{1.75}$ - and this is the opacity law for which the original threshold is applicable.  

For works that do not include this correction for their mass estimates \citep[such as the masses from][]{Battersby2011}, would therefore require to be compared to the equivalent threshold relation, without the 1.5 scaling, i.e. $M [$M$\odot]=580 (R [$pc$])^{1.33}$. The ATLASGAL works \citep[such as][]{urquhart2018_csc}, adopt a slightly different opacity law: they take the same specific opacities from \citet{Ossenkopf1994}, but without the 1.5 factor correction, and using a spectral index of 2 (instead of 1.75). With this change in the power law index plus the non-adoption of the 1.5 correction factor in the opacities, makes the ATLASGAL dust masses lower by roughly a factor 2 with respect to the masses from \citet{KauffmannPillai2010}. 

In our work, the gas masses were derived by calibrating the $^{13}$CO $(2-1)$ emission against the column density maps from the Herschel Hi-GAL survey \citep{Molinari2010}, which use a different opacity law from the above \citep{Elia2013}. It adopts the \citet{hildebrand1983} opacity at 250\,$\mu$m and a spectral index of 2, resulting in $k_{\nu} = 10$\,cm$^2$g$^{-1} (\nu / 1200 $GHz$)^2$. We note that our conversion factor from $^{13}$CO $(2-1)$ integrated intensities into H$_{2}$ column densities as per this comparison with the Hi-GAL data was also remarkably consistent with the conversion factor derived with a multi-transition modelling of the line emission \citep[combining SEDIGISM and THrUMMS data,][]{Schuller2017}.

At Herschel wavelengths, the difference between our opacity law and that used by \citet{KauffmannPillai2010} introduces only a small difference of $\sim20\%$ on the masses (well within the overall uncertainties in our mass estimates). Nevertheless, for consistency, we scale the \citet{KauffmannPillai2010} HMSF threshold line to match our particular opacity law, bringing the threshold to $M [$M$\odot]=1053 (R [$pc$])^{1.33}$. We note that the fraction of SEDIGISM clouds above and below the threshold line fluctuates only by $\sim 10\%$ if we were to adopt the original relation, thus not changing the global trends and results that we find.}


\bsp	
\label{lastpage}
\end{document}